\documentclass[11pt,a4paper]{article}
\usepackage[ansinew]{inputenc}
\usepackage{verbatim}
\usepackage{graphicx}
\usepackage{tabularx}
\usepackage{array}
\usepackage{delarray}
\usepackage{dcolumn}
\usepackage{float}
\usepackage{amsmath}
\usepackage{hhline}
\usepackage{amstext}
\usepackage{amssymb}
\usepackage{hyperref}
\usepackage{setspace}
\usepackage{amsthm}
\usepackage{makeidx}
\usepackage{url}

\RequirePackage[round,authoryear,longnamesfirst]{natbib}
\makeindex
\numberwithin{equation}{section}
\newtheorem{lem}{Lemma}[section]
\newtheorem{thm}{Theorem}[section]
\newtheorem{cor}{Corollary}[section]

\newtheorem{ass}{Assumption}

\newtheorem{ex}{Example}
\renewcommand{\citep}[1]{\citeauthor{#1}, \citeyear{#1}}

\renewcommand{\epsilon}{\varepsilon}
\DeclareMathOperator*{\argmax}{arg\,max}
\DeclareMathOperator*{\argmin}{arg\,min}
\makeatletter
\newcommand*{\rom}
[1]{\expandafter\@slowromancap\romannumeral #1@}
\makeatother
\numberwithin{equation}{section}
\allowdisplaybreaks

\usepackage{xr}
\begin{document}

\title{Detecting Latent Communities in Network Formation Models\thanks{%
		Ma's research was partially supported by NSF grants DMS 1712558 and DMS 2014221 and a UCR
		Academic Senate CoR grant. Zhang acknowledges the financial support from
		Singapore Ministry of Education Tier 2 grant under grant MOE2018-T2-2-169
		and the Lee Kong Chian fellowship. Any and all errors are our own.} \\
	\vspace{2mm} }
\author{Shujie Ma\thanks{%
		Department of Statistics, University of California, Riverside. \
		E-mail~address: shujie.ma@ucr.edu.} \and Liangjun Su\thanks{%
		School of Economics and Management, Tsinghua University.\ E-mail~address:
		sulj@sem.tsinghua.edu.cn.} \and Yichong Zhang\thanks{%
		School of Economics, Singapore Management University.\ E-mail~address:
		yczhang@smu.edu.sg.} }
\maketitle

\begin{abstract}
	This paper proposes a logistic undirected network formation model which
	allows for assortative matching on observed individual characteristics and
	the presence of edge-wise fixed effects. We model the coefficients of
	observed characteristics to have a latent community structure and the
	edge-wise fixed effects to be of low rank. We propose a multi-step
	estimation procedure involving nuclear norm regularization, sample
	splitting, iterative logistic regression and spectral clustering to detect
	the latent communities. We show that the latent communities can be exactly
	recovered when the expected degree of the network is of order $\log n$ or
	higher, where $n$ is the number of nodes in the network. The finite sample
	performance of the new estimation and inference methods is illustrated
	through both simulated and real datasets.\medskip
	
	 \bigskip
	
\end{abstract}

\section{Introduction}

In real world social and economic networks, individuals tend to form links
with someones who are alike to themselves, resulting in assortative matching
on observed individual characteristics (homophily). In addition, network
data often exhibit natural communities such that individuals in the same
community may share similar preferences for a certain type of homophily
while those in different communities tend to have quite distinctive
preferences. In many cases, such a community structure is latent and has to
be identified from the data. The detection of such community structures is
challenging yet crucial for network analyses. It prompts a couple of
important questions that need to be addressed: How do we formulate a network
formation model with individual characteristics, unobserved edge-wise fixed
effects, and latent communities?\ When the model is formulated, how do we
recover the community structure and estimate the community-specific
parameters effectively in the model?

To address the first issue above, we propose a logistic undirected network
formation model with observed measurements of homophily as regressors. We
allow the regression coefficients to have a latent community structure such
that the regression coefficient for covariate $l$ in the network formation
model is $B_{l,k_{1}k_{2}}$ for any nodes $i$ and $j$ in communities $k_{1}$
and $k_{2}$, respectively. The edge-wise fixed effects are assumed to have a
low-rank structure. This includes the commonly used discretized fixed
effects and additive fixed effects as special cases. To address the second
issue, we note that the estimation of this latent model is challenging, and
it has to involve a multi-step procedure. In the first step, we estimate the
coefficient matrices by a nuclear norm regularized logistic regression\
given their low-rank structures; we then obtain the estimators of their
singular vectors which contain information about the community memberships
via the singular value decomposition (SVD). Such singular vector estimates
are only consistent in Frobenius norms but not in\ uniform row-wise
Euclidean norm. A refined estimation is needed for accurate community
detection. In the second step, we use the singular vector estimates from the
first step as the initial values and iteratively run row-wise and
column-wise logistic regressions to reestimate the singular vectors.\textbf{%
	\ }The efficiency of the resulting estimator can be improved through this
iterative procedure. In the third step, we apply the standard K-means
algorithm to the singular vector estimates obtained in the second step. For
technical reasons, we have to resort to sample-splitting techniques to
estimate the singular vectors, and for numerical stability,\ both iterative
procedures and multiple-splits are called upon. We establish the exact
recovery of the latent community (strong consistency) under the condition
that the expected degree of the network diverges to infinity at the rate $%
\log n$ or higher order, where $n$ is the number of nodes. Under the exact
recovery property, we can treat the estimated community memberships as the
truth and further estimate the\ community-specific regression coefficients.

Our paper is closely related to three strands of literature in statistics
and econometrics. First, our paper is closely tied to the large literature
on the application of spectral clustering to detect communities in
stochastic block models (SBMs). Since the pioneering work of \cite{HLL83},
SBM has become the most popular model for community detection. The
statistical properties of spectral clustering in such models have been
studied by \cite{J15}, \cite{JY16}, \cite{LR15}, \cite{PC20}, \cite{QR13},
\cite{RCY11}, \cite{SB15}, \cite{ss15}, \cite{V18}, \cite{ww87}, \cite{YP14}%
, and \cite{YP16}, among others. From an information theory perspective,
\cite{AS15}, \cite{ABH2016}, \cite{MNS14}, and \cite{V18} establish the
phase transition threshold for the exact recovery of communities in SBMs,
which requires the expected degree to diverge to infinity at a rate no
slower than $\log n$. \cite{SWZ20} show that spectral clustering can achieve
this information-theoretical minimum rate for the exact recovery.
Nevertheless, most existing SBMs do not include covariates. A few exceptions
include \citep{B17}, \cite{Weng_Feng2016}, \cite{Yan_Sarkar2020} and \cite%
{Zhang_Levina_Zhu2016}, who consider covariates-assisted community detection
but not inferences on the underlying parameters. For more complicated models
that can incorporate both covariates and community structures, people often
resort to the variational EM algorithm, the performance of which highly
hinges on the proper choice of initial values. In contrast, the\ network
formation model proposed in this paper extends the SBM to a complex logistic
regression model with both latent community structures and covariates, and
our multi-step procedure provides an effective and reliable tool for the
estimation of such a complex network model. Despite the fact that the
regression coefficient matrices have to be estimated from the data in order
to obtain the associated singular vectors for spectral clustering, we are
able to obtain the exact recovery of the community structures at the minimal
rate on the expected node degree and to conduct inferences on the underlying
parameters in the model.%
\

Second, our paper is closely tied to the burgeoning literature on network
formation models and panel structure models. For the former, see \cite{CDS11}%
, \cite{G17}, \cite{Graham2019}, \cite{Graham2020}, \cite%
{Graham_de_Paula2019}, \cite{HL83}, \cite{Hoff_Raftery_Handcock2002}, \cite%
{J19}, \cite{L15}, \cite{M17}, \cite{RPF13}, and \cite{YX13}. We complement
these works by allowing for community structures on the regression
coefficients, which can capture a rich set of unobserved heterogeneity in
the network data. In a working paper, \cite{M2017b} also considers a network
formation model with heterogeneous players and latent community structure.
He\ assumes that the community structure follows an i.i.d. multinomial
distribution and imposes a prior distribution over communities and
parameters before conducting Bayesian estimation and inferences. In
contrast, we treat the community memberships as\ fixed parameters and aim to
recover them from a single observation of a large network. Our idea of
introducing the community structure into the network formation model is
mainly inspired by the recent works of \cite{BM15} and \cite{SSP16}, who
introduce latent group structures into panel data analyses. When the
community structure is unobserved, it is analogous to the latent group
structure in panel data models. For recent analyses of panel data models
with latent group structures, see \cite{Ando_Bai2016}, \cite{Chen2019}, \cite%
{CSS2019}, \cite{Dzemski_Okui2018}, \cite{HJS2020}, \cite{HJPS2021}, \cite%
{LSZZ2020}, \cite{Lu_Su2017}, \cite{Su_Ju2018}, \cite{SWJ2019}, \cite%
{Vogt_Linton2020}, \cite{Wang_Su2021}, and \cite{Xu_Yue_Zhang2020}, among
others. In particular, \cite{Wang_Su2021} establish the connection between
SBMs and panel data models with latent group structures and propose to adopt
the spectral clustering techniques to recover the latent group structures in
panel data models.

Last, our paper is related to the literature on the use of nuclear norm
regularization in various contexts; see \cite{aal20}, \cite{B19}, \cite{BN19}%
, \cite{CHLZ18}, \cite{FGZ2019}, \cite{F19}, \cite{KLT2011}, \cite{M18},
\cite{NW11}, \cite{NRWY2012}, and \cite{RT2011}, among others. All these
previous works focus on the error bounds (in Frobenius norm) for the nuclear
norm regularized estimates, except \cite{M18} and \cite{CHLZ18} who study
the inference problem in linear panel data models with a low-rank structure.
Like \cite{M18} and \cite{CHLZ18}, we simply use the nuclear norm
regularization to obtain consistent initial estimates. Unlike \cite{M18} and
\cite{CHLZ18}, we study a logistic network formation model with a latent
community structure and propose the iterative row- and column-wise logistic
regressions to improve the error bounds (in row-wise Euclidean norm) for the
singular vectors of the nuclear norm regularized estimates. Relying on such
an improvement, we can fully recover the community memberships. Then, we can
estimate the community specific parameters and make statistical inferences.

The rest of the paper is organized as follows. In Section \ref{sec:setup},
we introduce the model and basic assumptions. In Section \ref{sec:estimation}%
, we provide our multi-step estimation procedure. Section \ref{sec:sp}
establishes the statistical properties of the proposed estimators of the
singular vectors. Section \ref{sec:kmeans'} studies the K-means estimation
of the community memberships when the regression coefficient matrix is
assumed to exhibit some community structure. Section \ref{sec:inferb}
studies the asymptotic properties of the regression coefficient estimates in
the presence of latent community structures. Section \ref{sec:determine_Kl}
discusses the determination of the ranks of the regression coefficient
matrices. Section \ref{sec:sim} reports simulation results. In Section \ref%
{sec:app}, we apply the new methods to study the community structure of a
Facebook friendship networks at one hundred American colleges and
universities at a single time point. Section \ref{sec:concl} concludes. The
online supplement provides the proofs of all theoretical results and the
associated technical lemmas, and some additional technical details.

Notation. Throughout the paper, we write \textquotedblleft w.p.a.1" for
\textquotedblleft with probability approaching one," $M=\{M_{ij}\}$ as a
matrix with its $(i,j)$-th entry denoted as $M_{ij}$. We use $||\cdot
||_{op} $, $||\cdot ||_{F}$, and $||\cdot ||_{\ast }$ to denote matrix
spectral, Frobenius, and nuclear norms, respectively. We use $[n]$ to denote
$\{1,\cdots ,n\}$ for some positive integer $n$. For a vector $u$, $||u||$
and $u^{\top }$ denote its $L_{2}$ norm and transpose, respectively. For a
vector $a=(a_{1},\cdots ,a_{n})$, let $\text{diag}(a)$ be the diagonal
matrix whose diagonal is $a$. For a symmetric matrix $B\in \mathbb{R}%
^{K\times K}$, we define
\begin{equation*}
\text{vech}(B)=(B_{11},...,B_{1K},B_{22},...,B_{2K},\cdots
,B_{K-1,K-1},B_{K-1,K},B_{KK})^{\top }.
\end{equation*}%
We define $\max (u,v)=u\vee v$ and $\min (u,v)=u\wedge v$ for two real
numbers $u$ and $v$. We write $\mathbf{1}\{A\}$ to denote the usual
indicator function that takes value 1 if event $A$ happens and 0 otherwise.
Let $\odot $ denote Hadamard product.

%

\section{The Model and Basic Assumptions}

\label{sec:setup}

In this section, we introduce the model and basic assumptions.

\subsection{The Model}

For $i\neq j\in \left[ n\right] $, let $Y_{ij}$ denote the dummy variable
for a link between nodes $i$ and $j$. It takes value 1 if nodes $i$ and $j$
are linked and 0 otherwise. Let $W_{ij}=(W_{1,ij},...,W_{p,ij})^{\top }$
denote a $p$-vector of measurements of homophily between nodes $i$ and $j$.
Researchers observe the network adjacency matrix $\{Y_{ij}\}$ and covariates
$\{W_{ij}\}$. We model the link formation between $i$ and $j$ is as
\begin{equation}
Y_{ij}=\mathbf{1}\{\varepsilon _{ij}\leq \log \zeta
_{n}+\sum_{l=0}^{p}W_{l,ij}\Theta _{l,ij}^{\ast }\},\text{ }i<j,
\label{Model1}
\end{equation}%
where $\{\zeta _{n}\}_{n\geq 1}$ is a deterministic sequence that may decay
to zero and is used to control the expected degree in the network, $%
W_{0,ij}=1,$ and $W_{l,ij}=W_{l,ji}$ for $j\neq i$ and $l\in [ p]$. For
clarity, we consider undirected network so that $Y_{ij}=Y_{ji}$ and $\Theta
_{l,ij}^{\ast }=\Theta _{l,ji}^{\ast }$ $\forall l$ if $i\neq j,$ $%
\varepsilon _{ij}$ follows the standard logistic distribution for $i<j$, and
$\varepsilon _{ij}=\varepsilon _{ji}$. Let $Y_{ii}=0$ for all $i\in \left[ n%
\right] $.

Apparently, without making any assumptions on $\Theta _{l}^{\ast }=\{\Theta
_{l,ij}^{\ast }\}$ for $l\in \left[ p\right] \cup \{0\},$ one cannot
estimate all the parameters in (\ref{Model1}) as the number of parameters
can easily exceed the number of observations in the model. Specifically, we
will follow the literature on reduced rank regressions and assume that each $%
\Theta _{l}^{\ast }$ exhibits a certain low rank structure. Even so, it is
easy to see that our model in \eqref{Model1} is fairly general, and it
includes a variety of network formation models as special cases.

\begin{enumerate}
	\item If $\log(\zeta _{n})=2\bar{a} = \frac{2}{n}\sum_{i=1}^n a_i$, $%
	\alpha_i = a_i - \bar{a}$, $\Theta _{0,ij}^{\ast }=\alpha _{i}+\alpha _{j},$
	and $p=0,$ then
	\begin{equation}
	Y_{ij}=\mathbf{1}\{\varepsilon _{ij}\leq a _{i}+a _{j}\}.\text{ }
	\label{Model2}
	\end{equation}%
	Under the standard logistic\ distribution assumption on $\varepsilon _{ij},$
	$\mathbb{P}\left( Y_{ij}=1\right) =\frac{\exp \left( a _{i}+a _{j}\right) }{%
		1+\exp \left( a _{i}+a _{j}\right) }$ for all $i\neq j,$ and we have the
	simplest exponential graph model (Beta model) considered in the literature;
	see, e.g.,\ \cite{LKR2013book}.
	
	\item If $\log(\zeta _{n})$ and $\Theta _{0,ij}^{\ast }$ are defined as
	above and $\Theta _{l,ij}^{\ast }=\beta _{l}$ for $l\in [ p],$ then
	\begin{equation}
	Y_{ij}=\mathbf{1}\{\varepsilon _{ij}\leq a _{i}+a _{j}+W_{ij}^{\top }\beta
	\},\text{ }  \label{Model3}
	\end{equation}%
	where $\beta =(\beta _{1},...,\beta _{p})^{\top }$. Apparently, (\ref{Model3}%
	) is the undirected dyadic link formation model with degree heterogeneity
	studied in \cite{G17}. See also \cite{YJFL2019} for the case of a directed
	network.
	
	\item Let $\Theta _{0,ij}=\Theta _{0,ij}^{\ast }+\log \zeta _{n}.$ If $p=0,$
	and $\Theta _{0}=\{\Theta _{0,ij}\}\ $is assumed to exhibit a stochastic
	block structure such that $\Theta _{0,ij}=b_{kl}$ if nodes $i$ and $j$
	belong to communities $k$ and $l,$ respectively, then we have%
	\begin{equation}
	Y_{ij}=\mathbf{1}\{\varepsilon _{ij}\leq \Theta _{0,ij}\}.  \label{Model4}
	\end{equation}%
	Corresponding to the simple SBM with $K$ communities, the probability matrix
	$P=\left\{ P_{ij}\right\} $ with $P_{ij}=\mathbb{P}\left( Y_{ij}=1\right) \ $%
	can be written as $P=ZBZ^{\top }$ where $Z=\{Z_{ik}\}$ denotes an $n\times K$
	binary matrix providing the cluster membership of each node, i.e., $Z_{ik}=1$
	if node $i$ is in community $k$ and $Z_{ik}=0$ otherwise, and $B=\left\{
	B_{kl}\right\} $ denotes the block probability matrix that depends on $%
	b_{kl}.$ See \cite{HLL83} and the references cited in the introduction
	section.
	
	\item Let $\Theta _{0,ij}=\Theta _{0,ij}^{\ast }+\log \zeta _{n}.$ If $%
	\Theta _{0}=\{\Theta _{0,ij}\}\ $is assumed to exhibit the stochastic block
	structure such that $\Theta _{0,ij}=b_{kl}$ if nodes $i$ and $j$ belong to
	communities $k$ and $l,$ respectively, and $\Theta _{l,ij}^{\ast }=\beta
	_{l} $ for $l\in [ p],$ then
	\begin{equation}
	Y_{ij}=\mathbf{1}\{\varepsilon _{ij}\leq \Theta _{0,ij}+W_{ij}^{\top }\beta
	\}.\text{ }  \label{Model5}
	\end{equation}%
	Then (\ref{Model5}) defines a stochastic block model with covariates
	considered in \cite{Sweet2015}, \cite{Leger2016}, and \cite{RAM19}.
\end{enumerate}

Under the assumptions specified in the next subsection, it is easy to see
that the expected degree of the network is of order $n\zeta _{n}.$ In the
theory to be developed below, we allow $\zeta _{n}$ to shrink to zero at a
rate as slow as $n^{-1}\log n$, so that the expected degree can be as small
as $C\log n$ for some sufficiently large constant $C$ and the network is
semi-dense.\footnote{%
	A network is dense if the expected degree grows at rate-$n$ and semi-dense
	if it diverges to infinity at a rate slower than $n$.} Of course, if $\zeta
_{n}$ is fixed or convergent to a positive constant as $n\rightarrow \infty
, $ the network becomes dense.

To proceed, let $\tau _{n}=\log (\zeta _{n})$, $\Gamma _{0,ij}^{\ast }=\tau
_{n}+\Theta _{0,ij}^{\ast }$, $\Gamma _{ij}^{\ast }=(\Gamma _{0,ij}^{\ast
},\Theta _{1,ij}^{\ast },...,\Theta _{p,ij}^{\ast })^{\top }$, and $%
W_{ij}=(W_{0,ij},W_{1,ij},...,W_{p,ij})^{\top }$, where $W_{0,ij}=1$. Let $%
\Gamma ^{\ast }=(\Gamma _{0}^{\ast },\Theta _{1}^{\ast },...,\Theta
_{p}^{\ast }),$ where $\Gamma _{0}^{\ast }=\{\Gamma _{0,ij}^{\ast }\}$ and $%
\Theta _{l}^{\ast }=\{\Theta _{l,ij}^{\ast }\}$ for $l\in \left[ p\right] .$
Then, we can rewrite the model in (\ref{Model1}) as
\begin{equation}
Y_{ij}=\mathbf{1}\{\varepsilon _{ij}\leq W_{ij}^{\top }\Gamma _{ij}^{\ast
}\}.  \label{eq:Yij}
\end{equation}%
Below, we will let $\Gamma _{l}^{\ast }=\Theta _{l}^{\ast }$ for $l\in \left[
p\right] $ and impose some basic assumptions on the model in order to
propose a multiple-step procedure to estimate the parameters of interest in
the model.

\subsection{Basic Assumptions}

Now, we state a set of basic assumptions to characterize the model in %
\eqref{Model1}. The first assumption is about the data generating process
(DGP).

\begin{ass}
	\begin{enumerate}
		\item For $l\in \left[ p\right] $, there exists a function $g_{l}(\cdot )$
		such that $W_{l,ij}=g_{l}(X_{i},X_{j},e_{ij})$, where $g_{l}(\cdot ,\cdot
		,e) $ is symmetric in its first two arguments, $\{X_{i}\}_{i=1}^{n}$ and $%
		\{e_{ij}\}_{1\leq i<j\leq n}$ are two independent and identically
		distributed (i.i.d.) sequences of random variables, and $e_{ij}=e_{ji}$ for $%
		i\neq j$.
		
		\item $\{\varepsilon _{ij}\}_{1\leq i<j\leq n}$ is an i.i.d. sequence of
		logistic random variables. Moreover, $\{\varepsilon _{ij}\}_{1\leq i<j\leq
			n}\perp \!\!\!\perp (\{X_{i}\}_{i=1}^{n}\cup \{e_{ij}\}_{1\leq i<j\leq n})$.
		Let $\varepsilon _{ij}=\varepsilon _{ji}$ for $i>j.$
		
		\item $\max_{l\in \left[ p\right] }\max_{i\neq j\in \left[ n\right]
		}|W_{l,ij}|\leq M_{W}$ for some constant $M_{W}<\infty .$
	\end{enumerate}
	
	\label{ass:dgp}
\end{ass}

Assumption \ref{ass:dgp} specifies how the covariates and error terms are
generated. In some applications, $e_{ij}$ is absent and $W_{l,ij}$ depend on
$(X_{i},X_{j})$ only. For example, $W_{l,ij}=\left\Vert
X_{i}-X_{j}\right\Vert $ for some $l$. We further assume that it is
uniformly bounded to simplify the analysis. Assumption \ref{ass:dgp}.2 is
standard.

The next assumption imposes some structures on $\left\{ \Theta _{l}^{\ast
}\right\} _{0\leq l\leq p}.$

\begin{ass}
	\begin{enumerate}
		\item Suppose $\sum_{i,j\in \lbrack n]}\Theta _{0,ij}^{\ast }=0.$
		
		\item Suppose $\Theta _{l}^{\ast }$ is symmetric and of low rank $K_{l}$ for
		$l\in \lbrack p]\cup \{0\}$. The singular value decomposition of $%
		n^{-1}\Theta _{l}^{\ast }$ is $\mathcal{U}_{l}\Sigma _{l}\mathcal{V}%
		_{l}^{\top }$, where $\mathcal{U}_{l}$ and $\mathcal{V}_{l}$ are $n\times
		K_{l}$ matrices such that $\mathcal{U}_{l}^{\top }\mathcal{U}_{l}=I_{K_{l}}=%
		\mathcal{V}_{l}^{\top }\mathcal{V}_{l}$ and $\Sigma _{l}=\text{diag}(\sigma
		_{1,l},\cdots ,\sigma _{K_{l},l})$ with singular values $\sigma _{1,l}\geq
		\cdots \geq \sigma _{K_{l},l}$. We further denote $U_{l}=\sqrt{n}\mathcal{U}%
		_{l}\Sigma _{l}$ and $V_{l}=\sqrt{n}\mathcal{V}_{l}$. Then,
		\begin{equation}
		\Theta _{l}^{\ast }=n\mathcal{U}_{l}\Sigma _{l}\mathcal{V}_{l}^{\top
		}=U_{l}V_{l}^{\top }\text{ for }l=0,...,p.  \label{eq:Thetal}
		\end{equation}%
		Let $u_{i,l}^{\top }$ and $v_{i,l}^{\top }$ denote the $i$-th row of $U_{l}$
		and $V_{l}$, respectively for $l\in \left[ p\right] \cup \{0\}$. Then, $%
		\max_{i\in \lbrack n],l\in \lbrack p]}(||u_{i,l}||\vee ||v_{i,l}||)\leq M$
		for some constant $M<\infty $ and there are constants $C_{\sigma }$ and $%
		c_{\sigma }$ such that
		\begin{equation*}
		\infty >C_{\sigma }\geq \limsup_{n}\max_{l\in \left[ p\right] \cup
			\{0\}}\sigma _{1,l}\geq \liminf_{n}\min_{l\in \left[ p\right] \cup
			\{0\}}\sigma _{K_{l},l}\geq c_{\sigma }>0.
		\end{equation*}
	\end{enumerate}
	
	\label{ass:par}
\end{ass}

%
%


We note \eqref{eq:Thetal} implies that $\Theta _{l,ij}^{\ast }=u_{i,l}^{\top
}v_{j,l}.$ We view $\Theta _{0,ij}^{\ast }$ as the edge-wise fixed effects
for the network formation model. We impose the normalization that $%
\sum_{i,j\in \lbrack n]}\Theta _{0,ij}^{\ast }=0$ in the first part of
Assumption \ref{ass:par} because we have included the grand intercept term $%
\tau _{n}(\equiv \log (\zeta _{n}))$ in (\ref{Model1}). The low-rank
structure of $\Theta _{l}^{\ast }$ incorporates two special cases: (1)
additive structure and (2) latent community structure, as illustrated in
detail in Examples \ref{ex:aij} and \ref{ex:t0group} below, respectively.
When there are no covariates in regression and $\Theta _{0}$ belongs to the
two cases in Examples \ref{ex:aij} and \ref{ex:t0group}, the model becomes
the so-called Beta model and stochastic block model, respectively. We extend
these models to the scenario with edge-wise characteristics and latent
community structure for the slope coefficients.

\begin{ex}
	\label{ex:aij} Let $\Theta _{l,ij}^{\ast }=\alpha _{l,i}+\alpha _{l,j}$. In
	this case, $K_{l}=2$ and $n^{-1}\Theta _{l}^{\ast }=\mathcal{U}_{l}\Sigma
	_{l}\mathcal{V}_{l}^{T},$ where
	\begin{equation*}
	\mathcal{U}_{l}=%
	\begin{pmatrix}
	\frac{1}{\sqrt{2n}}(1+\frac{\alpha _{l,1}}{s_{l,n}}) & \frac{-1}{\sqrt{2n}}%
	(1-\frac{\alpha _{l,1}}{s_{l,n}}) \\
	\vdots & \vdots \\
	\frac{1}{\sqrt{2n}}(1+\frac{\alpha _{l,n}}{s_{l,n}}) & \frac{-1}{\sqrt{2n}}%
	(1-\frac{\alpha _{l,n}}{s_{l,n}})%
	\end{pmatrix}%
	,\text{ }\mathcal{V}_{l}=%
	\begin{pmatrix}
	\frac{1}{\sqrt{2n}}(1+\frac{\alpha _{l,1}}{s_{l,n}}) & \frac{1}{\sqrt{2n}}(1-%
	\frac{\alpha _{l,1}}{s_{l,n}}) \\
	\vdots & \vdots \\
	\frac{1}{\sqrt{2n}}(1+\frac{\alpha _{l,n}}{s_{l,n}}) & \frac{1}{\sqrt{2n}}(1-%
	\frac{\alpha _{l,n}}{s_{l,n}})%
	\end{pmatrix}%
	,\text{ }\Sigma _{l}=%
	\begin{pmatrix}
	s_{l,n} & 0 \\
	0 & s_{l,n}%
	\end{pmatrix}%
	,
	\end{equation*}%
	and $s_{l,n}^{2}=\frac{1}{n}\sum_{i=1}^{n}\alpha _{i}^{2}.$ Similarly, it is
	easy to verify that
	\begin{equation*}
	U_{l}=%
	\begin{pmatrix}
	\frac{1}{\sqrt{2}}(s_{l,n}+\alpha _{l,n}) & \frac{-1}{\sqrt{2}}%
	(s_{l,n}-\alpha _{l,n}) \\
	\vdots & \vdots \\
	\frac{1}{\sqrt{2}}(s_{l,n}+\alpha _{l,n}) & \frac{-1}{\sqrt{2}}%
	(s_{l,n}-\alpha _{l,n})%
	\end{pmatrix}%
	\text{ and }V_{l}=\sqrt{n}\mathcal{V}_{l}.
	\end{equation*}%
	When $l=0$, we further impose $\sum_{i,j\in \lbrack n]}\Theta _{0}^{\ast }=0$%
	, which implies $\sum_{i=1}^{n}\alpha _{0,i}=0$. We also allow $\{\alpha
	_{l,i}\}_{i=1}^{n}$ to depend on $\{W_{ij}\}_{1\leq i<j\leq n}$ so that $%
	\{\alpha _{l,i}\}_{i=1}^{n}$ are usually referred to as individual fixed
	effects in the literature.
\end{ex}

\begin{ex}
	\label{ex:t0group} Let $\Theta _{l}^{\ast }=Z_{l}B_{l}^{\ast }Z_{l}^{\top }$%
	, where $Z_{l}\in \mathbb{R}^{n\times K_{l}}$ is the membership matrix with
	one entry in each row taking value one and the rest taking value zero, $%
	K_{l} $ denotes the number of distinctive communities for $\Theta _{l}^{\ast
	} $, and $B_{l}^{\ast }\in \mathbb{R}^{K_{l}\times K_{l}}$ is symmetric with
	rank $K_{l}$. Let $p_{l}^{\top }=(\frac{n_{1,l}}{n},\cdots ,\frac{n_{K_{l},l}%
	}{n}) $ and $n_{k,l}$ denotes the size of $\Theta _{l}^{\ast }$'s $k$-th
	community for $k\in [ K_{l}]$. Then, as Lemma \ref{lem:theta} below shows,
	\begin{equation*}
	U_{l}=Z_{l}^{\top }(\Pi _{l,n})^{-1/2}S_{l}^{\prime }\Sigma _{l}\quad \text{%
		and}\quad V_{l}=Z_{l}^{\top }(\Pi _{l,n})^{-1/2}S_{l},
	\end{equation*}%
	where $S_{l}$ and $S_{l}^{\prime }$ are two $K_{l}\times K_{l}$ matrices
	such that $S_{l}^{\top }S_{l}=I_{K_{l}}=(S_{l}^{\prime })^{\top
	}S_{l}^{\prime }$, $\Pi _{l,n}=\text{diag}(p_{l})$, and $\Sigma _{l}$ is the
	singular value matrix of $\Pi _{l,n}^{1/2}B_{l}^{\ast }\Pi _{l,n}^{1/2}$.
	Let $\iota _{n}$ denote an $n\times 1$ vector of ones. If $l=0$, we further
	impose that $\iota _{n}^{\top }Z_{0}B_{0}^{\ast }Z_{0}^{\top }\iota
	_{n}=p_{0}^{\top }B_{0}^{\ast }p_{0}=0$.
\end{ex}

For classification and inference, we need to impose the latent community
structure as in Example \ref{ex:t0group}. This is summarized in the
following assumption.

\begin{ass}
	\begin{enumerate}
		\item $\Theta _{l}^{\ast }=Z_{l}B_{l}^{\ast }Z_{l}^{\top }$, where $Z_{l}\in
		\mathbb{R}^{n\times K_{l}}$.
		
		\item There exist some constants $C_{1}$ and $c_{1}$ such that
		\begin{equation*}
		\infty >C_{1}\geq \limsup_{n}\max_{k\in \left[ K_{l}\right] ,\text{ }l\in %
			\left[ p\right] }\pi _{l,kn}\geq \liminf_{n}\min_{k\in \left[ K_{l}\right] ,%
			\text{ }l\in \left[ p\right] }\pi _{l,kn}\geq c_{1}>0.
		\end{equation*}
	\end{enumerate}
	
	\label{ass:pi}
\end{ass}

Two remarks are in order. First, Assumption \ref{ass:pi} implies that if $%
\Theta_l^{\ast}$ has a latent community structure, the size of each
community must be proportional to the number of nodes $n$. Such an
assumption is common in the literature on network community detection and
panel data latent structure detection. Second, it is possible to allow for $%
\pi _{l,kn}$ and/or $\sigma _{k,l}$ to vary with $n$. In this case, one just
needs to keep track of all these terms in the proofs.

To proceed, we state a lemma that shows Assumption \ref{ass:pi} is a special
case of Assumption \ref{ass:par} and lays down the foundation for our
classification procedure Section \ref{sec:kmeans}.

\begin{lem}
	\label{lem:theta} Suppose Assumption \ref{ass:par} holds. Then,
	
	\begin{enumerate}
		\item $V_{l}=Z_{l}(\Pi _{l,n})^{-1/2}S_{l}$ and $U_{l}=Z_{l}(\Pi
		_{l,n})^{-1/2}S_{l}^{\prime }\Sigma _{l}$ for $l\in \left[ p\right] $, where
		$S_{l}$ and $S_{l}^{\prime }$ are two $K_{l}\times K_{l}$ matrices such that
		$S_{l}^{\top }S_{l}=I_{K_{l}}=(S_{l}^{\prime })^{\top }S_{l}^{\prime }$.
		
		\item $\max_{j\in [ n]}||v_{j,l}||\leq c_{1}^{-1/2}<\infty $ and $\max_{i\in
			[ n]}||u_{i,l}||\leq c_{1}^{-1/2}C_{\sigma }<\infty $.
		
		\item If $z_{i,l}\neq z_{j,l}$, then $\left\Vert \frac{v_{i,l}}{||v_{i,l}||}-%
		\frac{v_{j,l}}{||v_{j,l}||}\right\Vert =||(z_{i,l}-z_{j,l})S_{l}||=\sqrt{2}$.
	\end{enumerate}
\end{lem}

Lemma \ref{lem:theta} implies that, if $\Theta_l^{\ast}$ for some $l \in [p]
\cup \{0\}$ has the community structure, its singular vectors $%
\{v_{i,l}\}_{i \in [n]}$ contain information about the community structure.
A similar result has been established in the community detection literature;
see, e.g., \citet[Lemma 3.1]{RCY11} and \citet[Theorem II.1]{SWZ20}.

In Section \ref{sec:sp}, we only require $\Theta _{l}^{\ast }$, $l\in
\lbrack p]\cup \{0\}$ to be of low-rank and derive the uniform convergence
rate of the estimators of $(u_{i,l},v_{i,l})$ across $i\in \lbrack n]$. In
Section \ref{sec:kmeans'}, we further suppose that \textit{some} coefficient
$\Theta _{l}^{\ast }$ has a special community structure as in Assumption \ref%
{ass:pi} and apply the K-means algorithm to exactly recover their group
identities. Last, for inference in Section \ref{sec:inferb}, we impose that
\textit{all} coefficients $\Theta _{l}^{\ast }$, $l\in \lbrack p]$ have
(potentially different) community structures while $\Theta _{0}^{\ast }$
follows the structure in either Example \ref{ex:aij} or \ref{ex:t0group}.

\section{The Estimation Algorithm\label{sec:estimation}}

For notational simplicity, we will focus on the case of $p=1$. The general
case with multiple covariates involves fundamentally no new ideas but more
complicated notations.

First, we recognize that $\Gamma _{0}^{\ast }$ and $\Gamma _{1}^{\ast }$ are
both low rank matrices with ranks bounded from above by $K_{0}+1$ and $K_{1}$%
, respectively. We can obtain their preliminary estimates via the nuclear
norm penalized logistic regression. Second, based on the normalization
imposed in Assumption \ref{ass:par}.1, we can estimate $\tau _{n}$ and $%
\Theta _{0}^{\ast }$ separately. We then apply the SVD to the preliminary
estimates of $\Theta _{0}^{\ast }$ and $\Theta _{1}^{\ast }$ and obtain the
estimates of $U_{l}$, $\Sigma _{l}$, and $V_{l}$, $l=0,1$. Third, we plug
back the second step estimates of $\{V_{l}\}_{l=0,1}$ and re-estimate each
row of $U_{l}$ by a row-wise logistic regression. We can further iterate
this procedure and estimate $U_{l}$ and $V_{l}$ alternatively. Last, if we
further impose $\Theta_1^*$ has a community structure, then we can apply the
K-means algorithm to the final estimate of $V_{1}$ to recover the community
memberships. We rely on a sample splitting technique along with the
estimation. Throughout, we assume the ranks $K_{0}$ and $K_{1}$ are known.
We will propose an singular-value-ratio-based criterion to select them in
Section \ref{sec:determine_Kl}.

Below is an overview of the multi-step estimation procedure that we propose.

\begin{enumerate}
	\item Using the full sample, run the nuclear norm regularized estimation
	twice as detailed in Section \ref{sec:fslre} and obtain $\widehat{\tau }_{n}$
	and $\{\widehat{\Sigma }_{l}\}_{l=0,1}$, the preliminary estimates of $\tau
	_{n}$ and $\{\Sigma _{l}\}_{l=0,1}.$
	
	\item Randomly split the nodes into two subsets, denoted as $I_{1}$ and $%
	I_{2}$. Using edges $(i,j)\in I_{1}\times [ n]$, run the nuclear norm
	estimation twice as detailed in Section \ref{sec:sslre} and obtain $\{%
	\widehat{V}_{l}^{(1)}\}_{l=0,1},$ a preliminary estimate of $%
	\{V_{l}\}_{l=0,1}$, where the superscript $(1)$ means we use the first
	subsample to conduct the nuclear norm estimation. For $j\in [ n],$ denote
	the $j$-th row of $\widehat{V}_{l}^{(1)}$ as $(\widehat{v}%
	_{j,l}^{(1)})^{\top },$ which is a preliminary estimate of $v_{j,l}^{\top }.$
	
	\item For each $i\in I_{2},$ take $\{\widehat{v}_{j,l}^{(1)}\}_{j\in
		I_{2},l=0,1}$ as regressors and run the row-wise logistic regression to
	obtain $\{\widehat{u}_{i,l}^{(1)}\}_{l=0,1},$ the estimates of $%
	\{u_{i,l}\}_{l=0,1}$. For each $j\in [ n],$ take $\{\widehat{u}%
	_{i,l}^{(1)}\}_{i\in I_{2},l=0,1}$ as regressors and run the column-wise
	logistic regression to obtain updated estimates, $\{\dot{v}%
	_{j,l}^{(0,1)}\}_{l=0,1}$ of $\{v_{j,l}\}_{l=0,1}$, where $0$ in the
	superscript $(0,1)$ means it is the $0$-th step estimator for the full
	sample iteration below and $1$ in the superscript means it is computes when
	the first subsample is used for the nuclear norm estimation. See Section \ref%
	{sec:ssrwlr} for details.
	
	\item Based on $\{\dot{v}_{j,l}^{(0,1)}\}_{j\in \left[ n\right] ,l=0,1}$,
	obtain the iterative estimates $(\dot{u}_{i,0}^{(h,1)},\dot{u}%
	_{i,1}^{(h,1)})_{i\in [ n]}$ and $(\dot{v}_{j,0}^{(h,1)},\dot{v}%
	_{j,1}^{(h,1)})_{j\in [ n]}$ of the singular vectors as in Step 3 for $%
	h=1,2,\cdots ,H$. See Section \ref{sec:fsi} for details.
	
	\item Switch the roles of $I_{1}$ and $I_{2}$ and repeat Steps 2--4 to
	obtain $(\dot{u}_{i,0}^{(h,2)},\dot{u}_{i,1}^{(h,2)})_{i\in [ n]}$ and $(%
	\dot{v}_{j,0}^{(h,2)},\dot{v}_{j,1}^{(h,2)})_{j\in [ n]}$ for $h\in [ H]$,
	where $h$ in the superscript $(h,2)$ means it is the $h$-th step iteration
	of the full sample estimator and $2$ in the superscript means the second
	subsample is used for the nuclear norm estimation.
	
	\item Let $\overline{v}_{j,1}=\left( \frac{(\dot{v}_{j,1}^{(H,1)})^{\top }}{%
		||\dot{v}_{j,1}^{(H,1)}||},\frac{(\dot{v}_{j,1}^{(H,2)})^{\top }}{||\dot{v}%
		_{j,1}^{(H,2)}||}\right) ^{\top }$. Then, apply the K-means algorithm on $\{%
	\overline{v}_{j,1}\}_{j\in [ n]}$ to recover the community memberships in $%
	\Theta _{1}^{\ast }$ as detailed in Section \ref{sec:kmeans}.
\end{enumerate}

Several remarks are in order. First, $\widehat{\tau }_{n}$ and $\{\widehat{%
	\Sigma }_{l}\}_{l=0,1}$ obtained in Step 1 are used in Steps 3-5 and to
determine $\left\{ K_{l}\right\} _{l=0,1}$ in Section \ref{sec:determine_Kl}%
. Second, we employ the sample-splitting technique to create independence
between the edges used for Steps 2 and 3. As $\widehat{V}_{l}^{(1)}$ in Step
2 is estimated by the nuclear-norm regularized logistic regression, we can
only control the estimation error in Frobenius norm, as shown in Theorem \ref%
{thm:Frobeniusnorm}. On the other hand, to analyze the row-wise estimator,
we need to control for the\ estimation error of $\widehat{V}_{l}^{(1)}$ in
row-wise $L_{2}$ norm (denoted as $||\cdot ||_{2\rightarrow \infty }$). We
overcome the discrepancy between $||\cdot ||_{F}$ and $||\cdot
||_{2\rightarrow \infty }$ by the independence structure. Third, one may
propose to use each row of the full-sample lower-rank estimator $\widehat{V}%
_{l}$ as $\{\dot{v}_{j,l}^{0}\}_{j\in \lbrack n]}$, the initial estimates in
Step 4. However, as $\widehat{V}_{l}$ is estimated using the full sample, it
is not independent of, say, the $i$-th row of the edges if we want to
estimate $(u_{i,0}^{\top },u_{i,1}^{\top })$. Fourth, in the literature,
researchers overcome this difficulty by using the \textquotedblleft
leave-one-out\textquotedblright\ technique. See, for example, \cite{abbe2017}%
, \cite{B13}, \cite{JM15}, \cite{SWZ20}, and \cite{Z18}, among others.
Denote $\widehat{\Theta }_{l}^{(i)}$ as the low-rank estimator of $\Theta
_{l}^{\ast }$ using all the edges except those on the $i$-th row and column
and $\widehat{V}_{l}^{(i)}$ is obtained by applying the SVD on $\widehat{%
	\Theta }_{l}^{(i)}$. The key step for the \textquotedblleft
leave-one-out\textquotedblright\ technique is to establish a perturbation
theory to bound $\widehat{\Theta }_{l}^{(i)}-\widehat{\Theta }_{l}$, and
thus, $\widehat{V}_{l}^{(i)}-\widehat{V}_{l}$. However, unlike the community
detection literature, $\widehat{\Theta }_{l}$ and $\widehat{\Theta }%
_{l}^{(i)}$ are not directly observed but estimated by the nuclear-norm
regularized logistic regression. It is interesting but very challenging, if
possible, to establish such a perturbation theory. Fifth, although the
sample-splitting can result in information loss, we compensate it in three
aspects: (1) we just treat the sample-split estimator $\dot{v}_{j,l}^{(0,1)}$
as an initial value and in Step 4, we update it via an iterative algorithm
which uses all the edges; (2) we can switch the roles of $I_{1}$ and $I_{2}$
and obtain $\dot{v}_{j,l}^{(H,1)}$ and $\dot{v}_{j,l}^{(H,2)}$ after $H$
iterations; (3) to mitigate the concern of the randomness caused by a single
sample split, in Section \ref{sec:kmeans}, we propose to repeat the
sample-splitting $R$ times to obtain $R$ classifications, and select one of
them based on the maximum-likelihood principle.

\subsection{The Estimation of $(u_{i,l},v_{i,l})$}

In the estimation of $(u_{i,l},v_{i,l})$ (see Steps 1--5 in the above
procedure), we only require that $\Theta _{0}^{\ast }$ and $\Theta
_{1}^{\ast }$ be of low-rank.

\subsubsection{Full-Sample Low-Rank Estimation \label{sec:fslre}}

Recall that $\Gamma _{0}^{\ast }=\tau _{n}+\Theta _{0}^{\ast }$ and $\Gamma
_{1}^{\ast }=\Theta _{1}^{\ast }$. Let $\Gamma ^{\ast }=(\Gamma _{0}^{\ast
},\Gamma _{1}^{\ast })$, $\Lambda \left( u\right) =\frac{1}{1+\exp \left(
	-u\right) }$ denote the standard logistic probability density function,
\begin{equation*}
\ell _{ij}\left( \Gamma _{ij}\right) =Y_{ij}\log (\Lambda (W_{ij}^{\top
}\Gamma _{ij}))+(1-Y_{ij})\log (1-\Lambda (W_{ij}^{\top }\Gamma _{ij}))
\end{equation*}%
denote the conditional logistic log-likelihood function associated with
nodes $i$ and $j$, and
\begin{equation*}
\mathbb{T}\left( \tau ,c_{n}\right) =\{(\Gamma _{0},\Gamma _{1})\in \mathbb{R%
}^{n\times n}\times \mathbb{R}^{n\times n}:|\Gamma _{0,ij}-\tau |\leq
c_{n},|\Gamma _{1,ij}|\leq c_{n}\}.
\end{equation*}%
We propose to estimate $\Gamma ^{\ast }$ by $\widetilde{\Gamma }=(\widetilde{%
	\Gamma }_{0},\widetilde{\Gamma }_{1})$ via minimizing the negative logistic
log-likelihood function with the nuclear norm regularization:
\begin{equation}
\widetilde{\Gamma }=\argmin_{\Gamma \in \mathbb{T}\left( 0,\log n\right)
}Q_{n}(\Gamma )+\lambda _{n}\sum_{l=0}^{1}||\Gamma _{l}||_{\ast },
\label{eq:optimizationf}
\end{equation}%
where $Q_{n}(\Gamma )=\frac{-1}{n(n-1)}\sum_{i,j\in \lbrack n],i\neq j}\ell
_{ij}\left( \Gamma _{ij}\right) $ and $\lambda _{n}>0$ is a regularization
parameter. As mentioned above, we allow $\zeta _{n}$ to shrink to zero at a
rate as slow as $n^{-1}\log n$ so that $\tau _{n}=\log \left( \zeta
_{n}\right) $ is slightly smaller than $\log n$ in magnitude. So it is
sufficient to consider a parameter space $\mathbb{T}\left( 0,\log n\right) $
that expands at rate-$\log n.$ Later on, we specify $\lambda _{n}=\frac{%
	C_{\lambda }(\sqrt{\zeta _{n}n}+\sqrt{\log n})}{n(n-1)}$ for some constant
tuning parameter $C_{\lambda }$. Throughout the paper, we assume $W_{1,ij}$
has been rescaled so that its standard error is one. Therefore, we do not
need to consider different penalty loads for $||\Gamma _{0}||_{\ast }$ and $%
||\Gamma _{1}||_{\ast }$. Many statistical softwares automatically normalize
the regressors when estimating a generalized linear model. We recommend this
normalization in practice before using our algorithm.

Let $\widetilde{\tau }_{n}=\frac{1}{n(n-1)}\sum_{i\neq j}\widetilde{\Gamma }%
_{0,ij}$. We will show that $\widetilde{\tau }_{n}$ lies within $c_{\tau }%
\sqrt{\log n}$-neighborhood of the true value $\tau _{n},$ where $c_{\tau }$
can be made arbitrarily small provided that the expected degree is larger
than $C\log n$ for some sufficiently large $C.$\footnote{%
	Let $\eta _{0n}=\sqrt{\frac{\log n}{n\zeta _{n}}}$ and $\eta _{n}=\eta
	_{0n}+\eta _{0n}^{2}.$ The proof of Theorem \ref{thm:Frobeniusnorm}.1
	suggests that $\widetilde{\tau }_{n}-\tau _{n}=O_{p}(\eta _{n}\sqrt{\log n}%
	), $ which is\ $o_{p}(\sqrt{\log n})$ (resp. $o_{p}(1)$) if one assumes that
	the magnitude $n\zeta _{n}$ of the expected degree is of order higher than $%
	\log n$ (resp. $(\log n)^{2}$). But we will only assume that $\eta _{0n}\leq
	C_{F}\leq \frac{1}{4}$ for some sufficiently small constant $C_{F}$ below.}
This rate is insufficient and remains to be refined. Given $\widetilde{\tau }%
_{n}$, we propose to reestimate $\Gamma ^{\ast }$ by $\widehat{\Gamma }=(%
\widehat{\Gamma }_{0},\widehat{\Gamma }_{1}),$ where
\begin{equation*}
\widehat{\Gamma }=\argmin_{\Gamma \in \mathbb{T}(\widetilde{\tau }_{n},C_{M}%
	\sqrt{\log n})}Q_{n}(\Gamma )+\lambda _{n}\sum_{l=0}^{1}||\Gamma
_{l}||_{\ast },
\end{equation*}%
and $C_{M}$ is some constant to be specified later. Note that we now
restrict the parameter space to expand at rate-$\sqrt{\log n}$ only.

Let $\widehat{\tau }_{n}=\frac{1}{n(n-1)}\sum_{i\neq j}\widehat{\Gamma }%
_{0,ij}$. Since $\Theta _{l}^{\ast }=\{\Theta _{l,ij}^{\ast }\}$ are
symmetric, we define their preliminary low-rank estimators as $\widehat{%
	\Theta }_{l}=\{\widehat{\Theta }_{l,ij}\},$ where
\begin{equation*}
\widehat{\Theta }_{l,ij}=%
\begin{cases}
f_{M}((\widehat{\Gamma }_{l,ij}+\widehat{\Gamma }_{l,ji})/2-\widehat{\tau }%
_{n}\delta _{l0}) & \text{if}\quad i\neq j \\
0 & \text{if}\quad i=j%
\end{cases}%
\text{ for }l=0,1,
\end{equation*}%
$\delta _{l0}=\mathbf{1}\{l=0\},$ $f_{M}(u)=u\cdot \mathbf{1}\{|u|\leq
M\}+M\cdot \mathbf{1}\{u>M\}-M\cdot \mathbf{1}\{u<-M\}$ is the round
function, and $M$ is some positive constant. For $l=0,1$, we denote the SVD
of $n^{-1}\widehat{\Theta }_{l}$ as
\begin{equation*}
n^{-1}\widehat{\Theta }_{l}=\widehat{\widetilde{\mathcal{U}}}_{l}\widehat{%
	\widetilde{\Sigma }}_{l}(\widehat{\tilde{\mathcal{V}}}_{l})^{\top },
\end{equation*}%
where $\widehat{\widetilde{\Sigma }}_{l}=\text{diag}(\widehat{\sigma }%
_{1,l},...,\widehat{\sigma }_{n,l})$, $\widehat{\sigma }_{1,l}\geq \cdots
\geq \widehat{\sigma }_{n,l}\geq 0$, and both $\widehat{\widetilde{\mathcal{U%
}}}_{l}$ and $\widehat{\widetilde{\mathcal{V}}}_{l}$ are $n\times n$ unitary
matrices. Let $\widehat{\mathcal{V}}_{l}$ consist of the first $K_{l}$
columns of $\widehat{\widetilde{\mathcal{V}}}_{l}$, such that $(\widehat{%
	\mathcal{V}}_{l})^{\top }\widehat{\mathcal{V}}_{l}=I_{K_{l}}$ and $\widehat{%
	\Sigma }_{l}=\text{diag}(\widehat{\sigma }_{1,l},\cdots ,\widehat{\sigma }%
_{K_{l},l})$. Then $\widehat{V}_{l}=\sqrt{n}\widehat{\mathcal{V}}_{l}.$

\subsubsection{Split-Sample Low-Rank Estimation \label{sec:sslre}}

We divide the $n$ nodes into two roughly equal-sized subsets $(I_{1},I_{2})$%
. Let $n_{\ell }=\#I_{\ell }$ denote the cardinality of the set $I_{\ell }.$
If $n$ is even, one can simply set $n_{\ell }=n/2$ for $\ell =1,2.$

Now, we only use the pair of observations $(i,j)\in I_{1}\times [ n]$ to
conduct the low-rank estimation. Let $\Gamma _{l}^{\ast }(I_{1})$ consist
of\ the $i$-th row of $\Gamma _{l}^{\ast }$ for $i\in I_{1}$, $l=0,1.$ Let $%
\Gamma ^{\ast }(I_{1})=(\Gamma _{0}^{\ast }(I_{1}),\Gamma _{1}^{\ast
}(I_{1}))$. Define
\begin{equation*}
\mathbb{T}^{\left( 1\right) }\left( \tau ,c_{n}\right) =\{(\Gamma
_{0},\Gamma _{1})\in \mathbb{R}^{n_{1}\times n}\times \mathbb{R}%
^{n_{1}\times n}:|\Gamma _{0,ij}-\tau |\leq c_{n},|\Gamma _{1,ij}|\leq
c_{n}\}.
\end{equation*}%
We estimate $\Gamma ^{\ast }(I_{1})$ via the following nuclear-norm
regularized estimation
\begin{equation}
\widetilde{\Gamma }^{(1)}=\argmin_{\Gamma \in \mathbb{T}^{\left( 1\right)
	}\left( 0,\log n\right) }Q_{n}^{(1)}(\Gamma )+\lambda
_{n}^{(1)}\sum_{l=0}^{1}||\Gamma _{l}||_{\ast },  \label{eq:optimization}
\end{equation}%
where $Q_{n}^{(1)}(\Gamma )=\frac{-1}{n_{1}(n-1)}\sum_{i\in I_{1},j\in [
	n],i\neq j}\ell _{ij}\left( \Gamma _{ij}\right) $, $\lambda _{n}^{(1)}=\frac{%
	C_{\lambda }(\sqrt{\zeta _{n}n}+\sqrt{\log n})}{n_{1}(n-1)}$, and the
superscript $(1)$ means we use the first subsample ($I_1$) in this step.

Let $\widetilde{\tau }_{n}^{\left( 1\right) }=\frac{1}{n_{1}(n-1)}\sum_{i\in
	I_{1},j\in [ n],i\neq j}\widetilde{\Gamma }_{0,ij}^{(1)}$. As above, this
estimate lies within $c_{\tau }\sqrt{\log n}$-neighborhood of the true value
$\tau _{n}.$ To refine it, we can reestimate $\Gamma ^{\ast }(I_{1})$ by $%
\widehat{\Gamma }^{(1)}=(\widehat{\Gamma }_{0}^{(1)},\widehat{\Gamma }%
_{1j}^{(1)}):$
\begin{equation*}
\widehat{\Gamma }^{(1)}=\argmin_{\Gamma \in \mathbb{T}^{\left( 1\right) }(%
	\widetilde{\tau }_{n}^{\left( 1\right) },C_{M}\sqrt{\log n}%
	)}Q_{n}^{(1)}(\Gamma )+\lambda _{n}^{(1)}\sum_{l=0}^{1}||\Gamma _{l}||_{\ast
}.
\end{equation*}%
Let $\widehat{\tau }_{n}^{(1)}=\frac{1}{n_{1}(n-1)}\sum_{i\in I_{1},j\in [
	n],i\neq j}\widehat{\Gamma }_{0,ij}^{(1)}$. Noting that $\{\Gamma _{l}^{\ast
}\}_{l=0,1}$ are symmetric, we define the preliminary low-rank estimates for
the $n_{1}\times n$ matrices $\Theta _{l}^{\ast }(I_{1})$ by $\widehat{%
	\Theta }_{l}^{(1)}$ for $l=0,1$, where
\begin{equation*}
\widehat{\Theta }_{l,ij}^{(1)}=%
\begin{cases}
f_{M}((\widehat{\Gamma }_{l,ij}^{(1)}+\widehat{\Gamma }_{l,ji}^{(1)})/2-%
\widehat{\tau }_{n}^{(1)}\delta _{l0}) & \quad \text{if}\quad (i,j)\in
I_{1}\times I_{1},\text{ }i\neq j \\
0 & \quad \text{if}\quad (i,j)\in I_{1}\times I_{1},\text{ }i=j \\
f_{M}(\widehat{\Gamma }_{l,ij}^{(1)}-\widehat{\tau }_{n}^{(1)}\delta _{l0})
& \quad \text{if}\quad i\in I_{1},\text{ }j\notin I_{1}%
\end{cases}%
,
\end{equation*}%
and $\delta _{l0},$ $f_{M}(u)$ and $M$ are defined in\ Step 1. For $l=0,1$,
we denote the SVD of $n^{-1}\widehat{\Theta }_{l}^{(1)}$ as
\begin{equation*}
n^{-1}\widehat{\Theta }_{l}^{(1)}=\widehat{\widetilde{\mathcal{U}}}_{l}^{(1)}%
\widehat{\widetilde{\Sigma }}_{l}^{(1)}(\widehat{\widetilde{\mathcal{V}}}%
_{l}^{(1)})^{\top },
\end{equation*}%
where $\widehat{\widetilde{\Sigma }}_{l}^{(1)}$ is a rectangular ($%
n_{1}\times n$) diagonal matrix with $\widehat{\sigma }_{i,l}^{(1)}$
appearing in the $\left( i,i\right) $th position and zeros elsewhere, $%
\widehat{\sigma }_{1,l}^{(1)}\geq \cdots \geq \widehat{\sigma }%
_{n_{1},l}^{(1)}\geq 0$, and $\widehat{\tilde{\mathcal{U}}}_{l}^{(1)}$ and $%
\widehat{\tilde{\mathcal{V}}}_{l}^{(1)}$ are $n_{1}\times n_{1}$ and $%
n\times n$ unitary matrices, respectively. Let $\widehat{\mathcal{V}}%
_{l}^{(1)}$ consist of the first $K_{l}$ columns of $\widehat{\widetilde{%
		\mathcal{V}}}_{l}^{(1)}$ such that $(\widehat{\mathcal{V}}_{l}^{(1)})^{\top }%
\widehat{\mathcal{V}}_{l}^{(1)}=I_{K_{l}}.$ Let $\widehat{\Sigma }_{l}^{(1)}=%
\text{diag}(\widehat{\sigma }_{1,l}^{(1)},\cdots ,\widehat{\sigma }%
_{K_{l},l}^{(1)})$. Then $\widehat{V}_{l}^{(1)}=\sqrt{n}\widehat{\mathcal{V}}%
_{l}^{(1)},$ and $(\widehat{v}_{j,l}^{(1)})^{\top }$ is the $j$-th row of $%
\widehat{V}_{l}^{(1)}$ for $j\in \left[ n\right] $.

\subsubsection{Split-Sample Row- and Column-Wise Logistic Regressions \label%
	{sec:ssrwlr}}

We note that $\Theta_{l,ij}^{\ast} = u_{i,l}^{\top}v_{j,l}$ for $i \in I_2$
and $j \in [n]$. For the $i$-th row when $i \in I_2$, we can view $%
\{v_{j,l}\}_{j \in [n]}$ and $u_{i,l}$ as regressors and the parameter,
respectively, and estimate $u_{i,l}$ by the row-wise logistic regression.
Although $\{v_{j,l}\}_{j \in [n]}$ are unobservable, we can replace them by
their estimators obtained from the previous step.

Let $\mu =(\mu _{0}^{\top },\mu _{1}^{\top })^{\top }$ and $\Lambda _{ij}^{%
	\text{left}}(\mu )=\Lambda (\widehat{\tau }_{n}+\sum_{l=0}^{1}\mu _{l}^{\top
}\widehat{v}_{j,l}^{(1)}W_{l,ij})$ and $\ell _{ij}^{\text{left}}\left( \mu
\right) =Y_{ij}\log (\Lambda _{ij}^{\text{left}}(\mu ))$ $+(1-Y_{ij})\log
(1-\Lambda _{ij}^{\text{left}}(\mu ))$, where the superscript ''left" means
these functions are used to estimate the left singular vector $u_{i,l}$.
Given the preliminary estimate $\{\widehat{v}_{j,l}^{(1)}\}$ obtained in
Step 2, we can estimate the left singular vectors $\{u_{i,0},$ $u_{i,1}\}$
for each $i\in I_{2}$ by $\{\widehat{u}_{i,0}^{(1)},\widehat{u}%
_{i,1}^{(1)}\} $ via the row-wise logistic regression:
\begin{equation*}
((\widehat{u}_{i,0}^{(1)})^{\top },(\widehat{u}_{i,10}^{(1)})^{\top })^{\top
}=\argmin_{\mu =(\mu _{0}^{\top },\mu _{1}^{\top })^{\top }\in \mathbb{R}%
	^{K_{0}+K_{1}}}Q_{in,U}^{(0)}(\mu ),
\end{equation*}%
where $Q_{in,U}^{(0)}(\mu )=\frac{-1}{n_{2}}\sum_{j\in I_{2},j\neq i}\ell
_{ij}^{\text{left}}\left( \mu \right)$ and the superscript $(0)$ means it is
the initial step for the full sample iteration below. To keep the
independence between $\{\widehat{v}_{j,l}^{(1)}\}_{j\in [n]}$ and the data
in this regression, we only use $j \in I_2$ to run the regression.

Let $\nu =(\nu _{0}^{\top },\nu _{1}^{\top })^{\top }$ and $\Lambda _{ij}^{%
	\text{right}}(\nu )=\Lambda (\widehat{\tau }_{n}+\sum_{l=0}^{1}\nu
_{l}^{\top }\widehat{u}_{i,l}^{(1)}W_{l,ij})$ and $\ell _{ij}^{\text{right}%
}\left( \nu \right) =Y_{ij}\log (\Lambda _{ij}^{\text{right}}(\nu))$ $%
+(1-Y_{ij})\log (1-\Lambda _{ij}^{\text{right}}(\nu ))$, where the
superscript ''right" means the functions are used to estimate the right
singular vector $v_{j,l}$. Given $(\widehat{u}_{i,0}^{(1)},\widehat{u}%
_{i,1}^{(1)}),$ we update the estimate of the right singular vectors $%
\{v_{i,0},v_{i,1}\}$ for each $j\in [ n]$ by $\{\dot{v}_{j,0}^{(0,1)},\dot{v}%
_{j,1}^{(0,1)}\}$ via the column-wise logistic regression:
\begin{equation*}
((\dot{v}_{j,0}^{(0,1)})^{\top },(\dot{v}_{j,1}^{(0,1)})^{\top })^{\top }=%
\argmin_{\nu =(\nu _{0}^{\top },\nu _{1}^{\top })^{\top }\in \mathbb{R}%
	^{K_{0}+K_{1}}}Q_{jn,V}^{(0)}(\nu ),
\end{equation*}%
where $Q_{jn,V}^{(0)}(\nu )=\frac{-1}{n_{2}}\sum_{i\in I_{2},i\neq j}\ell
_{ij}^{\text{right}}\left( \nu \right) .$

Our final objective is to obtain accurate estimates of $\left\{
v_{j,l}\right\} _{j\in \left[ n\right] ,l=0,1}.$ To this end, we treat $\{%
\dot{v}_{j,0}^{(0,1)},\dot{v}_{j,1}^{(0,1)}\}_{j\in \left[ n\right] }$ as
the initial estimate in the following full-sample iteration procedure.

\subsubsection{Full-Sample Iteration \label{sec:fsi}}

Given the initial estimates, we use the full sample and iteratively run row-
and column-wise logistic regressions to estimate $\{u_{i,l},v_{i,l}\}_{i \in
	[n]}$. For $h=1,2,...,H,$ let $\Lambda _{ij}^{\text{left,}h}(\mu )=\Lambda (%
\widehat{\tau }_{n}+\sum_{l=0}^{1}\mu _{l}^{\top }\dot{v}%
_{j,l}^{(h-1,1)}W_{l,ij}))$ and $\ell _{ij}^{\text{left,}h}\left( \mu
\right) =Y_{ij}\log (\Lambda _{ij}^{\text{left,}h}(\mu ))$ $+(1-Y_{ij})\log
(1-\Lambda _{ij}^{\text{left,}h}(\mu )).$ Given $\{\dot{v}_{i,0}^{(h-1,1)},%
\dot{v}_{i,1}^{(h-1,1)}\}$, we can compute $\{\dot{u}_{i,0}^{(h,1)},\dot{u}%
_{i,1}^{(h,1)}\}$ via the row-wise logistic regression
\begin{equation*}
((\dot{u}_{i,0}^{(h,1)})^{\top },(\dot{u}_{i,01}^{(h,1)})^{\top })^{\top }=%
\argmin_{\mu =(\mu _{0}^{\top },\mu _{1}^{\top })^{\top }\in \mathbb{R}%
	^{K_{0}+K_{1}}}Q_{in,U}^{(h)}(\mu ),
\end{equation*}%
where\ $Q_{in,U}^{(h)}(\mu )=\frac{-1}{n}\sum_{j\in [ n],j\neq i}\ell _{ij}^{%
	\text{left,}h}\left( \mu \right) .$

Given $\{\dot{u}_{i,0}^{(h,1)},\dot{u}_{i,1}^{(h,1)}\}$, by letting $\Lambda
_{ij}^{\text{right,}h}(\nu )=\Lambda (\widehat{\tau }_{n}+\sum_{l=0}^{1}\nu
_{l}^{\top }\dot{u}_{i,l}^{(h,1)}W_{l,ij}))$ and $\ell _{ij}^{\text{right,}%
	h}\left( \nu \right) =Y_{ij}\log (\Lambda _{ij}^{\text{right,}h}(\nu ))$ $%
+(1-Y_{ij})\log (1-\Lambda _{ij}^{\text{right,}h}(\nu )),$ we compute $\{%
\dot{v}_{j,0}^{(h,1)},\dot{v}_{j,1}^{(h,1)}\}$ via the column-wise logistic
regression
\begin{equation*}
((\dot{v}_{j,0}^{(h,1)})^{\top },(\dot{v}_{j,1}^{(h,1)})^{\top })^{\top }=%
\argmin_{\nu =(\nu _{0}^{\top },\nu _{1}^{\top })^{\top }\in \mathbb{R}%
	^{K_{0}+K_{1}}}Q_{jn,V}^{(h)}(\nu ),
\end{equation*}%
where $Q_{jn,V}^{(h)}(\nu )=\frac{-1}{n}\sum_{i\in [ n],i\neq j}\ell _{ij}^{%
	\text{right,}h}\left( \nu \right) .$

We can stop iteration when certain convergence criterion is met for
sufficiently large $H.$ Switching the roles of $I_{1}$ and $I_{2}$ and
repeating the procedure in the last three steps, we can obtain the iterative
estimates $\{\dot{u}_{i,0}^{(h,2)},\dot{u}_{i,1}^{(h,2)}\}_{i\in [ n]}$ and $%
\{\dot{v}_{j,0}^{(h,2)},\dot{v}_{j,1}^{(h,2)}\}_{j\in [ n]}$ for $%
h=1,2,\cdots ,H$.

\subsection{K-means Classification \label{sec:kmeans}}

In this step, we further assume $\Theta _{1}^{\ast }$ has the latent
community structure and $\Theta _{0}^{\ast }$ remains to be of low-rank.
Recall that $\overline{v}_{j,1}=\left( \frac{(\dot{v}_{j,1}^{(H,1)})^{\top }%
}{||\dot{v}_{j,1}^{(H,1)}||},\frac{(\dot{v}_{j,1}^{(H,2)})^{\top }}{||\dot{v}%
	_{j,1}^{(H,2)}||}\right) ^{\top }$, a $2K_{1}\times 1$ vector. We now apply
the K-means algorithm to $\{\overline{v}_{j,1}\}_{j\in \lbrack n]}$. Let $%
\mathcal{B}=\{\beta _{1},\ldots ,\beta _{K_{1}}\}$ be a set of $K_{1}$
arbitrary $2K_{1}\times 1$ vectors: $\beta _{1},\ldots ,\beta _{K_{1}}$.
Define
\begin{equation*}
\widehat{Q}_{n}(\mathcal{B})=\frac{1}{n}\sum_{j=1}^{n}\min_{1\leq k\leq
	K_{1}}\left\Vert \overline{v}_{j,1}-\beta _{k}\right\Vert ^{2}
\end{equation*}%
and $\widehat{\mathcal{B}}_{n}=\{\widehat{\beta }_{1},\ldots ,\widehat{\beta
}_{K_{1}}\}$, where $\widehat{\mathcal{B}}_{n}=\argmin_{\mathcal{B}}\widehat{%
	Q}_{n}(\mathcal{B}).$ For each $j\in \lbrack n],$ we estimate the group
identity by
\begin{equation}
\hat{g}_{j}=\argmin_{1\leq k\leq K_{1}}\left\Vert \overline{v}_{j,1}-%
\widehat{\beta }_{k}\right\Vert ,  \label{eq:gi}
\end{equation}%
where if there are multiple $k$'s that achieve the minimum, $\hat{g}_{j}$
takes value of the smallest one.

As mentioned previously, we can repeat Steps 2--6 $R$ times to obtain $R$
membership estimates, denoted as $\{\hat{g}_{j,r}\}_{{j\in \lbrack n],r\in
		\lbrack R]}}$. Recall that
\begin{equation*}
\text{vech}(B_{1}^{\ast })=(B_{1,11}^{\ast },...,B_{1,1K_{1}}^{\ast
},B_{1,22}^{\ast },\cdots ,B_{1,2K_{1}}^{\ast },\cdots
,B_{1,K_{1}-1,K_{1}-1}^{\ast },B_{1,K_{1}-1,K_{1}}^{\ast
},B_{1,K_{1}K_{1}}^{\ast })^{\top },
\end{equation*}%
which is a $K_{1}(K_{1}+1)/2$-vector. In addition, let $\chi _{1,ij}$ be the
vectorization of the upper triangular part of the $K_{1}\times K_{1}$ matrix
whose $(g_{i}^{0},g_{j}^{0})$ and $(g_{j}^{0},g_{i}^{0})$ entries are one
and the rest entries are zero, i.e., $\chi _{1,ij}$ is a $K_{1}(K_{1}+1)/2$
vector such that the $((g_{i}^{0}\vee g_{j}^{0}-1)(g_{i}^{0}\vee
g_{j}^{0})/2+g_{i}^{0}\wedge g_{j}^{0})$-th element is one and the rest are
zeros, where $g_{i}^{0}\in \lbrack K_{1}]$ denotes the true group membership
of the $i$-th node in $\Theta _{1}^{\ast }$. By construction,
\begin{equation*}
\chi _{1,ij}^{\top }\text{vech}(B_{1}^{\ast
})=B_{1,g_{i}^{0}g_{j}^{0}}^{\ast }.
\end{equation*}%
Analogously, for the $r$-th split, denote $\hat{\chi}_{1r,ij}$ as a $%
K_{1}(K_{1}+1)/2$ vector such that the $((\hat{g}_{i,r}\vee \hat{g}_{j,r}-1)(%
\hat{g}_{i,r}\vee \hat{g}_{j,r})/2+\hat{g}_{i,r}\wedge \hat{g}_{j,r})$-th
element is one and the rest are zeros. We then estimate $B_{1}^{\ast }$ by $%
\widehat{B}_{1,r}$, a symmetric matrix constructed from\ $\widehat{b}_{r}$
by reversing the vech operator:
\begin{equation*}
\widehat{b}_{r}=\argmax_{b}\mathcal{L}_{n,r}(b),
\end{equation*}%
where $\mathcal{L}_{n,r}(b)=\sum_{i<j}[Y_{ij}\log (\hat{\Lambda}%
_{ij}(b))+(1-Y_{ij})\log (1-\hat{\Lambda}_{ij}(b)))]$ with $\hat{\Lambda}%
_{ij}(b)=\Lambda (\widehat{\tau }_{n}+\widehat{\Theta }_{0,ij}+W_{1,ij}\hat{%
	\chi}_{1r,ij}^{\prime }b),$ $\widehat{\tau }_{n}$ is obtained in Step 1, $%
\widehat{\Theta }_{0,ij}=[(\dot{u}_{i,0}^{(H,1)})^{\top }\dot{v}%
_{j,0}^{(H,1)}+(\dot{u}_{i,0}^{(H,2)})^{\top }\dot{v}_{j,0}^{(H,2)}]/2$, and
$(\dot{u}_{i,0}^{(H,1)},\dot{v}_{j,0}^{(H,1)},$ $\dot{u}_{i,0}^{(H,2)},\dot{v%
}_{j,0}^{(H,2)})$ are obtained in Step 5.\footnote{%
	If we have multiple covariates $W_{l}$, $l\in \lbrack p]$, to compute $%
	\mathcal{L}_{n,r}(b)$, we let $\widehat{\Theta }_{l,ij}=[(\dot{u}%
	_{i,l}^{(H,1)})^{\top }\dot{v}_{j,l}^{(H,1)}+(\dot{u}_{i,l}^{(H,2)})^{\top }%
	\dot{v}_{j,l}^{(H,2)}]/2$ when $\Theta _{l}^{\ast }$ is only assumed to be
	of low-rank. For those $\Theta _{l}^{*}$'s that have latent communities, for
	the $r$-th split, we can estimate their memberships by $\hat{g}_{i,l,r}$ and
	construct $\hat{\chi}_{lr,ij}^{\prime }$ similarly. Then, we can define $%
	\mathcal{L}_{n,r}(b)$ and $\widehat{\mathcal{L}}(r)$ in the same manner.}
Then, the likelihood of the $r$-th split is defined as $\widehat{\mathcal{L}}%
(r)=\mathcal{L}_{n,r}(\widehat{b}_{r}).$ Our final estimator $\{\hat{g}%
_{i,r^{\ast }}\}_{i\in \lbrack n]}$ of the membership corresponds to the $%
r^{\ast }$-th split, where
\begin{equation}
r^{\ast }=\argmax_{r\in \lbrack R]}\widehat{\mathcal{L}}(r).
\label{eq:rstar}
\end{equation}

\section{Statistical Properties of the Estimators of $(u_{i,l},v_{j,l})$%
	\label{sec:sp}}

In this section, we study the asymptotic properties of the estimators of $%
(u_{i,l},v_{j,l})$ proposed in the last section.

\subsection{Full- and Split-Sample Low-Rank Estimations}

Suppose the singular value decomposition of $\Gamma_l^*$ is $\Gamma_l^* =
\overline{U}_{l} \Sigma_l \overline{V}_l^{\top}$ for $l = 0,1$ and $%
\overline{U}_{l,c}$ and $\overline{V}_{l,c}$ are the left and right singular
matrices corresponding to the zero singular values. Let $\mathcal{P}%
_l(\Delta) = \overline{U}_{l,c}\overline{U}_{l,c}^{\top} \Delta \overline{V}%
_{l,c}\overline{V}_{l,c}^{\top}$ for some $n\times n$ matrix $\Delta$ and $%
\mathcal{M}_j(\Delta) = \Delta - \mathcal{P}_j(\Delta)$. Define the
restricted low-rank set as, for some $c_1>0$
\begin{align}
\mathcal{C}(c_1) =
\begin{Bmatrix}
(\Delta_0,\Delta_1): & ||\mathcal{P}_0(\Delta_0)||_{*}+||\mathcal{P}%
_1(\Delta_1)||_{*} \leq c_1||\mathcal{M}_0(\Delta_0)||_{*} + c_1||\mathcal{M}%
_1(\Delta_1)||_{*}%
\end{Bmatrix}%
.  \label{eq:rsc}
\end{align}

\begin{ass}
	For any $c_{1}>0$, there exist constants $\kappa ,c_{2},c_{3}>0$,
	\begin{eqnarray*}
		\mathcal{C}_{1}(c_{2}) &=&\{(\Delta _{0},\Delta _{1}):||\Delta
		_{0}||_{F}^{2}+||\Delta _{1}||_{F}^{2}\leq c_{2}\log (n)n/\zeta _{n}\},\text{
			and} \\
		\mathcal{C}_{2}(c_{3}) &=&\{(\Delta _{0},\Delta _{1}):||\Delta _{0}+\Delta
		_{1}\odot W_{1}||_{F}^{2}\geq \kappa (||\Delta _{0}||_{F}^{2}+||\Delta
		_{1}||_{F}^{2})-c_{3}\log (n)n/\zeta _{n}\},
	\end{eqnarray*}%
	such that
	\begin{equation*}
	\mathcal{C}(c_{1})\subset \mathcal{C}_{1}(c_{2})\cup \mathcal{C}%
	_{2}(c_{3})~w.p.a.1.
	\end{equation*}%
	The same condition holds when $(\Gamma _{0}^{\ast },\Gamma _{1}^{\ast })$
	are replaced by $(\Gamma _{0}^{\ast }(I_{1}),\Gamma _{1}^{\ast }(I_{1}))$
	and $(\Gamma _{0}^{\ast }(I_{2}),\Gamma _{1}^{\ast }(I_{2}))$. \label%
	{ass:RSC}
\end{ass}


Several remarks are in order. First, Assumption \ref{ass:RSC} is a slight
generalization of \citet[Assumption 3.1]{CHLZ18} where, in terms of our
notation, $\mathcal{C}_{1}(c_{2})$ and $\mathcal{C}_{2}(c_{3})$ take the
forms:%
\begin{eqnarray*}
	\mathcal{C}_{1}(c_{2}) &=&\{(\Delta _{0},\Delta _{1}):||\Delta
	_{0}||_{F}^{2}+||\Delta _{1}||_{F}^{2}\leq c_{2}n\}\text{ and} \\
	\mathcal{C}_{2}(c_{3}) &=&\{(\Delta _{0},\Delta _{1}):||\Delta _{0}+\Delta
	_{1}\odot W_{1}||_{F}^{2}\geq \kappa (||\Delta _{0}||_{F}^{2}+||\Delta
	_{1}||_{F}^{2})-nc_{3}\}.
\end{eqnarray*}%
Such a generalization is due to the fact that the network can be semi-dense,
and thus, the convergence rates of our estimators of the singular vectors
are slower than those of \citet{CHLZ18}'s estimators.

Second, suppose there are two sets of parameters $(\Gamma _{0}^{\ast
},\Gamma _{1}^{\ast })$ and $(\Gamma _{0}^{\dagger },\Gamma _{1}^{\dagger })$
such that $\Gamma _{l}^{\ast }\neq \Gamma _{l}^{\dagger }$ for some $l\in
\{0,1\}$. The singular value decomposition of $\Gamma _{l}^{\dagger }$ is $%
\Gamma _{l}^{\dagger }=\tilde{U}_{l}\tilde{\Sigma}_{l}\tilde{V}_{l}^{\top }$
for $l=0,1$ and $\tilde{U}_{l,c}$ and $\tilde{V}_{l,c}$ are the left and
right singular matrices corresponding to the zero singular values. Denote $%
\widetilde{\mathcal{P}}_{l}(\Delta )=\tilde{U}_{l,c}\tilde{U}%
_{l,c}^{T}\Delta \tilde{V}_{l_{c}}\tilde{V}_{l,c}^{T}$ and $\widetilde{%
	\mathcal{M}}_{l}(\Delta )=\Delta -\widetilde{\mathcal{P}}_{l}(\Delta )$.
Suppose that
\begin{equation*}
\begin{Bmatrix}
\tilde{\mathcal{C}}(c_{1})=(\Delta _{0},\Delta _{1}): & ||\widetilde{%
	\mathcal{P}}_{0}(\Delta _{0})||_{\ast }+||\widetilde{\mathcal{P}}_{1}(\Delta
_{1})||_{\ast }\leq c_{1}||\widetilde{\mathcal{M}}_{0}(\Delta _{0})||_{\ast
}+c_{1}||\widetilde{\mathcal{M}}_{1}(\Delta _{1})||_{\ast }%
\end{Bmatrix}%
,
\end{equation*}%
Assumption \ref{ass:RSC} holds for both $\mathcal{C}(c_{1})$ and $\tilde{%
	\mathcal{C}}(c_{1})$, and
\begin{equation*}
\Gamma _{0}^{\ast }+\Gamma _{1}^{\ast }\odot W_{1}=\Gamma _{0}^{\dagger
}+\Gamma _{1}^{\dagger }\odot W_{1}.
\end{equation*}%
Denote $\Delta _{l}=\Gamma _{l}^{\dagger }-\Gamma _{l}^{\ast }$, $l=0,1$.
Then $\Delta _{0}+\Delta _{1}\odot W_{1}=0$ and it is possible to show that
that $(\Delta _{0},\Delta _{1})$ belongs to either $\mathcal{C}(1)$ or $%
\tilde{\mathcal{C}}(1)$.\footnote{%
	Without loss of generality, we assume that $||\Gamma _{0}^{\dagger }||_{\ast
	}+||\Gamma _{1}^{\dagger }||_{\ast }\leq ||\Gamma _{0}^{\ast }||_{\ast
	}+||\Gamma _{1}^{\ast }||_{\ast }$. Noting that
	\begin{align*}
	||\Gamma _{l}^{\dagger }||_{\ast }=& ||\Gamma _{l}^{\ast }+\mathcal{M}%
	_{l}(\Delta _{l})+\mathcal{P}_{l}(\Delta _{l})||_{\ast } \\
	\geq & ||\Gamma _{l}^{\ast }+\mathcal{P}_{l}(\Delta _{l})||_{\ast }-||%
	\mathcal{M}_{l}(\Delta _{l})||_{\ast } \\
	=& ||\Gamma _{l}^{\ast }||_{\ast }+||\mathcal{P}_{l}(\Delta _{l})||_{\ast
	}-||\mathcal{M}_{l}(\Delta _{l})||_{\ast }\text{ for }l=0,1,
	\end{align*}%
	where the last equality holds due to \citet[Lemma D.2(i)]{CHLZ18}, we have
	\begin{align*}
	||\Gamma _{0}^{\ast }||_{\ast }+||\Gamma _{1}^{\ast }||_{\ast }\geq &
	||\Gamma _{0}^{\dagger }||_{\ast }+||\Gamma _{1}^{\dagger }||_{\ast } \\
	\geq & ||\Gamma _{0}^{\ast }||_{\ast }+||\mathcal{P}_{0}(\Delta
	_{0})||_{\ast }-||\mathcal{M}_{0}(\Delta _{0})||_{\ast }+||\Theta _{1}^{\ast
	}||_{\ast }+||\mathcal{P}_{1}(\Delta _{1})||_{\ast }-||\mathcal{M}%
	_{1}(\Delta _{1})||_{\ast },
	\end{align*}%
	which implies
	\begin{equation*}
	||\mathcal{P}_{0}(\Delta _{0})||_{\ast }+||\mathcal{P}_{1}(\Delta
	_{1})||_{\ast }\leq ||\mathcal{M}_{0}(\Delta _{0})||_{\ast }+||\mathcal{M}%
	_{1}(\Delta _{1})||_{\ast },
	\end{equation*}%
	i.e., $(\Delta _{0},\Delta _{1})\in \mathcal{C}(1)$.} If $(\Delta
_{0},\Delta _{1})\notin \mathcal{C}_{1}(c_{2})$, then Assumption \ref%
{ass:RSC} implies
\begin{equation*}
0=||\Delta _{0}+\Delta _{1}\odot W_{1}||_{F}^{2}\geq \kappa (||\Delta
_{0}||_{F}^{2}+||\Delta _{1}||_{F}^{2})-c_{3}\log (n)n/\zeta _{n},
\end{equation*}%
and thus,
\begin{equation*}
c_{3}\log (n)n/(\kappa \zeta _{n})\geq ||\Delta _{0}||_{F}^{2}+||\Delta
_{1}||_{F}^{2}>c_{2}\log (n)n/\zeta _{n}.
\end{equation*}%
Therefore,
\begin{equation*}
||\Delta _{0}||_{F}^{2}+||\Delta _{1}||_{F}^{2}\leq (c_{2}\vee \kappa
^{-1}c_{3})\log (n)n/\zeta _{n}.
\end{equation*}%
For any estimator $\hat{\Gamma}_{l}$ of $\Gamma _{l}^{\ast }$, $l=0,1$, we
have, w.p.a.1,
\begin{equation*}
\left\vert \frac{1}{n}(\sum_{l=0}^{1}||\hat{\Gamma}_{l}-\Gamma _{l}^{\ast
}||_{F})-\frac{1}{n}(\sum_{l=0}^{1}||\hat{\Gamma}_{l}-\Gamma _{l}^{\dagger
}||_{F})\right\vert \leq \frac{1}{n}(||\Delta _{0}||_{F}+||\Delta
_{1}||_{F})\leq \sqrt{\frac{2(c_{2}\vee \kappa ^{-1}c_{3})\log (n)}{n\zeta
		_{n}}}
\end{equation*}%
Based on Assumption \ref{ass:RSC} and other conditions in the paper, we can
show that (see Theorem \ref{thm:Frobeniusnorm} below)
\begin{equation*}
\frac{1}{n}(\sum_{l=0}^{1}||\hat{\Gamma}_{l}-\Gamma _{l}^{\ast }||_{F})\leq
48C_{F,1}\left( \sqrt{\frac{\log (n)}{n\zeta _{n}}}+\frac{\log (n)}{n\zeta
	_{n}}\right) ~w.p.a.1,
\end{equation*}%
where $C_{F,1}$ is some constant. This implies
\begin{equation*}
\frac{1}{n}(\sum_{l=0}^{1}||\hat{\Gamma}_{l}-\Gamma _{l}^{\dagger
}||_{F})\leq \left( 48C_{F,1}+\sqrt{2(c_{2}\vee \kappa ^{-1}c_{3})}\right)
\left( \sqrt{\frac{\log (n)}{n\zeta _{n}}}+\frac{\log (n)}{n\zeta _{n}}%
\right) ~w.p.a.1
\end{equation*}%
and vise versa. The same conclusion holds if $(\Delta _{0},\Delta _{1})\in
\mathcal{C}_{1}(c_{2})$. As a result, the ambiguity between $\Theta
_{l}^{\ast }$ and $\tilde{\Theta}_{l}$ is asymptotically negligible and will
not affect the convergence rates of their estimators.

Third, \citet[Appendix D.3]{CHLZ18} provide a sufficient condition for
Assumption \ref{ass:RSC}. Recall $W_{1,ij}=g_{1}(X_{i},X_{j},e_{ij})$.
Following the same arguments in \citet[Appendix D.3]{CHLZ18}, it is possible
to show that Assumption \ref{ass:RSC} holds if $W_{1,ij}$ is bounded and $%
Var(W_{1,ij}|X_{i},X_{j})>0$.\footnote{%
	In the general case with multiple covariates, they require $%
	\min_{i,j}\lambda _{\min }(\mathbb{E}W_{ij}W_{ij}^{\top }|X_{i},X_{j})\geq
	c>0$ where $\lambda _{\min }(A)$ is the minimum eigenvalue of matrix $A$ and
	$W_{ij}=(1,W_{1,ij},\cdots ,W_{p,ij})^{\top }$.} The sufficient condition
basically requires the existence of $e_{ij}$ in $g_{l}(\cdot )$ which is a
sequence of i.i.d. random variables across $i,j$.\footnote{%
	Note there are two key differences between the setups in our paper and \cite%
	{CHLZ18}. First, \cite{CHLZ18} consider the panel data with indexes $i\in
	\lbrack N]$ and $t\in \lbrack T]$ while we consider the network data with
	indexes $(i,j)\in \{1\leq i<j\leq n\}$. Second, \cite{CHLZ18} consider $%
	X_{it}=\mu _{it}+e_{it}$ such that given $\{\mu _{it}\}_{i\in \lbrack
		N],t\in \lbrack T]}$, $X_{it}$ is independent across both $t$ and $t$.
	Instead, we consider $W_{1,ij}=g_{1}(X_{i},X_{j},e_{ij})$ such that given $%
	\{X_{i}\}_{i\in \lbrack n]}$, $W_{1,ij}$ is independent across $1\leq
	i<j\leq n$. By examining the proofs of \citet[Lemmas D.3 and D.4]{CHLZ18},
	we note that their argument does not rely on the special structure of $%
	X_{it}=\mu _{it}+e_{it}$ and works if $X_{it}=f(\mu _{it},e_{it})$ for some
	non-additive function $f$.} Note that the presence of $e_{ij}$ is
sufficient, but may not be necessary. In our simulation, we generate $%
W_{1,ij}=|X_{i}-X_{j}|$ with $\{X_{i}\}_{i\in \lbrack n]}$ being a sequence
of i.i.d. standard normal random variables, and find that our method works
well.

Fourth, Assumption \ref{ass:RSC} rules out the case $%
W_{1,ij}=g_{1}(X_{i},X_{j})$ when $X_{i}$ is discrete, which is equivalent
to a community structure of $W_{1,ij}$. Suppose $W_{1,ij}=w_{k_{1}k_{2}}>0$ $%
\forall k_{1},k_{2}$ where $i,j$ are in groups $k_{1}$ and $k_{2}$. Then, we
can let $\Delta _{1}$ share the same community structure as $W_{1}$ and $%
\Delta _{1,ij}=w_{k_{1}k_{2}}^{-1}$. Let $\Delta _{0}=-\iota _{n}\iota
_{n}^{\top }$. Then we have
\begin{equation*}
\Delta _{0}+\Delta _{1}\odot W_{1}=0\quad \text{and}\quad ||\Delta
_{0}||_{F}^{2}+||\Delta _{1}||_{F}^{2}\geq ||\Delta _{0}||_{F}^{2}=n^{2}.
\end{equation*}%
Because both $\Delta _{1}$ and $\Delta _{0}$ are of low-rank, we have
\begin{equation*}
||\mathcal{P}_{0}(\Delta _{0})||_{\ast }+||\mathcal{P}_{1}(\Delta
_{1})||_{\ast }\leq ||\Delta _{0}||_{\ast }+||\Delta _{1}||_{\ast }\leq Cn,
\end{equation*}%
for some constant $C>0$. In addition, the singular value decomposition of $%
\Delta _{0}$ is $\Delta _{0}=(-\iota _{n}/\sqrt{n})\times n\times (\iota
_{n}/\sqrt{n})^{\top }$. It is possible to find some parameter $\Theta _{0}$
such that $||\mathcal{M}_{0}(\Delta _{0})||_{\ast }\geq cn$ for some $c>0$.%
\footnote{%
	This occurs, say, when $\iota _{n}/\sqrt{n}$ is in the spaces spanned by the
	left and right singular vectors of $\Theta _{0}$ that correspond to its
	nonzero singular values.} Then we can take $c_{1}=C/c$ so that
\begin{equation*}
||\mathcal{P}_{0}(\Delta _{0})||_{\ast }+||\mathcal{P}_{1}(\Delta
_{1})||_{\ast }\leq c_{1}||\mathcal{M}_{0}(\Delta _{0})||_{\ast }\leq c_{1}||%
\mathcal{M}_{0}(\Delta _{0})||_{\ast }+c_{1}||\mathcal{M}_{1}(\Delta
_{1})||_{\ast }.
\end{equation*}%
In this case, Assumption \ref{ass:RSC} does not hold because $||\Delta
_{0}||_{F}^{2}+||\Delta _{1}||_{F}^{2}\geq n^{2}>c_{2}\log (n)n/\zeta _{n}$%
\footnote{%
	When $\zeta _{n}=C_{\varsigma }n^{-1}\log (n),$ we require that $%
	C_{\varsigma }$ be sufficiently large.} and
\begin{equation*}
0=||\Delta _{0}+\Delta _{1}\odot W_{1}||_{F}^{2}<\kappa n^{2}-c_{3}\log
(n)n/\zeta _{n}\leq \kappa (||\Delta _{0}||_{F}^{2}+||\Delta
_{1}||_{F}^{2})-c_{3}\log (n)n/\zeta _{n}.
\end{equation*}

\begin{ass}
	\begin{enumerate}
		\item $C_{\lambda }>C_{\Upsilon }M_{W}$, where $C_{\Upsilon }$ is a constant
		defined in Lemma \ref{lem:op} in the online supplement.
		
		\item There exist constants $0<\underline{c}\leq \overline{c}<\infty $ such
		that $\zeta _{n}\underline{c}\leq \Lambda _{n,ij}\leq \zeta _{n}\overline{c}$%
		, where $\Lambda _{n,ij}\equiv \Lambda (W_{ij}^{\top }\Gamma _{ij}^{\ast }).$
		
		\item $\sqrt{\frac{\log n}{n\zeta _{n}}}\leq c_{F}\leq \frac{1}{4}$ for some
		sufficiently small constant $c_{F}$.
		
		
		\item $\sum_{i\in I_{1},j\in \left[ n\right] }\Theta^{\ast}_{0,ij} = o(\sqrt{%
			\frac{\log(n)}{n\zeta_n}})$.
	\end{enumerate}
	
	\label{ass:tune}
\end{ass}

Assumption \ref{ass:tune} is a regularity condition. In particular,
Assumptions \ref{ass:tune}.2 implies the order of the average degree in the
network is $n\zeta _{n}$. Assumption \ref{ass:tune}.3 means that the average
degree diverges to infinity at a rate that is not slower than $\log n$. Such
a rate is the slowest for exact recovery in the SBM, as established by \cite%
{ABH2016}, \cite{AS15}, \cite{MNS14}, and \cite{V18}. As our model
incorporates the SBM as a special case, the rate is also the minimal
requirement for the exact recovery of $Z_{1}$, which is established in
Theorem \ref{thm:kmeans} below. Assumption \ref{ass:tune}.4 usually holds as
the sample is split randomly and $\Theta _{0}^{\ast }$ satisfies the
normalization condition in Assumption \ref{ass:par}.1. If $\Theta _{0}^{\ast
}$ satisfies the additive structure as in Example \ref{ex:aij}, then
Assumption \ref{ass:tune}.4 provided that $\frac{1}{n_{1}}\sum_{i\in
	I_{1}}\alpha _{i}=o(\sqrt{\frac{\log (n)}{n\zeta _{n}}})$. Such a
requirement holds almost surely if $\alpha _{i}=a_{i}-\frac{1}{n}\sum_{i\in
	\lbrack n]}a_{i}$ and $\{a_{i}\}_{i=1}^{n}$ is a sequence of i.i.d. random
variables with finite second moments. If $\Theta _{0}^{\ast }$ has the
community structure as in Example \ref{ex:t0group}, then Assumption \ref%
{ass:tune}.4 holds provided that $p_{0}^{\top }(I_{1})B_{0}^{\ast }p_{0}=o(%
\sqrt{\frac{\log (n)}{n\zeta _{n}}})$, where $p_{0}^{\top }(I_{1})=(\frac{%
	n_{1,0}(I_{1})}{n_{1}},\cdots ,\frac{n_{K_{0},0}(I_{1})}{n_{1}})$ and $%
n_{k,0}(I_{1})$ denotes the size of $\Theta _{0}^{\ast }$'s $k$-th community
for the subsample of nodes with index $i\in I_{1}$. As $p_{0}^{\top
}B_{0}^{\ast }p_{0}=0$, the requirement holds almost surely if community
memberships are generated from a multinomial distribution so that $%
||p_{0}-p_{0}(I_{1})||=o_{a.s.}(\sqrt{\frac{\log (n)}{n\zeta _{n}}})$.


\begin{thm}
	\label{thm:Frobeniusnorm} Let Assumptions \ref{ass:dgp}, \ref{ass:par}, \ref%
	{ass:RSC}, and \ref{ass:tune} hold and $\eta _{n}=\sqrt{\frac{\log n}{n\zeta
			_{n}}}+\frac{\log n}{n\zeta _{n}}$. Then for $l=0,1$ and w.p.a.1, we have
	
	\begin{enumerate}
		\item $|\widehat{\tau }_{n}-\tau _{n}|\leq 30C_{F,1}\eta _{n},$ $|\widehat{%
			\tau }_{n}^{(1)}-\tau _{n}|\leq 30C_{F,1} \eta _{n},$
		
		\item $\frac{1}{n}||\widehat{\Theta }_{l}-\Theta _{l}^{\ast }||_{F}\leq
		48C_{F,1}\eta _{n},$ $\frac{1}{n}||\widehat{\Theta }_{l}^{(1)}-\Theta
		_{l}^{\ast }(I_{1})||_{F}\leq 48C_{F,1} \eta _{n},$
		
		\item $\max_{k\in [ K_{l}]}|\widehat{\sigma }_{k,l}-\sigma _{k,l}|\leq
		48C_{F,1}\eta _{n},$ $\max_{k\in [ K_{l}]}|\widehat{\sigma }%
		_{k,l}^{(1)}-\sigma _{k,l}|\leq 48C_{F,1} \eta _{n},$
		
		\item $||V_{l}-\widehat{V}_{l}\widehat{O}_{l}||_{F}\leq 136C_{F,2} \sqrt{n}%
		\eta _{n},$ and $||V_{l}-\widehat{V}_{l}^{(1)}\widehat{O}_{l}^{(1)}||_{F}%
		\leq 136C_{F,2} \sqrt{n}\eta _{n},$
		
		\noindent 
		where $\widehat{O}_{l}$ and $\widehat{O}_{l}^{(1)}$ are two $K_{l}\times
		K_{l}$ orthogonal matrices that depend on $(V_{l},\widehat{V}_{l})$ and $%
		(V_{l},\widehat{V}_{l}^{(1)})$, respectively, and $C_{F,1}$ and $C_{F,2}$
		are two constants defined respectively after (\ref{eq:CF1}) and (\ref{eq:CF2}%
		) in the Appendix.
	\end{enumerate}
\end{thm}

Part 1 of Theorem \ref{thm:Frobeniusnorm} indicates that despite the
possible divergence of the grand intercept $\tau _{n},$ we can estimate it
consistently up to rate $\eta _{n}.$ In the dense network, $\zeta _{n}\asymp
1$ where $a\asymp b$ denotes both $a/b$ and $b/a$ are stochastically
bounded. In this case, $\tau _{n}$ $\asymp 1$ and it can be estimated
consistently at rate-$\sqrt{(\log n)/n}.$ Note that the convergence rate of $%
\widehat{\Theta }_{l}$ and $\widehat{\Theta }_{l}^{(1)}$ in terms of the
Frobenius norm is also driven by $\eta _{n}.$ Similarly for $\widehat{\sigma
}_{k,l}$ $\widehat{\sigma }_{k,l}^{(1)},$ $\widehat{V}_{l}/\sqrt{n}$ and $%
\widehat{V}_{l}^{(1)}/\sqrt{n}.$ In part 4 of Theorem \ref{thm:Frobeniusnorm}%
, the orthogonal matrices $\widehat{O}_l$ and $\widehat{O}_l^{(1)}$ are
present because the singular values of $\Theta_l^*$ can be the same and its
singular vectors can only be identified up to some rotation.

\subsection{Split-Sample Row- and Column-Wise Logistic Regressions \label%
	{sec:row_theory}}

Define two $\left( K_{0}+K_{1}\right) \times \left( K_{0}+K_{1}\right) $
matrices:
\begin{equation*}
\Psi _{j}(I_{2})=\frac{1}{n_{2}}\sum_{i\in I_{2},i\neq j}%
\begin{bmatrix}
u_{i,0} \\
u_{i,1}W_{1,ij}%
\end{bmatrix}%
\begin{bmatrix}
u_{i,0} \\
u_{i,1}W_{1,ij}%
\end{bmatrix}%
^{\top }\text{ and }\Phi _{i}(I_{2})=\frac{1}{n_{2}}\sum_{j\in I_{2},j\neq i}%
\begin{bmatrix}
v_{j,0} \\
v_{j,1}W_{1,ij}%
\end{bmatrix}%
\begin{bmatrix}
v_{j,0} \\
v_{j,1}W_{1,ij}%
\end{bmatrix}%
^{\top }\text{.}
\end{equation*}%
To study the asymptotic properties of the third step estimator, we assume
that both matrices are well behaved uniformly in $i$ and $j$ in the
following assumption.

\begin{ass}
	There exist constants $C_{\phi }$ and $c_{\phi }$ such that w.p.a.1,
	\begin{eqnarray*}
		\infty &>&C_{\phi }\geq \limsup_{n}\max_{j\in [ n]}\lambda _{\max }(\Psi
		_{j}(I_{2}))\geq \liminf_{n}\min_{j\in [ n]}\lambda _{\min }(\Psi
		_{j}(I_{2}))\geq c_{\phi }>0\text{ and } \\
		\infty &>&C_{\phi }\geq \limsup_{n}\max_{i\in I_{2}}\lambda _{\max }(\Phi
		_{i}(I_{2}))\geq \liminf_{n}\min_{i\in I_{2}}\lambda _{\min }(\Phi
		_{i}(I_{2}))\geq c_{\phi }>0,
	\end{eqnarray*}%
	where $\lambda _{\max }(\cdot )$ and $\lambda _{\min }(\cdot )$ denote the
	maximum and minimum eigenvalues, respectively. \label{ass:phi}
\end{ass}

Assumption \ref{ass:phi} assumes that $\Phi _{i}(I_{2})$ and $\Psi
_{j}(I_{2})$ are positive definite (p.d.) uniformly in $i$ and $j$
asymptotically. Suppose $\Gamma _{1}$ follows the community structure as in
Example \ref{ex:t0group} with $K_{1}$ equal-sized communities and $%
B_{1}^{\ast }=I_{K_{1}}$, then $\Pi _{1,n}=\text{diag}(1/K_{1},\cdots
,1/K_{1})$. By Lemma \ref{lem:phi} in the online supplement, if node $j$ is
in community $k$, then $v_{j,1}=\sqrt{n}\sqrt{\frac{K_{1}}{n}}z_{j,1}=\sqrt{%
	K_{1}}e_{K_{1},k}$, where $e_{K_{1},k}$ denotes a $K_{1}\times 1$ vector
with the $k$-th unit being 1 and all other units being 0. In addition,
suppose $\Theta _{0}$ follows the specification in Example \ref{ex:aij}.
Then,
\begin{equation*}
\Phi _{i}(I_{2})=\frac{1}{n_{2}}\sum_{j\in I_{2}}%
\begin{pmatrix}
\frac{1}{\sqrt{2}}(1+\frac{\alpha _{0,j}}{s_{0,n}}) \\
\frac{1}{\sqrt{2}}(1-\frac{\alpha _{0,j}}{s_{0,n}}) \\
v_{j,1}W_{1,ij}%
\end{pmatrix}%
\begin{pmatrix}
\frac{1}{\sqrt{2}}(1+\frac{\alpha _{0,j}}{s_{0,n}}) \\
\frac{1}{\sqrt{2}}(1-\frac{\alpha _{0,j}}{s_{0,n}}) \\
v_{j,1}W_{1,ij}%
\end{pmatrix}%
^{\top }.
\end{equation*}%
Suppose that $\alpha _{0,i}=a_{i}-\bar{a}$ for some i.i.d. sequence $%
\{a_{i}\}_{i=1}^{n}$ with $\bar{a}=\frac{1}{n}\sum_{i=1}^{n}a_{i}$, and the
group identities of $\Theta _{1}^{\ast }$ ($\{z_{i}\}_{i\in \lbrack n]}$)
are independent of $\Theta _{0}^{\ast }$ and $\{X_{i}\}_{i\in \lbrack n]}$
and $\{e_{ij}\}_{i,j\in \lbrack n]}$. Further suppose $\mathbb{E}%
(W_{1,ij}a_{j}|X_{i})=0$, $\mathbb{E}(W_{1,ij}|X_{i})=0$, and $\mathbb{E}%
(W_{1,ij}^{2}|X_{i})\geq c>0$ for some constant $c$. Then, we can expect
that, uniformly over $i\in I_{2}$,
\begin{equation*}
\Phi _{i}(I_{2})\rightarrow \text{diag}(1,1,\mathbb{E}(W_{1,ij}^{2}|X_{i}),%
\cdots ,\mathbb{E}(W_{1,ij}^{2}|X_{i}))\text{ }a.s.,
\end{equation*}%
which implies Assumption \ref{ass:phi} holds.

If $\Theta_0^{\ast}$ has the community structure as in Example \ref%
{ex:t0group}. Further suppose $\Theta _{0}^{\ast }$ and $\Theta _{1}^{\ast }$
share the same community structure $Z_{1}$, which is independent of $W_{1}$,
$\mathbb{E}(W_{1,ij}|X_{i})=0$ and $\mathbb{E}(W_{1,ij}^{2}|X_{i})\geq c>0$
for some constant $c$, then one can expect that $\Phi _{i}(I_{2})$ has the
same limit as above uniformly over $i\in I_{2}$.

The following theorem studies the asymptotic properties of $\widehat{u}%
_{i,l}^{(1)}$ and $\dot{v}_{j,l}^{(0,1)}$ defined in Step 3.

\begin{thm}
	\label{thm:rowwisebound} Suppose that Assumptions \ref{ass:dgp}, \ref%
	{ass:par}, \ref{ass:RSC}--\ref{ass:phi} hold. Then,
	\begin{equation*}
	\max_{i\in I_{2}}||(\widehat{O}_{l}^{(1)})^{\top }\widehat{u}%
	_{i,l}^{(1)}-u_{i,l}||\leq C_{1}^{\ast }\eta _{n}\quad \text{and}\quad
	\max_{j\in [ n]}||(\widehat{O}_{l}^{(1)})^{\top }\dot{v}%
	_{j,l}^{(0,1)}-v_{j,l}||\leq C_{0,v}\eta _{n}~w.p.a.1,
	\end{equation*}%
	where $C_{1}^{\ast }$ and $C_{0,v}$ are some constants defined respectively
	in (\ref{eq:C1star}) and (\ref{eq:C0v}) in the Appendix.
\end{thm}

Theorem \ref{thm:rowwisebound} establishes the uniform bound for the
estimation error of $\dot{v}_{j,l}^{(0,1)}$ up to some rotation. However, we
only use half of the edges to estimate $\dot{v}_{j,l}^{(0,1)}$, which may
result in information loss. In the next section, we treat $\dot{v}%
_{j,l}^{(0,1)}$ as an initial value and iteratively re-estimate $%
\{u_{i,l}\}_{i\in [ n]}$ and $\{v_{j,l}\}_{i\in [ n]}$ using all the edges
in the network. We will show that the iteration can preserve the error bound
established in Theorem \ref{thm:rowwisebound}.

\subsection{Full-Sample Iteration}

Define two $\left( K_{0}+K_{1}\right) \times \left( K_{0}+K_{1}\right) $
matrices:
\begin{equation*}
\Psi _{j}=\frac{1}{n}\sum_{i\in [ n],i\neq j}%
\begin{bmatrix}
u_{i,0} \\
u_{i,1}W_{1,ij}%
\end{bmatrix}%
\begin{bmatrix}
u_{i,0} \\
u_{i,1}W_{1,ij}%
\end{bmatrix}%
^{\top }\quad \text{and}\quad \Phi _{i}=\frac{1}{n}\sum_{j\in [ n],j\neq i}%
\begin{bmatrix}
v_{j,0} \\
v_{j,1}W_{1,ij}%
\end{bmatrix}%
\begin{bmatrix}
v_{j,0} \\
v_{j,1}W_{1,ij}%
\end{bmatrix}%
^{\top }.
\end{equation*}%
To study the asymptotic properties of the fourth step estimators, we add an
assumption.

\begin{ass}
	There exist constants $C_{\phi }$ and $c_{\phi }$ such that $w.p.a.1$
	\begin{eqnarray*}
		\infty &>&C_{\phi }\geq \limsup_{n}\max_{j\in [ n]}\lambda _{\max }(\Psi
		_{j})\geq \liminf_{n}\min_{j\in [ n]}\lambda _{\min }(\Psi _{j})\geq c_{\phi
		}>0\text{ and} \\
		\infty &>&C_{\phi }\geq \limsup_{n}\max_{i\in [ n]}\lambda _{\max }(\Phi
		_{i})\geq \liminf_{n}\min_{i\in [ n]}\lambda _{\min }(\Phi _{i})\geq c_{\phi
		}>0.
	\end{eqnarray*}%
	\label{ass:phiprime}
\end{ass}

The above assumption parallels Assumption \ref{ass:phi} and is now imposed
for the full sample.

\begin{thm}
	\label{thm:iter} Suppose that Assumptions \ref{ass:dgp}, \ref{ass:par}, \ref%
	{ass:RSC}--\ref{ass:phiprime} hold. Then, for $h=1,\cdots ,H$ and $l=0,1$,
	\begin{equation*}
	\max_{i\in [ n]}||(\widehat{O}_{l}^{(1)})^{\top }\dot{u}%
	_{i,l}^{(h,1)}-u_{i,l}||\leq C_{h,u}\eta _{n}\quad \text{and}\quad
	\max_{i\in [ n]}||(\widehat{O}_{l}^{(1)})^{\top }\dot{v}%
	_{i,l}^{(h,1)}-v_{i,l}||\leq C_{h,v}\eta _{n}~w.p.a.1,
	\end{equation*}%
	where $\{C_{h,u}\}_{h=1}^{H}$ and $\{C_{h,v}\}_{h=1}^{H}$ are two sequences
	of constants defined in the proof of this theorem.
\end{thm}

Theorem \ref{thm:iter} establishes the uniform bound for the estimation
error in the iterated estimators $\{\dot{u}_{i,l}^{(h,1)}\}$ and $\{\dot{v}%
_{i,l}^{(h,1)}\}.$

By switching the roles of $I_{1}$ and $I_{2}$, we have, similar to Theorem %
\ref{thm:Frobeniusnorm}, that
\begin{equation*}
\Vert V_{l}-\widehat{V}_{l}^{(2)}\widehat{O}_{l}^{(2)}\Vert _{F}\leq
136C_{F,2}\sqrt{n}\eta _{n},
\end{equation*}%
where $\widehat{O}_{l}^{(2)}$ is a $K_{l}\times K_{l}$ rotation matrix that
depends on $V_{l}$ and $\widehat{V}_{l}^{(2)}$. Then, following the same
derivations of Theorems \ref{thm:rowwisebound} and \ref{thm:iter}, we have,
for $h=1,\cdots ,H$,
\begin{equation*}
\max_{i\in [ n]}||(\widehat{O}_{l}^{(2)})^{\top }\dot{u}%
_{i,l}^{(h,2)}-u_{i,l}||\leq C_{h,u}\eta _{n}\quad \text{and}\quad
\max_{i\in [ n]}||(\widehat{O}_{l}^{(2)})^{\top }\dot{v}%
_{i,l}^{(h,2)}-v_{i,l}||\leq C_{h,v}\eta _{n}~\text{\ }w.p.a.1
\end{equation*}

\section{K-means Classification \label{sec:kmeans'}}

If we further assume $\Theta_{1}^{\ast}$ has the community structure and
satisfies Assumption \ref{ass:pi}, then Lemma \ref{lem:theta} shows $%
\{v_{j,1}\}_{j\in [ n]}$ contains information about the community
memberships. It is intuitive to expect that we can use $\overline{v}_{j,l}$
defined in Section \ref{sec:kmeans} to recover the memberships as long as
the estimation error is sufficiently small.

Let $g_{i}^{0}\in \left[ K_{1}\right] $ denote the true group identity for
the $i$-th node in $\Theta_1^{\ast}$. To establish the strong consistency of
the membership estimator $\hat{g}_{i}$ defined in \eqref{eq:gi}, we add the
following side condition.

\begin{ass}
	Suppose $145K_{1}^{3/2}C_{H,v}C_{1}\eta _{n}\leq 1$, where $C_{H,v}$ is the
	constant defined in the proof of Theorem \ref{thm:iter}. \label{ass:kmeans}
\end{ass}

Apparently, Assumption \ref{ass:kmeans} is automatically satisfied in large
samples if $\eta _{n}=o\left( 1\right) .$ The constant in the statement is
not optimal.

\begin{thm}
	\label{thm:kmeans} If Assumptions \ref{ass:dgp}, \ref{ass:par}, \ref{ass:RSC}%
	--\ref{ass:kmeans} hold and $\Theta _{1}^{\ast }$ further satisfies
	Assumption \ref{ass:pi}, then up to some label permutation,
	\begin{equation*}
	\max_{1\leq i\leq n}\mathbf{1}\{\hat{g}_{i}\neq g_{i}^{0}\}=0~w.p.a.1.
	\end{equation*}
\end{thm}

Several remarks are in order. First, Theorem \ref{thm:kmeans} implies the
K-means algorithm can exactly recover the latent community structure of $%
\Theta _{1}^{\ast }$ $w.p.a.1$. Second, if we repeat the sample split $R$
times, we need to maintain Assumption \ref{ass:phi} for each split. Then, we
can show the exact recovery of $\hat{g}_{{i,r}}$ for $r\in [ R]$ in the
exact same manner, as long as $R$ is fixed. This implies $\hat{g}_{{i,r\ast }%
}$ for $r^{\ast }$ selected in \eqref{eq:rstar} also enjoys the property
that
\begin{equation*}
\max_{1\leq i\leq n}\mathbf{1}\{\hat{g}_{i,r^{\ast }}\neq
g_{i}^{0}\}=0~w.p.a.1.
\end{equation*}%
Third, if $\Theta _{0}^{\ast }$ also has the latent community structure as
in Example \ref{ex:t0group}, we can apply the same K-means algorithm to $\{%
\overline{v}_{j,0}\}_{j\in \left[ n\right] }$ with $\overline{v}_{j,0}\equiv
(\dot{v}_{j,0}^{(H,1)\top }/||\dot{v}_{j,0}^{(H,1)}||,\dot{v}%
_{j,0}^{(H,2)\top }/||\dot{v}_{j,0}^{(H,2)}||)^{\top }$ to recover the group
identities of $\Theta _{0}^{\ast }$. Last, if we further assume $%
Z_{0}=Z_{1}=Z$ (which implies $K_{0}=K_{1})$, then we can catenate $%
\overline{v}_{j,0}$ and $\overline{v}_{j,1}$ as a $4K_{1}\times 1$ vector
and apply the same K-means algorithm to this vector to recover the group
membership for each node.

\section{Inference for $B_{1}^{\ast }$ \label{sec:inferb}}

In this section, we maintain the assumption that $\Theta _{1}^{\ast }$ has a
latent community structure. In the general model with multiple covariates,
we allow $\{\Theta _{l}^{\ast }\}_{l\in \lbrack p]}$ to have potentially
different community structures. Note this includes the case that some of the
$\Theta _{l}^{\ast }$'s are homogeneous. We can recover the community
structures by applying the k-means algorithm in the previous section to each
$\Theta _{l}^{\ast }$.

For the rest of the section, for notation simplicity, we continue to
consider the case that there is only one covariate $W_1$ and $%
\Theta_1^{\ast} $ has a latent community structure, which is estimated by $\{%
\hat{g}_i\}_{i \in [n]}$ defined in the previous section. Given the exact
recovery of the community memberships asymptotically, we can just treat $%
\hat{g}_{i}$ as $g_{i}^{0}$.

We discuss the inference for $B_{1}^{\ast }$ for two specifications of $%
\Theta_0^{\ast}$: (1) $\Theta_{0,ij}^{\ast}$ has an additive structure as in
Example \ref{ex:aij} and (2) $\Theta_{0,ij}^{\ast}$ has a latent community
structure as in Example \ref{ex:t0group}. In the first model, once the group
membership of $\Theta_1^{\ast}$ is recovered, it boils to the one studied by
\cite{G17}. For the second model, when the memberships of both $%
\Theta_0^{\ast}$ and $\Theta_1^{\ast}$ are recovered, it boils down to the
standard logistic regression with finite-number of parameters.

\subsection{Additive Fixed Effects}

Suppose $\Gamma _{0,ij}^{\ast }=\tau _{n}+\alpha _{i}+\alpha _{j} $ and $%
\Gamma _{1}^{\ast }=\Theta _{1}^{\ast }=Z_{1}B_{1}^{\ast }Z_{1}^{\top }$.
Recall the definitions of $\chi _{1,ij},$ $\hat{\chi}_{1r,ij},$ and $\text{%
	vech}(B_{1}^{\ast })$ in Section \ref{sec:kmeans} such that $\chi
_{1,ij}^{\top }$vech$(B_{1}^{\ast })=B_{1,g_{i}^{0}g_{j}^{0}}^{\ast }.$ We
further denote $\hat{\chi}_{1,ij}$ as either $\hat{\chi}_{1,ij}$ if one
single split is used or $\hat{\chi}_{1r^{\ast },ij}$ if $R$ splits are used
and the $r^{\ast }$-th split is selected.

\begin{cor}
	\label{cor:iota} Suppose Assumptions \ref{ass:dgp}, \ref{ass:par}, \ref%
	{ass:RSC}--\ref{ass:kmeans} hold and $\Theta _{1}^{\ast }$ further satisfies
	Assumption \ref{ass:pi}. Then $\hat{\chi}_{1,ij}=\chi _{1,ij}~\forall
	i<j~w.p.a.1.$
\end{cor}

Corollary \ref{cor:iota} directly follows from Theorem \ref{thm:kmeans} and
implies that we can treat $\chi _{1,ij}$ as observed. Then, \eqref{eq:Yij}
can be written as
\begin{equation*}
Y_{ij}=\mathbf{1}\{\varepsilon _{ij}\leq \tau _{n}+\alpha _{i}+\alpha
_{j}+\omega _{1,ij}^{\top }\text{vech}(B_{1}^{\ast })\},
\end{equation*}%
where $\omega _{1,ij}=W_{1,ij}\chi _{1,ij}$. This model has already been
studied by \cite{G17}. We can directly apply his Tetrad logit regression to
estimate $\text{vec}(B_{1}^{\ast })$.

%

Let $S_{ij,i^{\prime }j^{\prime }}=Y_{ij}Y_{i^{\prime }j^{\prime
}}(1-Y_{ii^{\prime }})(1-Y_{jj^{\prime }})-(1-Y_{ij})(1-Y_{i^{\prime
	}j^{\prime }})Y_{ii^{\prime }}Y_{jj^{\prime }}.$ Then, for an arbitrary $%
K_{1}(K_{1}+1)/2$-vector $B$, the conditional likelihood of $S_{ij,i^{\prime
	}j^{\prime }}$ given $S_{ij,i^{\prime }j^{\prime }}\in \{-1,1\}$ is
\begin{equation*}
\ell _{ij,i^{\prime }j^{\prime }}(B)=|S_{ij,i^{\prime }j^{\prime }}|\left[
S_{ij,i^{\prime }j^{\prime }}\widetilde{\omega }_{1,ij,i^{\prime }j^{\prime
}}^{\top }B-\log \left( 1+\exp (S_{ij,i^{\prime }j^{\prime }}\widetilde{%
	\omega }_{1,ij,i^{\prime }j^{\prime }}^{\top }B)\right) \right] ,
\end{equation*}%
where $\widetilde{\omega }_{1,ij,i^{\prime }j^{\prime }}=\omega
_{1,ij}+\omega _{1,i^{\prime }j^{\prime }}-(\omega _{1,ii^{\prime }}+\omega
_{1,jj^{\prime }})$. Further denote
\begin{equation*}
\bar{\ell}_{ij,i^{\prime }j^{\prime }}(B)=\frac{1}{3}\left[ \ell
_{ij,i^{\prime }j^{\prime }}(B)+\ell _{ij,j^{\prime }i^{\prime }}(B)+\ell
_{ii^{\prime },j^{\prime }j}(B)\right] .
\end{equation*}%
Following \cite{G17}, we define the tetrad regression estimator $\widehat{B}$
for $\text{vech}(B^{\ast })$ as
\begin{equation*}
\widehat{B}=\argmax_{B}\sum_{i<i^{\prime }<j<j^{\prime }}\bar{\ell}%
_{ij,i^{\prime }j^{\prime }}(B).
\end{equation*}

Let
\begin{equation*}
\top _{iji^{\prime }j^{\prime }}=%
\begin{cases}
1 & \text{ if }S_{ij,i^{\prime }j^{\prime }}\in \{-1,1\}\cup S_{ij,j^{\prime
	}i^{\prime }}\in \{-1,1\}\cup S_{ii^{\prime },jj^{\prime }}\in \{-1,1\} \\
0 & \text{ otherwise}%
\end{cases}%
\end{equation*}%
be the indicator that the tetrad $\{i,j,i^{\prime },j^{\prime }\}$ take an
identifying configuration, and thus, contributes to the tetrad logit
regression. Further denote $t_{q,n}=\mathbb{P}(\top
_{i_{1}i_{2}i_{3}i_{4}}=1,\top _{jj_{2}j_{3}j_{4}}=1)$ as the probability
that tetrads $\{i_{1},i_{2},i_{3},i_{4}\}$ and $\{j,j_{2},j_{3},j_{4}\}$
both take an identifying configuration when sharing $q=0,1,2,3$, or 4 nodes
in common. Then, we make the following assumption on the Hessian matrix.

\begin{ass}
	\label{ass:hessian} Suppose that $\Upsilon _{0}\equiv \lim_{n\rightarrow
		\infty }t_{4,n}^{-1}\sum_{i<i^{\prime }<j<j^{\prime }}\nabla _{BB}\bar{\ell}%
	_{ij,i^{\prime }j^{\prime }}(B)$ is a finite nonsingular matrix.
\end{ass}

The following theorem reports the asymptotic normality of $\widehat{B}.$

\begin{thm}
	\label{thm:infer} Suppose that Assumptions \ref{ass:dgp}, \ref{ass:par}, \ref%
	{ass:RSC}--\ref{ass:hessian} hold. Suppose that $\Gamma _{0}^{\ast }=\tau
	_{n}+\alpha _{i}+\alpha _{j}$ and $\Theta _{1}^{\ast }$ satisfies Assumption %
	\ref{ass:pi}. Then $\widehat{B}\overset{p}{\longrightarrow }\text{vec}%
	(B^{\ast })$ and
	\begin{equation*}
	\left[ \frac{72}{(n-1)n}\hat{H}^{-1}\widehat{\Delta }_{2,n}\hat{H}^{-1}%
	\right] ^{-1/2}(\widehat{B}-\text{vech}(B^{\ast }))\rightsquigarrow \mathcal{%
		N}(0,I_{K_{1}(K_{1}+1)/2}),
	\end{equation*}%
	where
	\begin{equation*}
	\hat{H}=\binom{n}{4}^{-1}\sum_{i<j<i^{\prime }<j^{\prime }}\frac{\partial
		^{2}\bar{\ell}_{ij,i^{\prime }j^{\prime }}(\widehat{B})}{\partial B\partial
		B^{\top }},\quad \widehat{\Delta }_{2,n}=\frac{2}{n(n-1)}\sum_{i<j}\hat{\bar{%
			s}}_{ij}(\widehat{B})\hat{\bar{s}}_{ij}(\widehat{B})^{\top },
	\end{equation*}%
	$\hat{\bar{s}}_{ij}(B)=\frac{1}{n(n-1)/2-2(n-1)+1}\sum_{i^{\prime
		}<j^{\prime },\{i,j\}\cap \{i^{\prime },j^{\prime }\}=\emptyset
	}s_{ij,i^{\prime }j^{\prime }}(B)$, $s_{ij,i^{\prime }j^{\prime }}(B)=\nabla
	_{B}\bar{\ell}_{ij,i^{\prime }j^{\prime }}(B)$, and $I_{a}$ denotes an $%
	a\times a$ identity matrix.
\end{thm}

Theorem \ref{thm:infer} imposes two additional structures in order to make
the inferences on $B^{\ast }$ by borrowing the asymptotic results from \cite%
{G17}. One is that $\Gamma _{0}^{\ast }$ exhibits the usual additive fixed
effects structure (with $K_{0}=2$) and the other is $\Gamma _{1}^{\ast }$
has a latent community structure. The model reduces to that of \cite{G17} in
the special case of $K_{1}=1.$

\subsection{Latent Community Structure in the Fixed Effects}

Let $g_{i,0}^{0}$ be the true memberships of node $i$ for $\Theta _{0}^{\ast
}$ and $\hat{g}_{i,0}$ be its estimator which can be computed by applying
the K-means algorithm to $\{\overline{v}_{j,0}\}_{j\in \lbrack n]}$. Further
note $Z_{0}\iota _{K_{0}}=\iota _{n}$ where recall that $\iota _{b}$ denotes
a $b\times 1$ vector of ones. Therefore, $\Gamma _{0}^{\ast }=\tau _{n}\iota
_{n}\iota _{n}^{\top }+Z_{0}B_{0}^{\ast }Z_{0}^{\top }=Z_{0}(B_{0}^{\ast
}+\tau _{n}\iota _{K_{0}}\iota _{K_{0}}^{\top })Z_{0}^{\top }\equiv
Z_{0}B_{0}^{\ast \ast }Z_{0}^{\top }$, i.e., $\Gamma _{0}^{\ast }$ shares
the same community structure as $\Theta _{0}^{\ast }$. We then define $\chi
_{0,ij}$ be a $K_{0}(K_{0}+1)/2\times 1$ vector whose $((g_{i,0}^{0}\vee
g_{j,0}^{0}-1)(g_{i,0}^{0}\vee g_{j,0}^{0})/2+g_{i,0}^{0}\wedge g_{j,0}^{0})$%
-th element is one and the rest are zeros and $\hat{\chi}_{0,ij}$ be a $%
K_{0}(K_{0}+1)/2\times 1$ vector whose $((\hat{g}_{i,0}\vee \hat{g}_{j,0}-1)(%
\hat{g}_{i,0}\vee \hat{g}_{j,0})/2+\hat{g}_{i,0}\wedge \hat{g}_{j,0})$-th
element is one and the rest are zeros. Similar to Corollary \ref{cor:iota},
we have the following corollary.

\begin{cor}
	\label{cor:iota'} Suppose that Assumptions \ref{ass:dgp}, \ref{ass:par}, \ref%
	{ass:RSC}--\ref{ass:kmeans} hold. Suppose that $\Theta _{l}^{\ast },$ $l=0,1,
	$ further satisfy Assumption \ref{ass:pi}. Then, $\hat{\chi}_{l,ij}=\chi
	_{l,ij}~\forall i<j$ for $l=0,1~w.p.a.1.$
\end{cor}

We propose to estimate $\text{vech}(B^{\ast })\equiv (\text{vech}%
(B_{0}^{\ast \ast })^{\top },\text{vech}(B_{1}^{\ast })^{\top })^{\top }$ by
\begin{equation*}
\widehat{B}\equiv (\widehat{B}_{0}^{\top },\widehat{B}_{1}^{\top })^{\top }=%
\argmin_{b=(b_{0}^{\top },b_{1}^{\top })^{\top }\in \mathbb{R}%
	^{K_{0}(K_{0}+1)/2}\times \mathbb{R}^{K_{1}(K_{1}+1)/2}}Q_{n}(b),
\end{equation*}%
where
\begin{equation*}
Q_{n}(b)=\frac{-1}{n(n-1)}\sum_{1\leq i<j\leq n}[Y_{ij}\log (\hat{\Lambda}%
_{ij}(b))+(1-Y_{ij})\log (1-\hat{\Lambda}_{ij}(b))],
\end{equation*}
and
\begin{equation*}
\hat{\Lambda}_{ij}(b)=\Lambda (\hat{\chi}_{0,ij}^{\top }b_{0}+\hat{\chi}%
_{1,ij}^{\top }W_{1,ij}b_{1}).
\end{equation*}

Let $\Lambda _{n,ij}(u)=\Lambda (\omega _{ij}^{\top }[$vech$(B^{\ast
})+u(n^{2}\zeta _{n})^{-1/2}])$ and $\Lambda _{n,ij}\equiv \Lambda
_{n,ij}(0),$ where\ $\omega _{ij}=(\chi _{0,ij}^{\top },\chi _{1,ij}^{\top
}W_{1,ij})^{\top }$ is an\ $\mathcal{K}$-vector with $\mathcal{K}%
=\sum_{l=0}^{1}K_{l}(K_{l}+1)/2.$ Note that $\Lambda _{n,ij}=\Lambda
(W_{ij}^{\top }\Gamma _{ij}^{\ast }).$

\begin{ass}
	\label{ass:H} $\sup_{\left\Vert u\right\Vert \leq C}\frac{1}{n^{2}\zeta _{n}}
	\sum_{1\leq i<j\leq n}\Lambda _{n,ij}\left( u\right) (1-\Lambda
	_{n,ij}(u))\omega _{ij}\omega _{ij}^{\top }\overset{p}{\longrightarrow }
	\mathcal{H}$ for some positive-definite matrix $\mathcal{H}$ and large but
	fixed constant $C.$
\end{ass}

\begin{thm}
	\label{thm:e2} Suppose that Assumptions \ref{ass:dgp}, \ref{ass:par}, \ref%
	{ass:RSC}--\ref{ass:kmeans}, \ref{ass:H} hold and $\Theta _{l}^{\ast },$ $%
	l=0,1,$ further satisfy Assumption \ref{ass:pi}. Let $\widehat{\mathcal{H}}%
	_{n}=\sum_{1\leq i<j\leq n}\Lambda (\omega _{ij}^{\top }\hat{B})(1-\Lambda
	(\omega _{ij}^{\top }\hat{B}))\omega _{ij}\omega _{ij}^{\top }.$ Then
	\begin{equation*}
	\widehat{\mathcal{H}}_{n}^{-1/2}(\widehat{B}-\text{vech}(B^{\ast
	}))\rightsquigarrow \mathcal{N}(0,I_{\mathcal{K}}).
	\end{equation*}
\end{thm}

Although in theory, the inference for $B_{1}^{\ast }$ in the above two cases
is straightforward, there are two finite-sample issues. First, the tetrad
logistic regression does not scale with the number of nodes $n$ because the
algorithm scans over all four-nodes figurations, which contains a total of $%
O(n^{4})$ operations in a brutal force implementation. Although the Python
code by \cite{G17} incorporates a number of computational speed-ups by
keeping careful track of non-contributing configurations as the estimation
proceeds, we still find in our simulations that the implementation turns
extremely hard for networks with over 1000 nodes. One can, instead, use
subsampling or divide-and-conquer algorithm for estimation. To establish the
theoretical properties of such an estimator is an important and interesting
topic for future research.

Second, for the specification in the second example, based on unreported
simulation results, we find that $\widehat{B}_{1}$ has a small bias if there
are some misclassified nodes. However, as the standard error of our
estimator is even smaller, such a small bias may not be ignored in making
inferences. If we further increase the sample size, then the classification
indeed achieves exact recovery and such a bias vanishes quickly. However, in
practice, researchers cannot know whether their sample size is sufficiently
large. It is interesting to further investigate such a bias issue and make
proper bias-corrections. This is, again, left as a topic for future research.

\section{Determination of $K_{0}$ and $K_{1}$ \label{sec:determine_Kl}}

In practice, $K_{0}$ and $K_{1}$ are unknown and need to be estimated from
the data. In this case, we propose to replace them by a large but fixed
integer $K_{\max }$ in the first step estimation to obtain the singular
value estimates $\left\{ \hat{\sigma}_{k,l}\right\} _{k\in \left[ K_{\max }%
	\right] ,l=0,1}.$ We propose a version of singular-value ratio (SVR)
statistic in the spirit of the eigenvalue-ratio statistics of \cite{AH2013}
and \cite{LY2012}. That is, for $l=0,1,$ we estimate $K_{l}$ by
\begin{equation}
\widehat{K}_{l}=\arg \max_{1\leq k\leq K_{\max }-1}\frac{\widehat{\sigma }%
	_{k,l}}{\widehat{\sigma }_{k+1,l}}\mathbf{1}\left\{ \widehat{\sigma }%
_{k,l}\geq c_{l}\left( \sqrt{\frac{\log n}{n\overline{Y}}}+\frac{\log n}{n%
	\overline{Y}}\right) \right\}   \label{eq:Kl}
\end{equation}%
where $\bar{Y}=\frac{2}{n(n-1)}\sum_{1\leq i<j\leq n}Y_{ij},$ and $c_{l}$ is
a tuning parameter to be specified. Without the indicator function in the
above definition, $\widehat{K}_{l}$ is nothing but the SVR statistic. The
use of the indicator function helps to avoid the overestimation of the
ranks. Apparently, $n\bar{Y}$ consistently estimate the expected degree that
is of order $n\zeta _{n}.$ By using Assumption \ref{ass:pi} and the results
in Theorem \ref{thm:Frobeniusnorm}, we can readily establish the consistency
of $\hat{K}_{l}$.

\section{Monte Carlo Simulations \label{sec:sim}}

In this section, we conduct some simulations to evaluate the performance of
our procedure.

\subsection{Data generation mechanisms}

We generate data from the following two models.

\textbf{Model 1. }We simulate the responses $Y_{ij}$ from the Bernoulli
distribution with mean $\Lambda (\log (\zeta _{n})+\Theta _{0,ij}^{\ast
}+W_{1,ij}\Theta _{1,ij}^{\ast })$ for $i<j$, where $\Theta _{0,ij}^{\ast
}=\alpha _{i}+\alpha _{j}$ and $\Theta _{1}^{\ast }=ZB_{1}^{\ast }Z^{\top }$%
. We generate $\alpha _{i}\overset{i.i.d}{\sim }$ $\mathcal{U}(-1/2,1/2)$
for $i=1,...,n$, and $W_{1,ij}=|X_{i}-X_{j}|$ for $i\neq j$, where $X_{i}$ $%
\overset{i.i.d}{\sim }\mathcal{N}(0,1)$. For the $i^{\text{th}}$ row of the
membership matrix $Z\in \mathbb{R}^{n\times K_{1}}$, the $C_{i}^{\text{th}}$
component is $1$ and other entries are $0$, where $C=(C_{1},...,C_{n})^{\top
}\in \mathbb{R}^{n}$ is the membership vector with $C_{i}\in \left[ K_{1}%
\right] $.

\textbf{Case 1.} Let $K_{1}=2$ and $B_{1}^{\ast }=((0.6,0.2)^{\top
},(0.2,0.7)^{\top })^{\top }$. The membership vector $C=(C_{1},...,C_{n})^{%
	\top }$ is generated by sampling each entry independently from $\{1,2\}$
with probabilities $\{0.4,0.6\}$. Let $\zeta _{n}=0.7n^{-1/2}\log n$.

\textbf{Case 2. }Let $K_{1}=3$ and $B_{1}^{\ast }=((0.8,0.4,0.3)^{\top
},(0.4,0.7,0.4)^{\top },(0.3,0.4,0.8)^{\top })^{\top }$. The membership
vector $C=(C_{1},...,C_{n})^{\top }$ is generated by sampling each entry
independently from $\{1,2,3\}$ with probabilities $\{0.3,0.3,0.4\}$. Let $%
\zeta _{n}=1.5n^{-1/2}\log n$.\bigskip

\textbf{Model 2. }We simulate the responses $Y_{ij}$ from the Bernoulli
distribution with mean $\Lambda (\log (\zeta _{n})+\Theta _{0,ij}^{\ast
}+W_{1,ij}\Theta _{1,ij}^{\ast })$ for $i<j$, where $\Theta _{0}^{\ast
}=ZB_{0}^{\ast }Z^{\top }$, $\Theta _{1}^{\ast }=ZB_{1}^{\ast }Z^{\top }$,
and $W_{1,ij}$ is simulated in the same way as in Model 1. Note here we
impose that the latent community structures for $\Theta _{0}^{\ast }$ and $%
\Theta _{1}^{\ast }$ are the same. We then apply the K-means algorithm to
the $4K_{1}\times 1$ vector $\{\overline{v}_{j,0}^{\top },\overline{v}%
_{j,1}^{\top }\}_{j\in [ n]}$ to recover the community membership, as
described in Section \ref{sec:kmeans'}.

\textbf{Case 1.} Let $K_{0}=K_{1}=2$ and $B_{0}^{\ast }=((0.6,0.2)^{\top
},(0.2,0.7)^{\top })^{\top }$, $B_{1}^{\ast }=((0.6,0.2)^{\top
},(0.2,0.5)^{\top })^{\top }$. The membership vector $C=(C_{1},...,C_{n})^{%
	\top }$ is generated by sampling each entry independently from $\{1,2\}$
with probabilities $\{0.3,0.7\}$. Let $\zeta _{n}=0.5n^{-1/2}\log n$.

\textbf{Case 2. }Let $K_{0}=K_{1}=3$ and $B_{0}^{\ast
}=((0.7,0.2,0.2)^{\top},(0.2,0.6,0.2)^{\top },(0.2,0.2,0.7)^{\top })^{\top }$%
, $B_{1}^{\ast }=((0.7,0.3,0.2)^{\top },(0.3,0.7,0.2)^{\top
},(0.2,0.2,0.6)^{\top })^{\top }$. The membership vector is generated in the
same way as given in Case 2 of Model 1. Let $\zeta _{n}=1.5n^{-1/2}\log n$.

We consider $n=500,$ $1000,$ and $1500$. All simulation results are based on
200 realizations.

\subsection{Simulation Results}

We select the number of communities $K_{1}$ by an eigenvalue ratio method
given as follows. Let $\widehat{\sigma }_{1,1}\geq \cdots \geq \widehat{%
	\sigma }_{K_{\max },1}$ be the first $K_{\max }$ singular values of the SVD
decomposition of $\widehat{\Theta }_{1}$ from the nuclear norm penalization
method given in Section \ref{sec:fslre}. We estimate $K_{1}$ by $\widehat{K}%
_{1}$ defined in (\ref{eq:Kl}) by setting $c_{1}=0.1$ and $K_{\max }=10$. We
set the tuning parameter $\lambda _{n}=C_{\lambda }\{\sqrt{n\overline{Y}}+%
\sqrt{\log n}\}/\{n(n-1)\}$ with $C_{\lambda }=2$ and similarly for $\lambda
_{n}^{\left( 1\right) }$. To require that the estimator of $\widehat{\Theta }%
_{l,ij}$ is bounded by finite constants, we let $M=2$ and $C_{M}=2$. The
performance of the method is not sensitive to the choice of these finite
constants. Define the mean squared error (MSE) of the nuclear norm estimator
$\widehat{\Theta }_{l}$ for $\Theta _{l}$ as $\sum\nolimits_{i\neq j}(%
\widehat{\Theta }_{l,ij}-\Theta _{l,ij}^{\ast })^{2}/\{n(n-1)\}$ for $l=0,1$.

Table \ref{Table:K} reports the MSEs for $\widehat{\Theta } _{l}$, the mean
of $\widehat{K}_{1}$ and the percentage of correctly estimating $K_{1}$
based on the 200 realizations. We observe that the mean value of $\widehat{K}%
_{1}$ gets closer to the true number of communities $K_{1}$ and, the
percentage of correctly estimating $K_{1}$ approaches to 1, as the samples
size $n$ increases. When $n$ is large enough ($n=1500$), the mean value of $%
\widehat{K}_{1}$ is the same as $K_{1}$ and the percentage of correctly
estimating $K$ is exactly equal to 1.
\begin{table}[tbph]
	\caption{{}The MSEs for $\protect\widehat{\Theta} _{l}$, the mean of $%
		\protect\widehat{K}_{1}$ and the percentage of correctly estimating $K_{1}$
		based on the 200 realizations for Models 1 and 2.}
	\label{Table:K}
	\begin{center}
		{{\
				\begin{tabular*}{1\textwidth}{l|@{\extracolsep{\fill}}cccccc}
					\hline\hline
					& \multicolumn{3}{c}{$K_{1}=2$} & \multicolumn{3}{c}{$K_{1}=3$} \\ \hline
					$n$ & $500$ & $1000$ & $1500$ & $500$ & $1000$ & $1500$ \\ \hline
					\multicolumn{7}{l}{\ \ \ \ \ \ \ \ \ \ \ \ \ \ \ \ \ \ \ \ \ \ \ \ \ \ \ \ \
						\ \ \ \ \ \ \ \ \ \ \ \ \ \ \ \ \ \ \ \ \ \ \ \ \ \ \ \ \ \ \ \ \ \ \ \ \ \
						\ Model 1} \\ \hline
					MSE for $\widehat{\Theta} _{0}$ & 0.083 & 0.079 & 0.092 & 0.112 & 0.091 &
					0.088 \\
					MSE for $\widehat{\Theta} _{1}$ & 0.226 & 0.215 & 0.211 & 0.256 & 0.263 &
					0.265 \\
					\multicolumn{1}{l|}{mean of $\widehat{K}_{1}$} & 1.990 & 2.000 & 2.000 &
					2.990 & 3.000 & 3.000 \\
					\multicolumn{1}{l|}{percentage} & 0.990 & 1.000 & 1.000 & 0.990 & 1.000 &
					1.000 \\ \hline
					\multicolumn{7}{l}{\ \ \ \ \ \ \ \ \ \ \ \ \ \ \ \ \ \ \ \ \ \ \ \ \ \ \ \ \
						\ \ \ \ \ \ \ \ \ \ \ \ \ \ \ \ \ \ \ \ \ \ \ \ \ \ \ \ \ \ \ \ \ \ \ \ \ \
						\ Model 2} \\ \hline
					MSE for $\widehat{\Theta} _{0}$ & 0.304 & 0.318 & 0.328 & 0.173 & 0.184 &
					0.196 \\
					MSE for $\widehat{\Theta} _{1}$ & 0.150 & 0.157 & 0.170 & 0.153 & 0.155 &
					0.151 \\
					mean of $\widehat{K}_{1}$ & 1.980 & 2.005 & 2.000 & 2.725 & 3.000 & 3.000 \\
					percentage & 0.980 & 0.995 & 1.000 & 0.705 & 1.000 & 1.000 \\ \hline\hline
				\end{tabular*}
		}}
	\end{center}
\end{table}

Next, we use three commonly used criteria for evaluating the accuracy of
membership estimation for our proposed method. These criteria include the
Normalized Mutual Information (NMI), the Rand Index (RI) and the proportion
(PROP) of nodes whose memberships are correctly identified. They all give a
value between 0 and 1, where 1 means a perfect membership estimation. Table %
\ref{Table:membership} presents the mean of the NMI, RI and PROP values
based on the 200 realizations for Models 1 and 2. The values of NMI, RI and
PROP increase to 1 as the sample size increases for all cases. These results
demonstrate that our method is quite effective for membership estimation in
both models, and corroborate our large-sample theory.
\begin{table}[tbph]
	\caption{{}The means of the NMI, RI and PROP values based on the 200
		realizations for Models 1 and 2.}
	\label{Table:membership}
	\begin{center}
		{{\
				\begin{tabular*}{1\textwidth}{l|@{\extracolsep{\fill}}cccccc}
					\hline\hline
					& \multicolumn{3}{c}{$K_{1}=2$} & \multicolumn{3}{c}{$K_{1}=3$} \\ \hline
					$n$ & $500$ & $1000$ & $1500$ & $500$ & $1000$ & $1500$ \\ \hline
					\multicolumn{7}{l}{\ \ \ \ \ \ \ \ \ \ \ \ \ \ \ \ \ \ \ \ \ \ \ \ \ \ \ \ \
						\ \ \ \ \ \ \ \ \ \ \ \ \ \ \ \ \ \ \ \ \ \ \ \ \ \ \ \ \ \ \ \ \ \ \ \ \
						Model 1} \\ \hline
					NMI & 0.9247 & 0.9976 & 0.9978 & 0.5494 & 0.7867 & 0.8973 \\
					RI & 0.9807 & 0.9995 & 0.9996 & 0.7998 & 0.9062 & 0.9593 \\
					PROP & 0.9903 & 0.9999 & 0.9999 & 0.8063 & 0.9089 & 0.9670 \\ \hline
					\multicolumn{7}{l}{\ \ \ \ \ \ \ \ \ \ \ \ \ \ \ \ \ \ \ \ \ \ \ \ \ \ \ \ \
						\ \ \ \ \ \ \ \ \ \ \ \ \ \ \ \ \ \ \ \ \ \ \ \ \ \ \ \ \ \ \ \ \ \ \ \ \
						Model 2} \\ \hline
					NMI & 0.9488 & 0.9977 & 0.9984 & 0.9664 & 0.9843 & 0.9977 \\
					RI & 0.9881 & 0.9966 & 0.9998 & 0.9790 & 0.9909 & 0.9987 \\
					PROP & 0.9940 & 0.9978 & 0.9999 & 0.9838 & 0.9928 & 0.9988 \\ \hline\hline
				\end{tabular*}
		}}
	\end{center}
\end{table}
\ \

Last, we estimate the parameters $B_{0}^{\ast }$ and $B_{1}^{\ast }$ by our
proposed method given in Section \ref{sec:inferb} for Model 2. Tables \ref%
{Table:B0} and \ref{Table:B1} show the empirical coverage rate (coverage) of
the $95\%$ confidence intervals, the absolute value of bias (bias), the
empirical standard deviation (emp\_sd), and the average value of the
estimated asymptotic standard deviation (asym\_sd) of the estimates for $%
B_{0}^{\ast }$ and $B_{1}^{\ast }$ in cases 1 and 2 of model 2,
respectively, based on 200 realizations. We observe that the emp\_sd and
asym\_sd decrease and the empirical coverage rate gets close to the nominal
level $0.95$, as the sample size increases. Moreover, the value of emp\_sd
is similar to that of asym\_sd for each parameter. This result confirms our
established formula (in the online supplement) for the asymptotic variances
of the estimators for the parameters. When the sample size is large enough $%
(n=1500) $, the value of bias is very small compared to asym\_sd, so that it
can be negligible for constructing confidence intervals of the parameters.

\begin{table}[tbph]
	\caption{The empirical coverage rate (coverage), the absolute bias (bias),
		empirical standard deviation (emp\_sd) and asymptotic standard deviation
		(asym\_sd) of the estimators for $B_{0}^{\ast }$ and $B_{1}^{\ast }$ in case
		1 of Model 2 based on 200 realizations. }
	\label{Table:B0}
	\begin{center}
		{{\
				\begin{tabular*}{1\textwidth}{ll|@{\extracolsep{\fill}}ccc|ccc}
					\hline\hline
					$n$ &  & $B_{0,11}^{\ast }$ & $B_{0,12}^{\ast }$ & $B_{0,22}^{\ast }$ & $%
					B_{1,11}^{\ast }$ & $B_{1,12}^{\ast }$ & $B_{1,22}^{\ast }$ \\ \hline
					& coverage & 0.880 & 0.860 & 0.975 & 0.960 & 0.915 & 0.955 \\
					$500$ & bias & 0.023 & 0.020 & 0.003 & 0.002 & 0.007 & 0.001 \\
					& emp\_sd & 0.042 & 0.036 & 0.014 & 0.021 & 0.018 & 0.009 \\
					& asym\_sd & 0.035 & 0.029 & 0.015 & 0.020 & 0.017 & 0.009 \\ \hline
					& coverage & 0.960 & 0.940 & 0.945 & 0.945 & 0.945 & 0.940 \\
					$1000$ & bias & 0.004 & 0.001 & $<0.001$ & 0.002 & 0.002 & $<0.001$ \\
					& emp\_sd & 0.017 & 0.016 & 0.008 & 0.010 & 0.009 & 0.005 \\
					& asym\_sd & 0.018 & 0.015 & 0.008 & 0.011 & 0.008 & 0.005 \\ \hline
					& coverage & 0.945 & 0.955 & 0.945 & 0.945 & 0.945 & 0.940 \\
					$1500$ & bias & $<0.001$ & 0.001 & 0.001 & 0.001 & 0.001 & $<0.001$ \\
					& emp\_sd & 0.014 & 0.011 & 0.006 & 0.008 & 0.006 & 0.003 \\
					& asym\_sd & 0.013 & 0.011 & 0.005 & 0.007 & 0.006 & 0.003 \\ \hline\hline
				\end{tabular*}
		}}
	\end{center}
\end{table}

\begin{table}[tbph]
	\caption{The empirical coverage rate (coverage), the absolute bias (bias),
		empirical standard deviation (emp\_sd) and asymptotic standard deviation
		(asym\_sd) of the estimators for $B_{0}^{\ast }$ and $B_{1}^{\ast }$ in case
		of Model 2 based on 200 realizations. }
	\label{Table:B1}
	\begin{center}
		{\
			\begin{tabular*}{1\textwidth}{ll|@{\extracolsep{\fill}}cccccc}
				\hline\hline
				$n$ &  & $B_{0,11}^{\ast }$ & $B_{0,12}^{\ast }$ & $B_{0,13}^{\ast }$ & $%
				B_{0,22}^{\ast }$ & $B_{0,23}^{\ast }$ & $B_{0,33}^{\ast }$ \\ \hline
				& coverage & 0.910 & 0.920 & 0.900 & 0.875 & 0.925 & 0.960 \\
				$500$ & bias & 0.018 & 0.025 & $<0.001$ & 0.008 & 0.002 & 0.009 \\
				& emp\_sd & 0.033 & 0.029 & 0.035 & 0.030 & 0.028 & 0.032 \\
				& asym\_sd & 0.033 & 0.031 & 0.032 & 0.028 & 0.027 & 0.032 \\ \hline
				& coverage & 0.915 & 0.935 & 0.955 & 0.930 & 0.950 & 0.925 \\
				$1000$ & bias & 0.005 & 0.005 & 0.001 & 0.004 & 0.006 & 0.006 \\
				& emp\_sd & 0.018 & 0.016 & 0.015 & 0.014 & 0.014 & 0.017 \\
				& asym\_sd & 0.017 & 0.015 & 0.017 & 0.013 & 0.014 & 0.016 \\ \hline
				& coverage & 0.940 & 0.945 & 0.940 & 0.960 & 0.940 & 0.955 \\
				$1500$ & bias & 0.001 & 0.001 & $<0.001$ & 0.001 & 0.002 & $<0.001$ \\
				& emp\_sd & 0.012 & 0.010 & 0.012 & 0.008 & 0.009 & 0.011 \\
				& asym\_sd & 0.011 & 0.010 & 0.011 & 0.009 & 0.010 & 0.011 \\ \hline
				$n$ &  & $B_{1,11}^{\ast }$ & $B_{1,12}^{\ast }$ & $B_{1,13}^{\ast }$ & $%
				B_{1,22}^{\ast }$ & $B_{1,23}^{\ast }$ & $B_{1,33}^{\ast }$ \\ \hline
				& coverage & 0.885 & 0.900 & 0.915 & 0.900 & 0.960 & 0.925 \\
				$500$ & bias & 0.020 & 0.005 & 0.001 & 0.016 & $<0.001$ & 0.005 \\
				& emp\_sd & 0.023 & 0.019 & 0.020 & 0.021 & 0.017 & 0.022 \\
				& asym\_sd & 0.025 & 0.019 & 0.019 & 0.020 & 0.016 & 0.022 \\ \hline
				& coverage & 0.930 & 0.905 & 0.945 & 0.925 & 0.940 & 0.930 \\
				$1000$ & bias & 0.003 & 0.001 & 0.006 & 0.007 & 0.002 & 0.002 \\
				& emp\_sd & 0.011 & 0.011 & 0.011 & 0.009 & 0.008 & 0.011 \\
				& asym\_sd & 0.012 & 0.009 & 0.010 & 0.009 & 0.008 & 0.011 \\ \hline
				& coverage & 0.940 & 0.955 & 0.940 & 0.960 & 0.960 & 0.950 \\
				$1500$ & bias & $<0.001$ & $<0.001$ & $<0.001$ & 0.001 & $<0.001$ & 0.001 \\
				& emp\_sd & 0.009 & 0.006 & 0.007 & 0.005 & 0.005 & 0.007 \\
				& asym\_sd & 0.008 & 0.006 & 0.007 & 0.006 & 0.006 & 0.007 \\ \hline\hline
			\end{tabular*}
		}
	\end{center}
\end{table}

\section{Empirical applications \label{sec:app}}

In this section, we apply the proposed method to study the community
structure of social network datasets

\subsection{Pokec social network}

\subsubsection{The dataset and model}

\begin{figure}[tbp]
	\caption{left panel depicts the number of nodes in different age groups;
		right panel shows the boxplots of degrees by age groups. }
	\label{Fig:Pokec_age}\centering
	\vspace{0.5cm} $%
	\begin{array}{cc}
	\includegraphics[width = 0.45\linewidth]{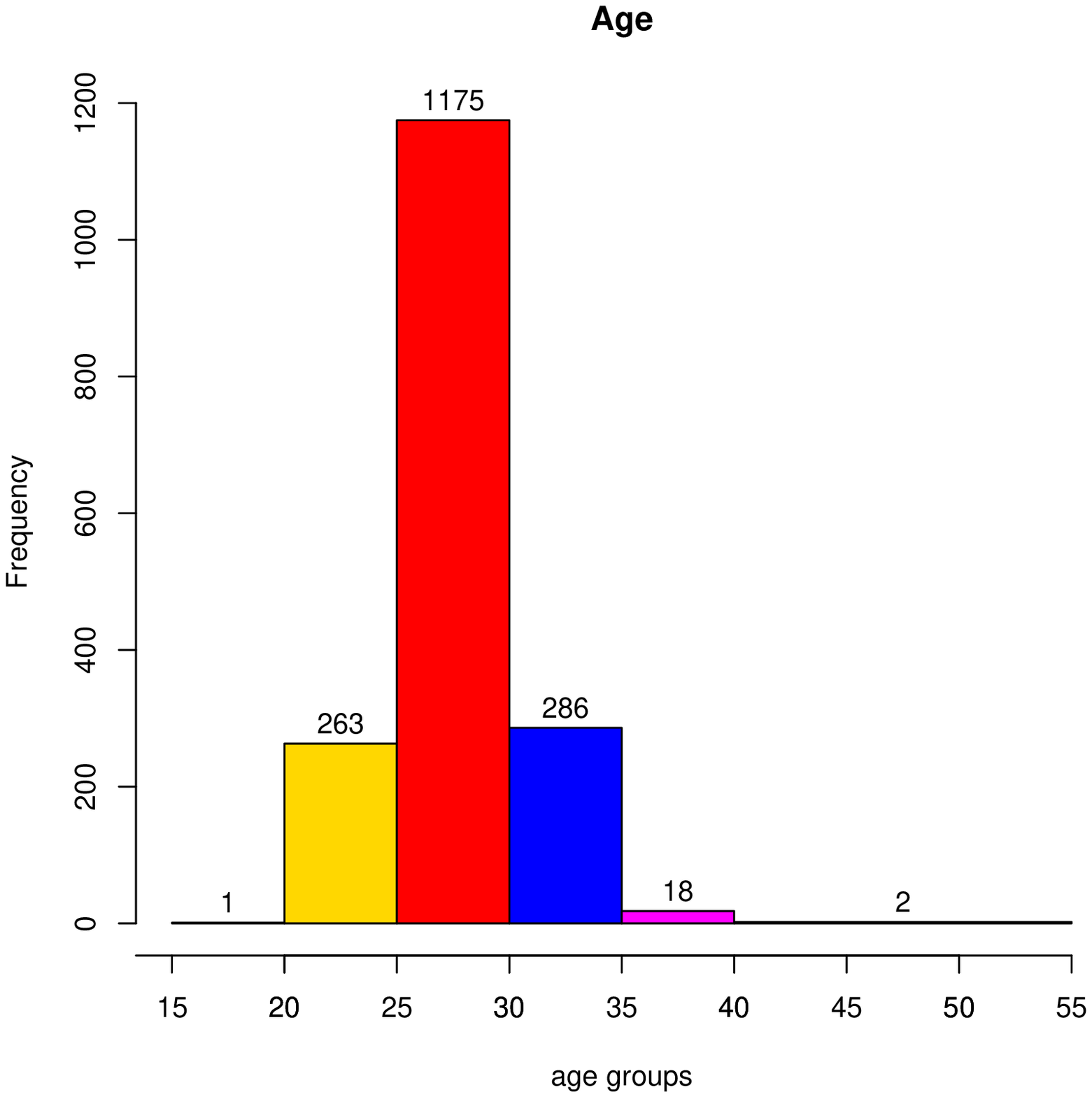} &
	\includegraphics[width
	=0.45\linewidth]{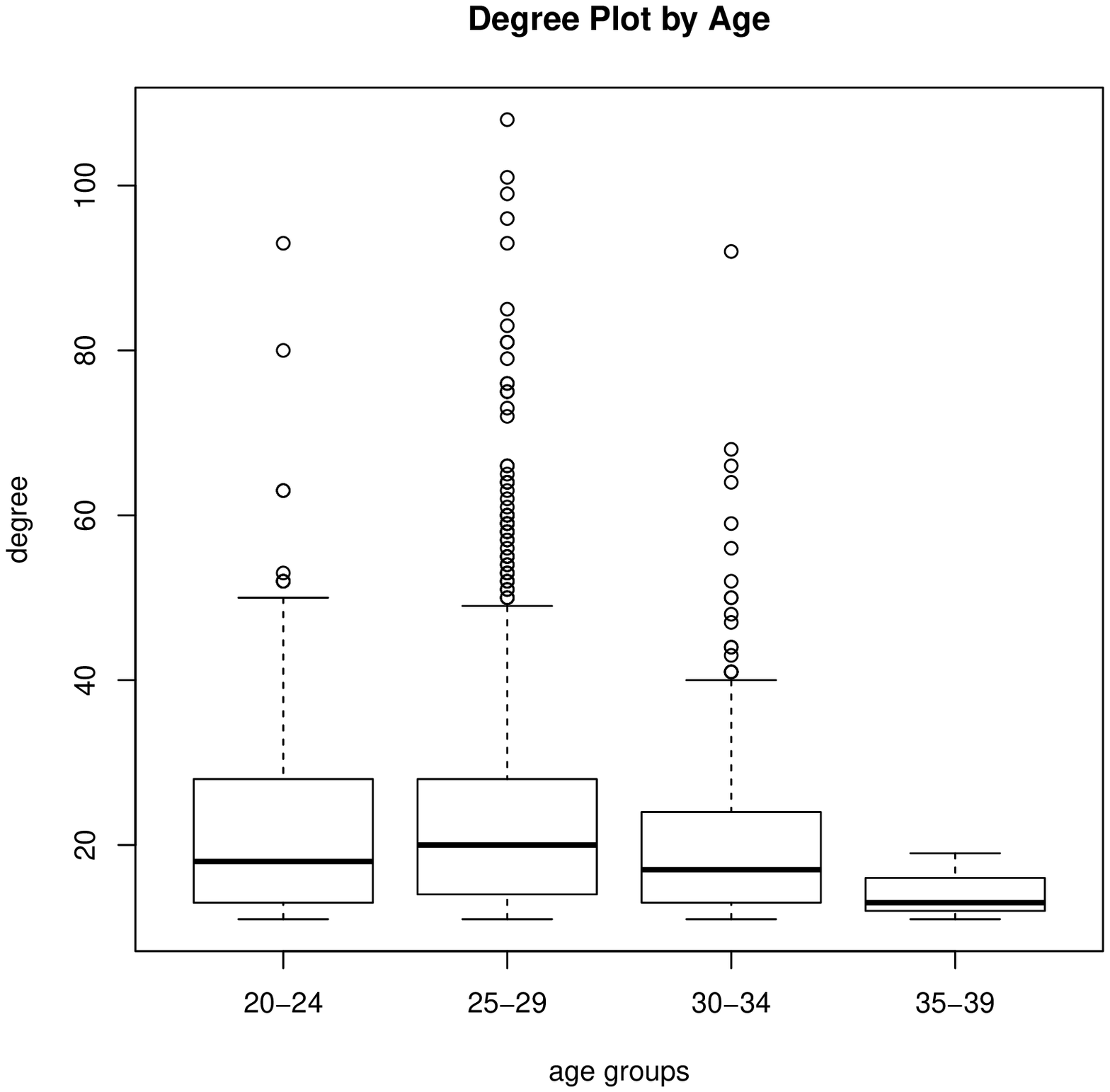}%
	\end{array}
	$%
\end{figure}

Pokec is a popular on-line social network in Slovakia. The whole dataset has
more than 1.6 million users, and it can be downloaded from
https://snap.stanford.edu/data/soc-Pokec.html. In this social network, nodes
are anonymized users of Pokec and edges represent friendships. Moreover,
demographical features of the users are provided, including gender, age,
hobbies, interest, education, etc. To illustrate our method, we select the
first 10000 users. Each user is a node in the graph. After deleting the
nodes with missing values in age and with degree less than 10, we have $1745$
nodes in our dataset. We use the continuous variable, age, as the covariate
in our model, and use the friendship network to create an undirected
adjacency matrix which has $1745$ nodes and $39650$ edges. The average
degree in this dataset is $22.72$. The left panel of Figure \ref%
{Fig:Pokec_age} shows the number of nodes in different age groups. We see
that the age group of 25-29 is the largest group with 1175 users and the age
groups of 20-24 and 30-34 have similar number of users. Around 98.8\% of
users are between the ages of 20 and 35 years old. Moreover, in the right
panel of Figure \ref{Fig:Pokec_age}, we depict the boxplots of degrees (the
number of users connected to each user) for the four age groups 20-24,
25-29, 30-34 and 35-39 that include most users. The plots of degrees vary
across different age groups, indicating that age may play a role in the
prediction of connections between users.
\

We consider fitting the model:
\begin{equation}
Y_{ij}=\mathbf{1}\{\varepsilon _{ij}\leq \tau _{n}+\Theta _{0,ij}^{\ast
}+W_{1,ij}\Theta _{1,ij}^{\ast }\},\quad i>j,  \label{MOD1}
\end{equation}%
for $i=1,...,1745$, where $Y_{ij}$ is the observed value ($0$ or $1$) of the
adjacency matrix in our dataset, and $W_{1,ij}=|X_{i}-X_{j}|/(\sqrt{%
	X_{i}^{2}+X_{j}^{2}})$, in which $X_{i}$ is the normalized age of the $i^{%
	\text{th}}$ customer.\footnote{%
	The variable $W_{1,ij}$ takes 1444 distinctive values. Given there are only
	1745 nodes in our dataset, we can view $W_{ij}$ as continuous.} In this
model, $(\tau _{n},\Theta _{0,ij}^{\ast },\Theta _{1,ij}^{\ast })$ are
unknown parameters, and $\Theta _{0,ij}^{\ast }$ and $\Theta _{1,ij}^{\ast }$
have the latent group structures $\Theta _{0}^{\ast }=ZB_{0}^{\ast }Z^{\top }
$ and $\Theta _{1}^{\ast }=ZB_{1}^{\ast }Z^{\top }$, respectively. Model (%
\ref{MOD1}) considered for this real application is similar to Model 2 in
the simulation, and it allows for not only the main effect but also possible
interaction effects of age and the latent community structure.

\subsubsection{Estimation results}

We first use the singular-value ratio method to obtain the estimated number
of groups for $\Theta _{0}^{\ast }$ and $\Theta _{1}^{\ast }$: $\widehat{K}%
_{0}=2$ and $\widehat{K}_{1}=2$, i.e., we identify two subgroups in the
friendship network.

\begin{figure}[tbp]
	\caption{left panel depicts the friendship network with two communities;
		right panel shows the adjacency matrix reorganized according to the node's
		memberships.}
	\label{Fig:friendship}\centering
	\vspace{0.5cm} $%
	\begin{array}{cc}
	\mathbf{{\tiny {Adjacency Matrix}}} \vspace{-0.4cm} & \mathbf{{\tiny {%
				Boxplot of Age by Group}}} \vspace{-0.2cm} \\
	\includegraphics[width = 0.45\linewidth]{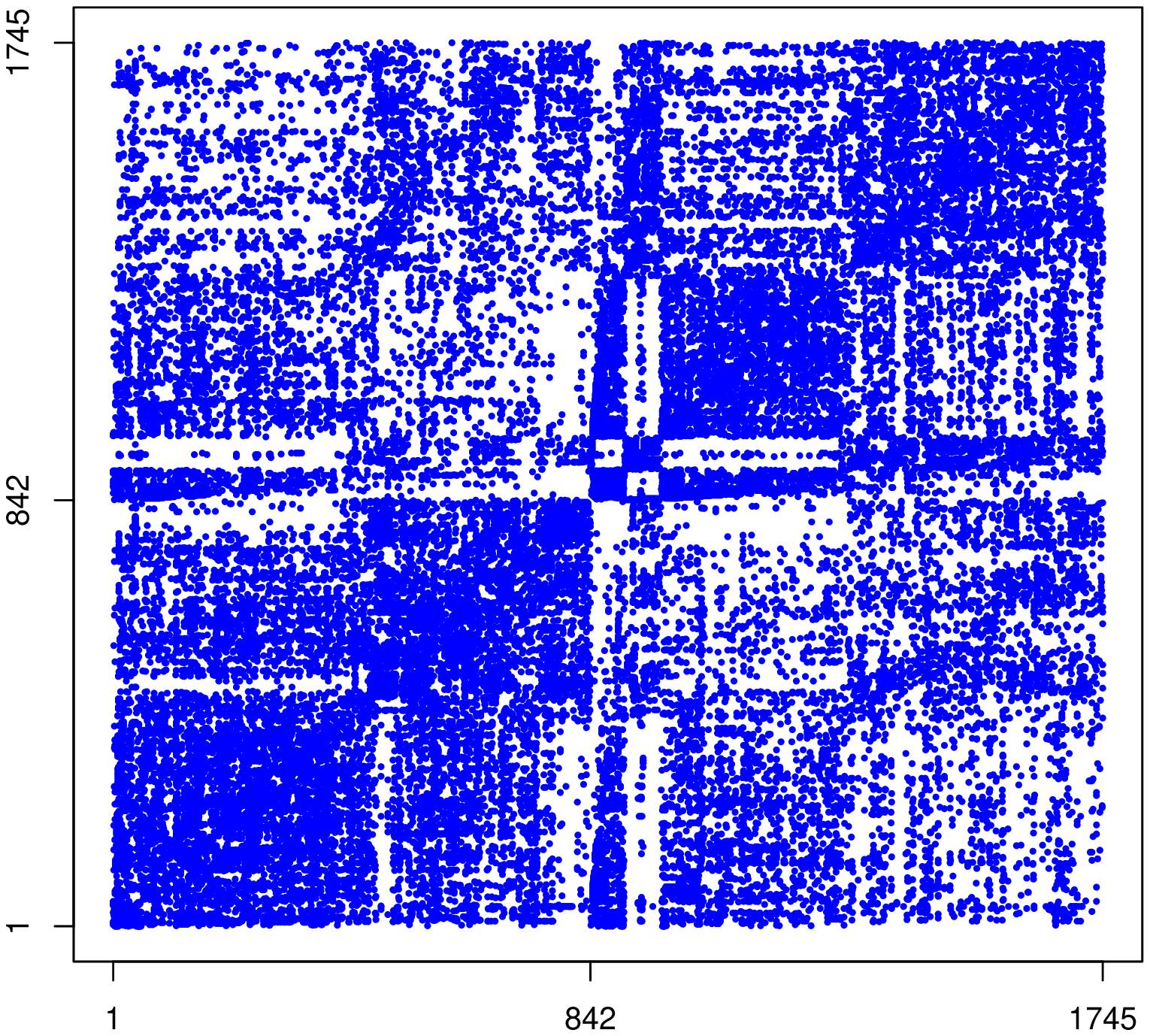} & %
	\includegraphics[width =0.45\linewidth]{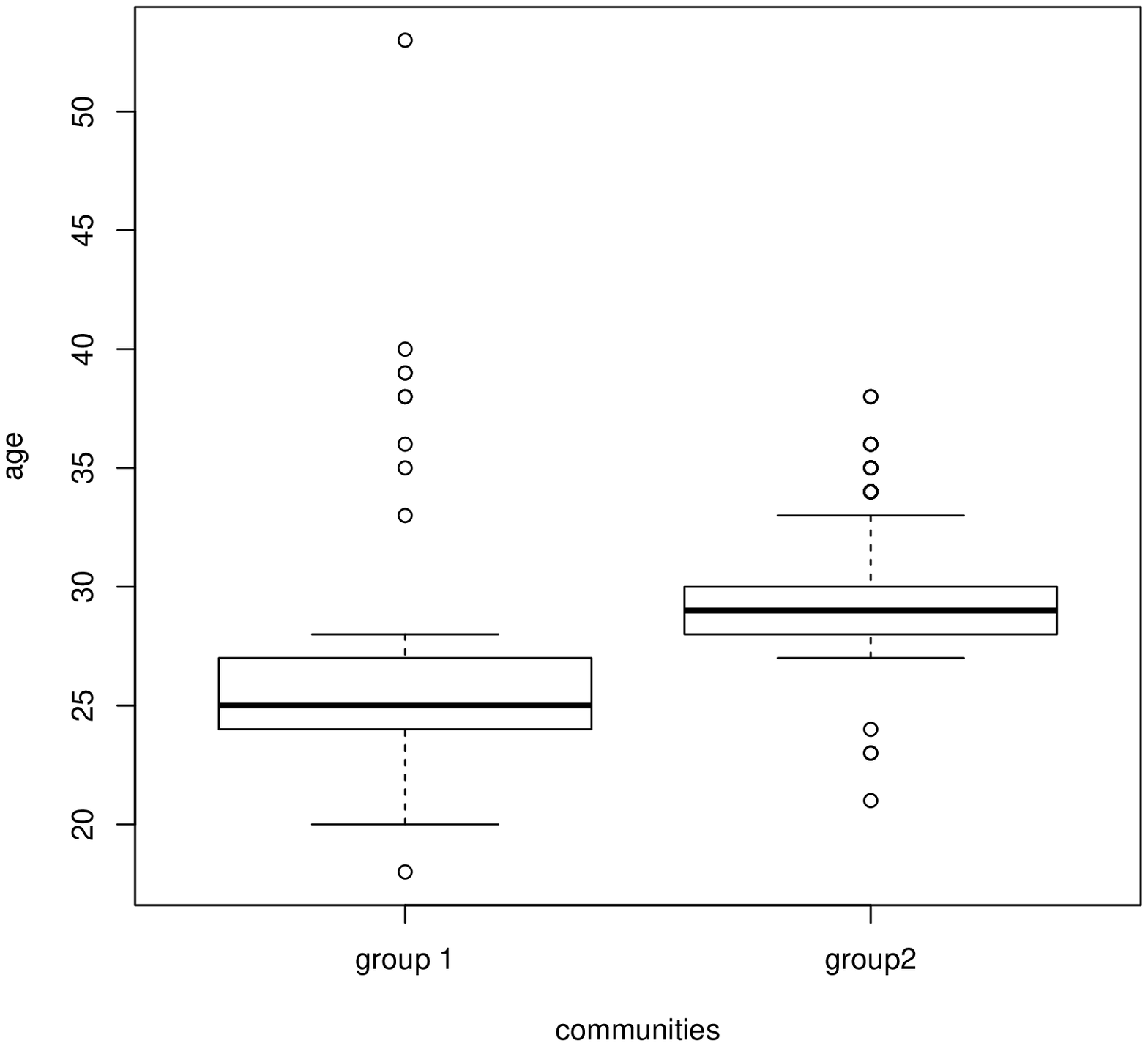}%
	\end{array}
	$%
\end{figure}

Next, we use our proposed method to obtain the estimated membership for each
node. As a result, we have identified $842$ nodes in one community and $903$
nodes in the other community. We reorganize the observed adjacency matrix
according to the estimated memberships of the nodes, i.e., the nodes in the
same estimated community are put together in the adjacency matrix. We use
blue dots to represent the edges between nodes. The left panel of Figure \ref%
{Fig:friendship} displays the reorganized adjacency. We see that nodes
within each community are generally more densely connected than nodes
between communities. In the right panel of Figure \ref{Fig:friendship}, we
show the boxplots of age for the two identified subgroups. We can observe
that in general, the values of age in group 1 are smaller than those in
group 2.

Last, Table \ref{Bestimate_Pokec} shows the estimates of $B_{0}^{\ast }$ and
$B_{1}^{\ast }$ and their standard errors (s.e.). We obtain the p-value$%
<0.01 $ for testing each coefficient in $B_{1}^{\ast }$ equal to zero,
indicating that the covariate age has a significant effect on the prediction
of the friendships between users.
\begin{table}[tbph]
	\caption{The estimates of $B_{0}^{\ast }$ and $B_{1}^{\ast }$ and their
		standard errors (s.e.).}
	\label{Bestimate_Pokec}
	\begin{center}
		{{\
				\begin{tabular*}{1\textwidth}{l|@{\extracolsep{\fill}}cccccc}
					\hline\hline
					& $B_{0,11}^{\ast }$ & $B_{0,12}^{\ast }$ & $B_{0,22}^{\ast }$ & $%
					B_{1,11}^{\ast }$ & $B_{1,12}^{\ast }$ & $B_{1,22}^{\ast }$ \\ \hline
					estimate & -3.922 & -4.119 & -3.425 & -0.444 & -0.518 & -0.477 \\
					s.e. & 0.017 & 0.025 & 0.017 & 0.027 & 0.019 & 0.016 \\ \hline\hline
				\end{tabular*}
		}}
	\end{center}
\end{table}

\subsection{Facebook friendship network}

\subsubsection{The dataset and model}

The dataset contains Facebook friendship networks at one hundred American
colleges and universities at a single point in time. It was provided and
analyzed by \cite{TMP12}, and can be downloaded from
https://archive.org/details/oxford-2005-facebook-matrix. \cite{TMP12} used
the dataset to illustrate the relative importance of different
characteristics of individuals across different institutions, and showed
that gender, dormitory residence and class year may play a role in network
partitions by using assortativity coefficients. We, therefore, use these
three user attributes as the covariates $X_{i}=(X_{i1},X_{i2},X_{i3})^{\top
} $, where $X_{i1}=$binary indicator for gender, $X_{i2}=$multi-category
variable for dorm number (e.g., \textquotedblleft 202\textquotedblright ,
\textquotedblleft 203\textquotedblright , etc.), and $X_{i3}=$integer valued
variable for class year (e.g., \textquotedblleft 2004\textquotedblright ,
\textquotedblleft 2005\textquotedblright , etc.). We use the dataset of Rice
University to identify the latent community structure interacted with the
covariates by our proposed method. \

We use the dataset to fit the model:
\begin{equation}
Y_{ij}=\mathbf{1}\{\varepsilon _{ij}\leq \tau _{n}+\Theta _{0,ij}^{\ast
}+W_{1,ij}\Theta _{1,ij}^{\ast }\},\quad i>j,  \label{MOD}
\end{equation}%
where $Y_{ij}$ is the observed value ($0$ or $1$) of the adjacency matrix in
the dataset, and $W_{1,ij}=\{\sum\nolimits_{k=1}^{3}(2D_{ij,k}/\Delta
_{k})^{2}\}^{1/2}$, where $\Delta _{k}=\max (D_{ij,k})-\min (D_{ij,k})$ and $%
D_{ij,k}=X_{ik}-X_{jk}$ for $k=1,2,3$.\footnote{%
	We note that $W_{1,ij}$ takes 1512 distinctive values. Given there are just
	3073 nodes in the dataset, we can view $W_{1,ij}$ as continuous.} In this
model, $(\tau _{n},\Theta _{0,ij}^{\ast },\Theta _{1,ij}^{\ast })$ are
unknown parameters, and $\Theta _{0,ij}^{\ast }$ and $\Theta _{1,ij}^{\ast }$
have the latent group structures $\Theta _{0}^{\ast }=ZB_{0}^{\ast }Z^{\top
} $ and $\Theta _{1}^{\ast }=ZB_{1}^{\ast }Z^{\top }$, respectively.
Following model 2 in the simulation, we impose that $\Theta _{0}^{\ast }$
and $\Theta _{1}^{\ast } $ share the same community structure. It is worth
noting that \cite{RAM19} fit a similar regression model as (\ref{MOD}) but
let the coefficient of the pairwise covariate be an unknown constant with
respect to $(i,j)$ such that $\Theta _{1,ij}^{\ast }=\Theta _{1}^{\ast }$.
Although \citeauthor{RAM19}'s \citeyear{RAM19} model can take into account
the covariate effect for community detection, it does not consider possible
interaction effects of the observed covariates and the latent community
structure. As a result, it may cause the number of estimated groups to be
inflated. In the dataset of Rice University, we delete the nodes with
missing values and with degree less than 10, and consider the class year
from 2004 to 2009. After the cleanup, there are $n=3073$ nodes and 279916
edges in the dataset for our analysis.

\subsubsection{Estimation results}

We first use the eigenvalue ratio method to obtain the estimated number of
groups for $\Theta^{\ast } _{0}$ and $\Theta^{\ast } _{1}$: $\widehat{K}%
_{0}=4$ and $\widehat{K}_{1}=4.$

Next, we use our proposed method to obtain the estimated membership for each
node. Table \ref{TAB:counts1} presents the number of students in each
estimated group for female and male, for different class years, and for
different dorm numbers. It is interesting to observe that most female
students belong to either group 2 or group 4, and most male students belong
to either group 1 or group 3. There is a clear community division between
female and male; within each gender category, the students are further
separated into two large groups. Moreover, most students in the class years
of 2004 and 2005 are in either group 1 or group 2, while most students in
the class years of 2008 and 2009 are in either group 3 or group 4. Students
in the class years of 2006 and 2007 are almost evenly distributed across the
four groups, with a tendency that more students will join groups 3 and group
4 when they are in later class years. This result indicates that students
tend to be in different groups as the gap between their class years becomes
larger. Last, Table \ref{Bestimate} shows the estimates of $B^{\ast }_{0}$
and $B^{\ast }_{1} $ and their standard errors (s.e.). We obtain the p-value$%
<0.01$ for testing each coefficient in $B^{\ast }_{1}$ equal to zero,
indicating that the three covariates are useful for identifying the
community structure.
\begin{table}[tbph]
	\caption{The number of persons\ in each estimated group for female and male,
		for different class years, and for different dorm numbers. }
	\label{TAB:counts1}
	\begin{center}
		{{\
				\begin{tabular*}{1\textwidth}{l|@{\extracolsep{\fill}}ccccccccc}
					\hline\hline
					& \multicolumn{2}{c}{gender} & \multicolumn{7}{c}{class year} \\
					& female & male &  & 2004 & 2005 & 2006 & 2007 & 2008 & 2009 \\ \hline
					group 1 & 1 & 515 &  & 112 & 139 & 147 & 110 & 37 & 1 \\
					group 2 & 540 & 4 &  & 103 & 135 & 116 & 165 & 50 & 2 \\
					\multicolumn{1}{l|}{group 3} & 4 & 1050 &  & 38 & 79 & 152 & 178 & 277 & 300
					\\
					\multicolumn{1}{l|}{group 4} & 958 & 1 &  & 30 & 62 & 125 & 156 & 288 & 271
					\\ \hline
					& \multicolumn{9}{c}{dorm number} \\
					& 202 & 203 & 204 & 205 & 206 & 207 & 208 & 209 & 210 \\ \hline
					group 1 & 71 & 67 & 36 & 42 & 41 & 50 & 57 & 59 & 93 \\
					group 2 & 65 & 98 & 53 & 46 & 20 & 63 & 56 & 56 & 84 \\
					group 3 & 94 & 116 & 142 & 138 & 129 & 130 & 121 & 101 & 83 \\
					group 4 & 92 & 72 & 124 & 125 & 139 & 95 & 122 & 110 & 83 \\ \hline\hline
				\end{tabular*}
		}}
	\end{center}
\end{table}
\begin{table}[tbph]
	\caption{The estimates of $B^{\ast }_{0}$ and $B^{\ast }_{1}$ and their
		standard errors (s.e.).}
	\label{Bestimate}
	\begin{center}
		{{\
				\begin{tabular*}{1\textwidth}{l|@{\extracolsep{\fill}}cccccccccc}
					\hline\hline
					& $B^{\ast }_{0,11}$ & $B^{\ast }_{0,12}$ & $B^{\ast }_{0,13}$ & $B^{\ast
					}_{0,14}$ & $B^{\ast }_{0,22}$ & $B^{\ast }_{0,23}$ & $B^{\ast }_{0,24}$ & $%
					B^{\ast }_{0,33}$ & $B^{\ast }_{0,34}$ & $B^{\ast }_{0,44}$ \\ \hline
					estimate & -0.730 & 4.912 & -1.543 & 6.197 & -0.751 & 4.123 & -1.624 & -1.702
					& 5.933 & -1.419 \\
					s.e. & 0.018 & 0.112 & 0.024 & 0.171 & 0.017 & 0.195 & 0.024 & 0.017 & 0.207
					& 0.016 \\ \hline
					& $B^{\ast }_{1,11}$ & $B^{\ast }_{1,12}$ & $B^{\ast }_{1,13}$ & $B^{\ast
					}_{1,14}$ & $B^{\ast }_{1,22}$ & $B^{\ast }_{1,23}$ & $B^{\ast }_{1,24}$ & $%
					B^{\ast }_{1,33}$ & $B^{\ast }_{1,34}$ & $B^{\ast }_{1,44}$ \\ \hline
					estimate & -3.397 & -6.381 & -4.398 & -5.656 & -3.600 & -5.628 & -4.387 &
					-6.384 & -6.704 & -7.567 \\
					s.e. & 0.042 & 0.102 & 0.057 & 0.155 & 0.042 & 0.180 & 0.059 & 0.059 & 0.196
					& 0.060 \\ \hline\hline
				\end{tabular*}
		}}
	\end{center}
\end{table}

\section{Conclusion \label{sec:concl}}

In this paper, we proposed a network formation model which can capture
heterogeneous effects of homophily via a latent community structure. When
the expected degree diverges at a rate no slower than rate-$\log n$, we
established that the proposed method can exactly recover the latent
community memberships almost surely. By treating the estimated community
memberships as the truth, we can then estimate the regression coefficients
in the model by existing methods in the literature.

\pagebreak

\appendix%
\linespread{1.1}%

\begin{center}
{\Huge Appendix}
\end{center}

\section{Proofs of the Main Results}

\label{sec:proof_main} In this appendix, we prove the main results in the
paper. Given the fact that our proofs involve a lot of constants defined in
the assumptions and proofs, we first provide a list of these constants in
Appendix \ref{sec:const}. Then we prove Lemma \ref{lem:theta} and Theorems %
\ref{thm:Frobeniusnorm}--\ref{thm:kmeans} in Appendices \ref%
{sec:prooflemtheta}--\ref{sec:thmkmeans}, respectively.

\subsection{List of constants \label{sec:const}}

Before we prove the main results, we first list the frequently used
constants in Table \ref{Table:constants}. We specify each constant to
illustrate that all our results hold as long as $\sqrt{\log n/(n\zeta _{n})}%
\leq c_{F}\leq \frac{1}{4}$ for some sufficiently small\ constant $c_{F}$.
Apparently, if $\log n/(n\zeta _{n})\rightarrow 0$, $c_{F}$ can be
arbitrarily small as long as $n$ is sufficiently large. Then all the rate
requirements in the proof hold automatically. However, $\log n/(n\zeta
_{n})\rightarrow 0$ is sufficient but not necessary.

\begin{table}[tbph]
	\caption{Table of constants}
	\label{Table:constants}
	\begin{tabular}{c|l}
		\hline\hline
		Name & Description \\ \hline
		$M_{W}$ & $|W_{1,ij}|\leq M_{W}$. \\
		$M$ & $\max_{i\in [ n],l=0,1}|\Theta _{l,ij}^{\ast }|\leq M,$ used in the
		definition of $f_{M}(\cdot )$ and Assumption \ref{ass:par}. \\
		$C_{\lambda }$ & Used in the definition of $\lambda _{n}^{(1)}$. \\
		$C_{M}$ & Used in the definition of $\mathbb{T}^{(1)}$. \\
		$C_{\sigma },c_{\sigma },C_{1},c_{1}$ & Defined in Assumption \ref{ass:pi}.
		\\
		$\kappa $ & Defined in Assumption \ref{ass:RSC}. \\
		$\overline{c},\underline{c},c_{F}$ & Defined in Assumption \ref{ass:tune}.
		\\
		$C_{\phi },$ $c_{\phi }$ & Defined in Assumption \ref{ass:phi}. \\
		$C_{F},C_{F,1},C_{F,2}$ & Defined in Theorem \ref{thm:Frobeniusnorm}. \\
		$C_{1}^{\ast }$ & Defined in Theorem \ref{thm:rowwisebound}. \\
		$C_{h,u},C_{h,v}$ & Defined in Theorem \ref{thm:iter}. \\
		$C_{\Upsilon }$ & Defined in Lemma \ref{lem:op}. \\ \hline\hline
	\end{tabular}
\end{table}

\subsection{Proof of Lemma \protect\ref{lem:theta}}

\label{sec:prooflemtheta} We prove the results for $U_{l}$ first. Let $\Pi
_{l,n}=Z_{l}^{\top }Z_{l}/n=\text{diag}(\pi _{l,1n},\cdots ,\pi
_{l,K_{l}n}). $ Then,
\begin{equation*}
(n^{-1}\Theta _{l}^{\ast })(n^{-1}\Theta _{l}^{\ast })^{\top
}=n^{-1}Z_{l}B_{l}^{\ast }\Pi _{l,n}B_{l}^{\ast }Z_{l}^{\top }.
\end{equation*}%
Consider the spectral decomposition of $\chi \equiv \Pi
_{l,n}^{1/2}B_{l}^{\ast }\Pi _{l,n}B_{l}^{\ast }\Pi _{l,n}^{1/2}:\chi
=S_{l}^{\prime }\tilde{\Omega}_{l}^{2}(S_{l}^{\prime })^{\top }.$ Let $%
\mathcal{U}_{l}=Z_{l}(Z_{l}^{\top }Z_{l})^{-1/2}S_{l}^{\prime }$, where $%
S_{l}$ is a $K_{l}\times K_{l}$ matrix such that $(S_{l}^{\prime })^{\top
}S_{l}^{\prime }=I_{K_{l}}$. Then, we have
\begin{equation*}
\mathcal{U}_{1}\tilde{\Omega}_{l}^{2}\mathcal{U}_{l}^{\top }=n^{-1}Z_{l}\Pi
_{n}^{-1/2}S_{l}\tilde{\Omega}_{l}^{2}S_{l}^{\top }\Pi
_{n}^{-1/2}Z_{l}^{\top }=n^{-1}Z_{l}B_{l}^{\ast }\Pi _{n}B_{l}^{\ast
}Z_{l}^{\top }=(n^{-1}\Theta _{l}^{\ast })^{2}.
\end{equation*}%
In addition, note that $\mathcal{U}_{l}^{\top }\mathcal{U}_{l}=I_{K_{l}}$
and $\tilde{\Omega}_{l}^{2}$ is a diagonal matrix. This implies $\tilde{
	\Omega}_{l}^{2}=\Sigma _{l}^{2}$ (after reordering the eigenvalues) and $%
\mathcal{U}_{l}$ is the corresponding singular vector matrix. Then, by
definition,
\begin{equation*}
U_{l}=\sqrt{n}\mathcal{U}_{l}\Sigma _{l}=Z_{l}(\Pi
_{l,n})^{-1/2}S_{l}^{\prime }\Sigma _{l}.
\end{equation*}%
Similarly, by considering the spectral decomposition of $(n^{-1}\Theta
_{l}^{\ast })^{\top }(n^{-1}\Theta _{l}^{\ast })$, we can show that $%
V_{l}=Z_{l}(\Pi _{l,n})^{-1/2}S_{l}$ for some rotation matrix $S_{l}$. Parts
(2) and (3) can be verified directly by noting that $S_{l}$ and $%
S_{l}^{\prime }$ are orthonormal, $\Pi _{l,n}$ is diagonal, and Assumption %
\ref{ass:pi} holds.

\subsection{Proof of Theorem \protect\ref{thm:Frobeniusnorm}}

We focus on the split-sample low-rank estimators. The full-sample results
can be derived in the same manner. Denote $Q_{n,ij}(\Gamma
_{ij})=-[Y_{ij}\log (\Lambda (W_{ij}^{\top }\Gamma _{ij}))+(1-Y_{ij})\log
(1-\Lambda (W_{ij}^{\top }\Gamma _{ij}))]$, which is a convex function for
each element in $\Gamma _{ij}=(\Gamma _{0,ij},\Gamma _{1,ij})^{\top }$. In
addition, we note that the true parameter $\Gamma ^{\ast }(I_{1})\in \mathbb{%
	\ T}^{(1)}(0,\log n)$. Denote $\widetilde{\Gamma }^{(1)}=\{\widetilde{\Gamma
}_{ij}^{(1)}\}_{i\in I_{1},j\in [ n]},$ $\widetilde{\Gamma }_{ij}^{(1)}=(%
\widetilde{\Gamma }_{0,ij}^{(1)},\widetilde{\Gamma }_{1,ij}^{(1)})^{\top }$
and $\Delta _{ij}=\widetilde{\Gamma }_{ij}^{(1)}-\Gamma _{ij}^{\ast }\equiv
(\Delta _{0,ij},\Delta _{1,ij})^{\top }$, for $i\in I_{1}$, $j\in [ n]$.
Then, we have
\begin{align}
\lambda _{n}^{(1)}\sum_{l=0}^{1}\left( ||\Gamma _{l}^{\ast }(I_{1})||_{\ast
}-||\widetilde{\Gamma }_{l}^{(1)}||_{\ast }\right) \geq & \frac{1}{n_{1}(n-1)%
}\sum_{i\in I_{1},j\in \left[ n\right] ,i\neq j}\left( Q_{n,ij}(\widetilde{
	\Gamma }_{ij}^{(1)})-Q_{n,ij}(\Gamma _{ij}^{\ast })\right)  \notag
\label{eq:1} \\
\geq & \frac{1}{n_{1}(n-1)}\sum_{i\in I_{1},j\in \left[ n\right] ,i\neq
	j}\left( \partial _{\Gamma _{ij}}Q_{n,ij}^{\top }(\Gamma _{ij}^{\ast
})\right) ^{\top }\Delta _{ij}  \notag \\
=& \frac{-1}{n_{1}(n-1)}\sum_{i\in I_{1},j\in \left[ n\right] ,i\neq
	j}\left( Y_{ij}-\Lambda (W_{ij}^{\top }\Gamma _{ij}^{\ast })\right)
W_{ij}^{\top }\Delta _{ij}  \notag \\
\equiv & \frac{-1}{n_{1}(n-1)}\sum_{l=0}^{1}\text{trace}(\Upsilon _{l}^{\top
}\Delta _{l}),
\end{align}%
where $\partial _{\Gamma _{ij}}Q_{n,ij}^{\top }(\Gamma _{ij}^{\ast
})=\partial Q_{n,ij}(\Gamma _{ij}^{\ast })/\partial \Gamma _{ij},$ $\Upsilon
_{l}$ is an $n_{1}\times n$ matrix with $(i,j)$-th entry
\begin{equation*}
\Upsilon _{l,ij}=%
\begin{cases}
\left( Y_{ij}-\Lambda (W_{ij}^{\top }\Gamma _{ij}^{\ast })\right) W_{l,ij} &
\text{if}\quad i\in I_{1},j\in \left[ n\right] ,\text{ }j\neq i \\
0 & \text{if}\quad i=j\in I_{1}%
\end{cases}%
,
\end{equation*}%
and $\text{trace}(\cdot )$ is the trace operator. By \eqref{eq:1}, we have
\begin{align}
0\leq & \lambda _{n}^{(1)}\sum_{l=0}^{1}\left( ||\Gamma _{l}^{\ast
}(I_{1})||_{\ast }-||\widetilde{\Gamma }_{l}^{(1)}||_{\ast }\right) +\frac{1%
}{n_{1}(n-1)}\left\vert \sum_{l=0}^{1}\text{trace}(\Upsilon _{l}^{\top
}\Delta _{l})\right\vert  \notag  \label{eq:lambda1} \\
\leq & \lambda _{n}^{(1)}\sum_{l=0}^{1}\left( ||\Gamma _{l}^{\ast
}(I_{1})||_{\ast }-||\widetilde{\Gamma }_{l}^{(1)}||_{\ast }\right) +\frac{1%
}{n_{1}(n-1)}\sum_{l=0}^{1}||\Upsilon _{l}||_{op}||\Delta _{l}||_{\ast }.
\end{align}

For some generic $n_{1}\times n$ matrix $\Delta $, let $\mathcal{M}%
_{l}^{(1)}(\Delta )$ and $\mathcal{P}_{l}^{(1)}(\Delta )$ be the residual
and projection matrices of $\Delta $ with respect to $\Gamma _{l}^{\ast
}(I_{1})$, as defined in Assumption \ref{ass:RSC}. By \citet[Lemma
D.2]{CHLZ18} and the fact that $\Gamma _{0}^{\ast }(I_{1})$ and $\Gamma
_{1}^{\ast }(I_{1})$ are exact low-rank matrices with ranks upper bounded by
$K_{0}+1$ and $K_{1}$, respectively, we have $\Delta _{l}=\mathcal{M}%
_{l}^{(1)}(\Delta _{l})+\mathcal{P}_{l}^{(1)}(\Delta _{l})$, $\text{rank}(%
\mathcal{M}_{0}^{(1)}(\Delta _{0}))\leq 2K_{0}+2$, $\text{rank}(\mathcal{M}%
_{1}^{(1)}(\Delta _{1}))\leq 2K_{1}$, and for $l=0,1$,
\begin{equation}
||\Delta _{l}||_{F}^{2}=||\mathcal{M}_{l}^{(1)}(\Delta _{l})||_{F}^{2}+||%
\mathcal{P}_{l}^{(1)}(\Delta _{l})||_{F}^{2}\text{ and }||\Gamma _{l}^{\ast
}(I_{1})+\mathcal{P}_{l}^{(1)}(\Delta _{l})||_{\ast }=||\Gamma _{l}^{\ast
}(I_{1})||_{\ast }+||\mathcal{P}_{l}^{(1)}(\Delta _{l})||_{\ast }.
\label{eq:deltaF}
\end{equation}%
This implies that
\begin{align}
||\Gamma _{l}^{\ast }(I_{1})||_{\ast }-||\widetilde{\Gamma }%
_{l}^{(1)}||_{\ast }=& ||\Gamma _{l}^{\ast }(I_{1})||_{\ast }-||\Gamma
_{l}^{\ast }(I_{1})+\mathcal{M}_{l}^{(1)}(\Delta _{l})+\mathcal{P}%
_{l}^{(1)}(\Delta _{l})||_{\ast }  \notag \\
\leq & ||\mathcal{M}_{l}^{(1)}(\Delta _{l})||_{\ast }-||\mathcal{P}%
_{l}^{(1)}(\Delta _{l})||_{\ast },\quad l=0,1.  \label{eq:Gamma}
\end{align}%
Therefore, combining \eqref{eq:lambda1}, Lemma \ref{lem:op}, and %
\eqref{eq:Gamma}, we have
\begin{equation*}
0\leq \lambda _{n}^{(1)}\sum_{l=0}^{1}\left( ||\mathcal{M}_{l}^{(1)}(\Delta
_{l})||_{\ast }-||\mathcal{P}_{l}^{(1)}(\Delta _{l})||_{\ast }\right) +\frac{%
	C_{\Upsilon }M_{W}(\sqrt{\zeta _{n}n}+\sqrt{\log n})}{n_{1}(n-1)}%
\sum_{l=0}^{1}\left( ||\mathcal{M}_{l}^{(1)}(\Delta _{l})||_{\ast }+||%
\mathcal{P}_{l}^{(1)}(\Delta _{l})||_{\ast }\right) .
\end{equation*}%
Noting that $\lambda _{n}^{(1)}=\frac{C_{\lambda }(\sqrt{\zeta _{n}n}+\sqrt{%
		\log n})}{n_{1}(n-1)}$ and $C_{\lambda }>C_{\Upsilon }M_{W}$, the last
inequality implies that
\begin{equation}
(C_{\lambda }-C_{\Upsilon }M_{W})\sum_{l=0}^{1}||\mathcal{P}%
_{l}^{(1)}(\Delta _{l})||_{\ast }\leq (C_{\lambda }+C_{\Upsilon
}M_{W})\sum_{l=0}^{1}||\mathcal{M}_{l}^{(1)}(\Delta _{l})||_{\ast },
\label{eq:cone}
\end{equation}%
and that $(\Delta _{0},\Delta _{1})\in \mathcal{C}(\tilde{c})$ for $\tilde{c}%
=\frac{C_{\lambda }+C_{\Upsilon }M_{W}}{C_{\lambda }-C_{\Upsilon }M_{W}}>0,$
with a slight abuse of notation.

Next, we first aim to show
\begin{align*}
\frac{1}{n}(\sum_{l=0}^{1}||\Delta_l||_F^2)^{1/2} \leq 17 C_F\left(\frac{%
	\log(n)}{\sqrt{n\zeta_n}} + \frac{(\log n)^{3/2}}{n \zeta_n}\right),
\end{align*}
where $C_{F}=\frac{\sqrt{\bar{K}}(M_{W}+1)(C_{\lambda }+C_{\Upsilon }M_{W})}{
	\underline{c}\kappa } + \sqrt{\frac{c_3}{\kappa}} + \sqrt{c_2}$. We suppose $%
(\Delta_0,\Delta_1) \notin \mathcal{C}_1(c_2)$, i.e.,
\begin{align}
\sum_{l=0}^{1}||\Delta_l||_F^2 > c_2 n \log(n)/\zeta_n,  \label{eq:c1}
\end{align}
otherwise,
\begin{align*}
\frac{1}{n}(\sum_{l=0}^{1}||\Delta_l||_F^2)^{1/2} \leq \sqrt{\frac{c_2
		\log(n)}{ n \zeta_n}} < 17 C_F\left(\frac{\log(n)}{\sqrt{n\zeta_n}} + \frac{%
	(\log n)^{3/2}}{n \zeta_n}\right),
\end{align*}
and we are done.

Now we consider the second-order Taylor expansion of $Q_{n,ij}(\Gamma _{ij})$%
, following the argument in \cite{BCFH13}. Let $f_{ij}(t)=\log \{1+\exp
(W_{ij}^{\top }(\Gamma _{ij}^{\ast }+t\Delta _{ij}))\},$ where $\Delta
_{ij}=(\Delta _{0,ij},\cdots ,\Delta _{p,ij})^{\top }$. Note
\begin{equation*}
Q_{n,ij}(\widetilde{\Gamma }_{ij}^{(1)})-Q_{n,ij}(\Gamma _{ij}^{\ast
})-\partial _{\Gamma _{ij}}Q_{n,ij}^{\top }(\Gamma _{ij}^{\ast })\Delta
_{ij}=f_{ij}(1)-f_{ij}(0)-f_{ij}^{\prime }(0)
\end{equation*}%
and that $f_{ij}(\cdot )$ is a three times differentiable convex function
such that for all $t\in \mathbb{R}$,
\begin{align*}
|f_{ij}^{^{\prime \prime \prime }}(t)|=& |W_{ij}^{\top }\Delta
_{ij}|^{3}\Lambda (W_{ij}^{\top }(\Delta _{ij}+t\Delta _{ij}))(1-\Lambda
(W_{ij}^{\top }(\Delta _{ij}+t\Delta _{ij})))|1-2\Lambda (W_{ij}^{\top
}(\Delta _{ij}+t\Delta _{ij}))| \\
\leq & |W_{ij}^{\top }\Delta _{ij}|f_{ij}^{^{\prime \prime }}(t).
\end{align*}%
Then, by \citet[Lemma 1]{B10} we have
\begin{align}
f_{ij}(1)-f_{ij}(0)-f_{ij}^{\prime }(0)\geq & \frac{f_{ij}^{^{\prime \prime
	}}(0)}{(W_{ij}^{\top }\Delta _{ij})^{2}}\left[ \exp (-|W_{ij}^{\top }\Delta
_{ij}|)+|W_{ij}^{\top }\Delta _{ij}|-1\right]  \notag  \label{eq:f} \\
=& \Lambda (W_{ij}^{\top }\Gamma _{ij}^{\ast })(1-\Lambda (W_{ij}^{\top
}\Gamma _{ij}^{\ast }))\left[ \exp (-|W_{ij}^{\top }\Delta
_{ij}|)+|W_{ij}^{\top }\Delta _{ij}|-1\right]  \notag \\
\geq & \underline{c}\zeta _{n}\left[ \exp (-|W_{ij}^{\top }\Delta
_{ij}|)+|W_{ij}^{\top }\Delta _{ij}|-1\right]  \notag \\
\geq & \underline{c}\zeta _{n}\left( \frac{(W_{ij}^{\top }\Delta _{ij})^{2}}{
	4(\max_{i,j}|W_{ij}^{\top }\Delta _{ij}|\vee \log (2))}\right)  \notag \\
\geq & \frac{\zeta _{n}\underline{c}(W_{ij}^{\top }\Delta _{ij})^{2}}{
	8(M_{W}+1)\log n},
\end{align}%
where the third inequality holds by Lemma \ref{lem:e} and the last
inequality holds because of Assumption \ref{ass:tune} and the fact that $%
|W_{ij}^{\top }\Delta _{ij}|\leq |\widetilde{\Gamma }_{0,ij}-\Gamma
_{0,ij}|+M_{W}|\widetilde{\Gamma }_{1,ij}-\Gamma _{1,ij}|\leq 2(M_{W}+1)\log
n.$ Therefore, w.p.a.1,
\begin{align}
F_{n}(\Delta _{0},\Delta _{1})\equiv & \frac{1}{n_{1}(n-1)}\sum_{i\in
	I_{1},j\in \left[ n\right] ,j\neq i}\left[ Q_{n,ij}(\widetilde{\Gamma }%
_{ij}^{(1)})-Q_{n,ij}(\Gamma _{ij}^{\ast })-\partial _{\Gamma
	_{ij}}Q_{n,ij}^{\top }(\Gamma _{ij}^{\ast })\Delta _{ij}\right]  \notag
\label{eq:F} \\
\geq & \frac{\zeta _{n}\underline{c}}{8n_{1}(n-1)(M_{W}+1)\log n}\sum_{i\in
	I_{1},j\in \left[ n\right] ,j\neq i}(W_{ij}^{\top }\Delta _{ij})^{2}  \notag
\\
\geq & \frac{\zeta _{n}\underline{c}}{8n_{1}(n-1)(M_{W}+1)\log n}\left[
\kappa \sum_{l=0}^{1}||\Delta _{l}||_{F}^{2}-4(M_{W}+1)^{2}(\log n)^{2}n_{1}
- c_3 n \log (n)/\zeta_n\right] ,
\end{align}%
where the last inequality holds by Assumption \ref{ass:RSC}, \eqref{eq:c1},
and the fact that $|\Delta _{l,ii}|\leq 2\log n$, $i\in I_{1}$.

On the other hand, by \eqref{eq:1},
\begin{align}
F_{n}(\Delta _{0},\Delta _{1})\leq & \lambda _{n}^{(1)}\sum_{l=0}^{1}\left(
||\Gamma _{l}^{\ast }(I_{1})||_{\ast }-||\widetilde{\Gamma }%
_{l}^{(1)}||_{\ast }\right) +\left\vert \frac{1}{n_{1}(n-1)}\sum_{l=0}^{1}%
\text{trace}(\Upsilon _{l}^{\top }\Delta _{l})\right\vert  \notag
\label{eq:Fupper} \\
\leq & \lambda _{n}^{(1)}\sum_{l=0}^{1}\left( ||\mathcal{M}_l^{(1)}(\Delta
_{l})||_{\ast }-||\mathcal{P}_l^{(1)}(\Delta_l)||_{\ast }\right) +\frac{1}{
	n_{1}(n-1)}\sum_{l=0}^{1}||\Upsilon _{l}||_{op}||\Delta _{l}||_{\ast }
\notag \\
\leq & \frac{\sqrt{\zeta _{n}n}+\sqrt{\log n}}{n_{1}(n-1)}\left[
\sum_{l=0}^{1}(C_{\lambda }+C_{\Upsilon }M_{W})||\mathcal{M}_l^{(1)}(\Delta
_{l})||_{\ast }-\sum_{l=0}^{1}(C_{\lambda }-C_{\Upsilon }M_{W})||\mathcal{P}%
_l^{(1)}(\Delta _{l})||_{\ast }\right]  \notag \\
\leq & \frac{\sqrt{\zeta _{n}n}+\sqrt{\log n}}{n_{1}(n-1)}(C_{\lambda
}+C_{\Upsilon }M_{W})(\sum_{l=0}^{1}||\mathcal{M}_l^{(1)}(\Delta
_{l})||_{\ast })  \notag \\
\leq & \frac{\sqrt{\zeta _{n}n}+\sqrt{\log n}}{n_{1}(n-1)}(C_{\lambda
}+C_{\Upsilon }M_{W})\sqrt{2\bar{K}}(\sum_{l=0}^{1}||\mathcal{M}%
_l^{(1)}(\Delta _{l})||_{F})  \notag \\
\leq & \frac{\sqrt{\zeta _{n}n}+\sqrt{\log n}}{n_{1}(n-1)}(C_{\lambda
}+C_{\Upsilon }M_{W})\sqrt{2\bar{K}}(\sum_{l=0}^{1}||\Delta _{l}||_{F})
\notag \\
\leq & \frac{\sqrt{\zeta _{n}n}+\sqrt{\log n}}{n_{1}(n-1)}(C_{\lambda
}+C_{\Upsilon }M_{W})2\sqrt{\bar{K}}(\sum_{l=0}^{1}||\Delta
_{l}||_{F}^{2})^{1/2},
\end{align}%
where $\bar{K}=\max (K_{0}+1,K_{1})$, the first inequality is due to %
\eqref{eq:1}, the second inequality is due to \eqref{eq:Gamma} and the trace
inequality, the third inequality holds by the definition of $\lambda
_{n}^{(1)}$ and Lemma \ref{lem:op}, the fourth inequality is due to the fact
that $C_{\lambda }-C_{\Upsilon }M_{W}>0$, the fifth inequality is due to the
fact that $\text{rank}(\mathcal{M}_l^{(1)}(\Delta _{l}))\leq 2\bar{K}$, the
second last inequality is due to \eqref{eq:deltaF}, and the last inequality
is due to the Cauchy's inequality.

Combining \eqref{eq:F} and \eqref{eq:Fupper}, we have
\begin{align*}
& \left[ (\sum_{l=0}^{1}||\Delta _{l}||_{F}^{2})^{1/2}-\frac{8\sqrt{\bar{K}}
	(M_{W}+1)(C_{\lambda }+C_{\Upsilon }M_{W})}{\underline{c}\kappa }\frac{\log
	n[\sqrt{n\zeta _{n}}+\sqrt{\log n}]}{\zeta _{n}}\right] ^{2} \\
\leq & \bar{K}\left[ \frac{8(M_{W}+1)(C_{\lambda }+C_{\Upsilon }M_{W})}{
	\underline{c}\kappa }\right] ^{2}\left( \frac{\log n[\sqrt{n\zeta _{n}}+
	\sqrt{\log n}]}{\zeta _{n}}\right) ^{2}+\frac{4n_{1}(M_{W}+1)^{2}(\log n)^{2}%
}{\kappa } + \frac{c_3 n \log(n)}{\kappa \zeta_n},
\end{align*}%
and thus,
\begin{equation}
\frac{1}{n}(\sum_{l=0}^{1}||\Delta _{l}||_{F}^{2})^{1/2}\leq 17C_{F}\left(
\frac{\log n}{\sqrt{n\zeta _{n}}}+\frac{(\log n)^{3/2}}{n\zeta _{n}}%
\right)~w.p.a.1.  \label{eq:CF}
\end{equation}%
Then,
\begin{align}
|\widetilde{\tau }_{n}^{\left( 1\right) }-\tau _{n}|=& \left\vert \frac{1}{
	n_{1}n}\sum_{i\in I_{1},j\in \left[ n\right] }(\widetilde{\Gamma }%
_{0,ij}-\tau _{n})\right\vert \leq \left\vert \frac{1}{n_{1}n}\sum_{i\in
	I_{1},j\in \left[ n\right] }(\widetilde{\Gamma }_{0,ij}-\Gamma _{0,ij}^{\ast
})\right\vert +\left\vert \frac{1}{n_{1}n}\sum_{i\in I_{1},j\in \left[ n %
	\right] }\Theta _{0,ij}^{\ast }\right\vert  \notag \\
\leq & \frac{1}{\sqrt{n_{1}n}}||\Delta _{0}||_{F}+M\leq 30C_{F}\left( \frac{
	\log n}{\sqrt{n\zeta _{n}}}+\frac{(\log n)^{3/2}}{n\zeta _{n}}\right)  \notag
\\
\leq & 30C_{F}(c_{F}+c_{F}^{2})\sqrt{\log n}~w.p.a.1,  \label{eq:alphatilde1}
\end{align}%
where the last inequality follows Assumption \ref{ass:tune}.3.

Next, we rerun the nuclear norm regularized logistic regression with the
parameter space restriction $\mathbb{T}^{(1)}(0,\log n)$ replaced by $%
\mathbb{T}^{(1)}(\widetilde{\tau }_{n}^{\left( 1\right) },C_{M}\sqrt{\log n}%
) $. First, we note that the true parameter $\Gamma ^{\ast }(I_{1})\in
\mathbb{T}^{(1)}(\widetilde{\tau }_{n}^{\left( 1\right) },C_{M}\sqrt{\log n}%
) $ because $|\Gamma _{1,ij}^{\ast }| \leq C_M \sqrt{\log n}$ and
\begin{equation}
|\Gamma _{0,ij}^{\ast }-\widetilde{\tau }_{n}^{\left( 1\right) }|\leq
|\Theta _{0,ij}^{\ast }|+|\widetilde{\tau }_{n}^{\left( 1\right) }-\tau
_{n}|\leq |\Theta _{0,ij}^{\ast }|+30C_{F}(c_{F}+c_{F}^{2})\sqrt{\log n}\leq
C_{M}\sqrt{\log n},  \label{eq:cm}
\end{equation}%
where we use the fact that $c_{F}$, and thus, $30(c_{F}+c_{F}^{2})C_{F}$ is
sufficiently small.

Therefore, following the same arguments used to obtain \eqref{eq:cone}, we
can show that $\widehat{\Delta }\equiv (\widehat{\Delta }_{0},\widehat{
	\Delta }_{1})\in \mathcal{C}(\tilde{c}),$ where $\widehat{\Delta }_{l}=%
\widehat{\Gamma }_{l}^{(1)}-\Gamma _{l}^{\ast }(I_{1})$. Let $\widehat{
	\Delta }_{ij}=(\widehat{\Delta }_{0,ij},\widehat{\Delta }_{1,ij})^{\top }$.
Now let $f_{ij}(t)=\log (1+\exp (W_{ij}^{\top }(\Gamma _{ij}^{\ast }+t%
\widehat{\Delta }_{ij}))).$ We aim to show that
\begin{align}
\frac{1}{n}\left( \sum_{l=0}^{1}||\widehat{\Delta }_{l}||_{F}^{2}\right)
^{1/2}\leq 17C_{F,1} \eta _{n}~w.p.a.1,  \label{eq:CF1}
\end{align}
where with $C_{F,1}=\frac{\sqrt{\bar{K}}(M_{W}+C_{M})(C_{\lambda
	}+C_{\Upsilon }M_{W})}{\underline{c}\kappa }+ \sqrt{\frac{c_3}{\kappa}}+%
\sqrt{c_2}$ and $\eta _{n}=\sqrt{\frac{\log n}{n\zeta _{n}}}+\frac{\log n}{%
	n\zeta _{n}}$. Following the same argument as before, we can suppose that $(%
\widehat{\Delta }_{0},\widehat{\Delta }_{1}) \notin \mathcal{C}_1(c_2)$.
Then, following \eqref{eq:f},
\begin{equation*}
f_{ij}(1)-f_{ij}(0)-f_{ij}^{\prime }(0)\geq \underline{c}\zeta _{n}\left(
\frac{(W_{ij}^{\top }\widehat{\Delta }_{ij})^{2}}{4(\max_{i,j}|W_{ij}^{\top
	} \widehat{\Delta }_{ij}|\vee \log (2))}\right) \geq \frac{\zeta _{n}
	\underline{c}(W_{ij}^{\top }\widehat{\Delta }_{ij})^{2}}{8(C_{M}+M_{W})\sqrt{
		\log n}},
\end{equation*}%
where the last inequality holds because of \eqref{eq:cm} and uniformly in $%
\left( i,j\right) $
\begin{align*}
|W_{ij}^{\top }\widehat{\Delta }_{ij}|\leq & |\widehat{\Gamma }%
_{0,ij}^{(1)}-\Gamma _{0,ij}^{\ast }|+M_{W}|\hat{\Gamma}_{1,ij}^{(1)}-\Theta
_{1,ij}^{\ast }| \\
\leq & |\widehat{\Gamma }_{0,ij}^{(1)}-\widetilde{\tau }_{n}^{\left(
	1\right) }|+|\widetilde{\tau }_{n}^{\left( 1\right) }-\Gamma _{0,ij}^{\ast
}|+M_{W}(\sqrt{\log n}+M)\leq 2(C_{M}+M_{W})\sqrt{\log n}.
\end{align*}%
Then, similar to \eqref{eq:F} and \eqref{eq:Fupper},
\begin{align*}
F_{n}(\widehat{\Delta }_{0},\widehat{\Delta }_{1})\equiv & \frac{1}{
	n_{1}(n-1)}\sum_{i\in I_{1},j\in \left[ n\right] ,j\neq i}\left( Q_{n,ij}(%
\widehat{\Gamma }_{ij}^{(1)})-Q_{n,ij}(\Gamma _{ij}^{\ast })-\partial
_{\Gamma _{ij}}Q_{n,ij}^{\top }(\Gamma _{ij}^{\ast })\widehat{\Delta }%
_{ij}\right) \\
\geq & \frac{\zeta _{n}\underline{c}}{8n_{1}(n-1)(M_{W}+C_{M})\sqrt{\log n}}%
\left[ \kappa \left( \sum_{l=0}^{1}||\widehat{\Delta }_{l}||_{F}^{2}\right)
-4(M_{W}+C_{M})^{2}\log(n)n_{1} - c_3 \log(n) n/\zeta_n\right]
\end{align*}%
and
\begin{equation*}
F_{n}(\widehat{\Delta }_{0},\widehat{\Delta }_{1})\leq \frac{\sqrt{\zeta
		_{n}n}+\sqrt{\log n}}{n_{1}(n-1)}(C_{\lambda }+C_{\Upsilon }M_{W})2\sqrt{
	\bar{K}}(\sum_{l=0}^{1}||\widehat{\Delta }_{l}||_{F}^{2})^{1/2}.
\end{equation*}%
Therefore, we have
\begin{align*}
& \left[ \left( \sum_{l=0}^{1}||\widehat{\Delta }_{l}||_{F}^{2}\right)
^{1/2}-\frac{8\sqrt{\bar{K}}(M_{W}+C_{M})(C_{\lambda }+C_{\Upsilon }M_{W})}{
	\underline{c}\kappa }\left( \frac{\sqrt{\log n}(\sqrt{n\zeta _{n}}+\sqrt{
		\log n})}{\zeta _{n}}\right) \right] ^{2} \\
\leq & \bar{K}\left[ \frac{8(M_{W}+C_{M})(C_{\lambda }+C_{\Upsilon }M_{W})}{
	\underline{c}\kappa }\right] ^{2}\left( \frac{\sqrt{\log n}(\sqrt{n\zeta
		_{n} }+\sqrt{\log n})}{\zeta _{n}}\right) ^{2}+\frac{4(M_{W}+C_{M})^{2}%
	\log(n)n_{1}}{\kappa} + \frac{c_3 \log(n) n}{\kappa \zeta_n},
\end{align*}%
and thus, \eqref{eq:CF1} holds. Then, similar to \eqref{eq:alphatilde1} and
by Assumption \ref{ass:tune}.4, we have $|\widehat{\tau }_{n}^{(1)}-\tau
_{n}|\leq \frac{1}{\sqrt{n_{1}n}}||\widehat{\Delta }_{0}||_{F}+o(\eta
_{n})\leq 30C_{F,1}\eta _{n}.$ This establishes the first result in Theorem %
\ref{thm:Frobeniusnorm}.

In addition,
\begin{align*}
& \frac{1}{n}||\widehat{\Theta }_{1}^{(1)}-\Theta _{1}^{\ast }(I_{1})||_{F}
\\
& \leq \frac{1}{n}\left[ \sum_{(i,j)\in I_{1}\times I_{1},i\neq j}\left(
\frac{1}{2}(\widehat{\Gamma }_{1,ij}^{(1)}+\widehat{\Gamma }%
_{1,ji}^{(1)})-\Theta _{1,ij}^{\ast }\right) ^{2}+\sum_{(i,j):i\in
	I_{1},j\notin I_{1}}(\widehat{\Gamma }_{1,ij}^{(1)}-\Theta _{1,ij}^{\ast
})^{2}\right] ^{1/2}+\frac{1}{n}\left( \sum_{i\in I_{1}}\Theta _{1,ii}^{\ast
	2}\right) ^{1/2} \\
& \leq \frac{1}{n}\left[ \sum_{i\in I_{1},j\in \left[ n\right] ,i\neq j}(%
\widehat{\Gamma }_{1,ij}^{(1)}-\Theta _{1,ij}^{\ast })^{2}\right] ^{1/2}+%
\frac{1}{n}\left( \sum_{i\in I_{1}}\Theta _{1,ii}^{\ast 2}\right) ^{1/2} \\
& \leq \frac{1}{n}\left( \sum_{l=0}^{1}||\widehat{\Delta }%
_{l}||_{F}^{2}\right) ^{1/2}+\sqrt{\frac{M^{2}}{3n}}\leq 18C_{F,1}\eta
_{n}~w.p.a.1,
\end{align*}%
where the first inequality holds due to the facts that $f_{M}(\cdot )$ is
1-Lipschitz continuous, $\Theta _{1}^{\ast }=(\Theta _{1}^{\ast })^{\top }$,
and $|\Theta _{1,ij}^{\ast }|\leq M$. Similarly,
\begin{align*}
& \frac{1}{n}||\widehat{\Theta }_{0}^{(1)}-\Theta _{0}^{\ast }(I_{1})||_{F}
\\
\leq & \frac{1}{n}\left[ \sum_{(i,j)\in I_{1}\times I_{1},i\neq j}\left(
\frac{1}{2}(\widehat{\Gamma }_{0,ij}^{(1)}\text{+}\widehat{\Gamma }%
_{0,ji}^{(1)})-\Theta _{0,ij}^{\ast }-\widehat{\tau }_{n}^{(1)}\right) ^{2}%
\text{+}\sum_{(i,j):i\in I_{1},j\notin I_{1}}(\Gamma _{0,ij}^{\ast }-%
\widehat{\tau }_{n}^{(1)}-\Theta _{0,ij}^{\ast })^{2}\right] ^{1/2} \\
& +\frac{1}{n}\left( \sum_{i\in I_{1}}\Theta _{0,ii}^{\ast 2}\right) ^{1/2}
\\
\leq & \frac{1}{n}\left[ \sum_{i\in I_{1},j\in \left[ n\right] ,i\neq j}(%
\widetilde{\Gamma }_{0,ij}^{(1)}-\Gamma _{0,ij}^{\ast })^{2}\right] ^{1/2}+|%
\widehat{\tau }_{n}^{(1)}-\tau _{n}|+\sqrt{\frac{M^{2}}{3n}}\leq
48C_{F,1}\eta _{n}~w.p.a.1
\end{align*}%
Then, by the Weyl's inequality, $\max_{k=1,\cdots ,K_{l}}|\widehat{\sigma }%
_{k,l}^{(1)}-\sigma _{k,l}|\leq 48C_{F,1}\eta _{n}~w.p.a.1$ for $l=0,1.$

Last, noting that $\widehat{V}_{l}^{(1)}$ consists of the first $K_{l}$
eigenvectors of $(\frac{1}{n}\widehat{\Theta }_{l}^{(1)})^{\top }(\frac{1}{n}%
\widehat{\Theta }_{l}^{(1)})$, we have%
\begin{equation*}
\left\Vert \frac{1}{n}\widehat{\Theta }_{l}^{(1)\top }\left( \frac{1}{n}%
\widehat{\Theta }_{l}^{(1)}\right) -\frac{1}{n}\Theta _{l}^{\ast \top
}(I_{1})\left( \frac{1}{n}\Theta _{l}^{\ast }(I_{1})\right) \right\Vert
_{op}\leq \frac{2C_{\sigma }}{n}||\widehat{\Theta }_{l}^{(1)}-\Theta
_{l}^{\ast }(I_{1})||_{F}\leq 96C_{F,1}C_{\sigma }\eta _{n}.
\end{equation*}%
Then by the Davis-Kahan $\sin \Theta $ Theorem (\citet[Lemma C.1]{SWZ20}),
we have
\begin{eqnarray}
||\mathcal{V}_{l}-\widehat{\mathcal{V}}_{l}^{(1)}\widehat{O}_{l}^{(1)}||_{F}
&\leq &\sqrt{K_{l}}||\mathcal{V}_{l}-\widehat{\mathcal{V}}_{l}^{(1)}\widehat{
	O}_{l}^{(1)}||_{op}\leq \frac{96\sqrt{2K_{l}}C_{F,1}C_{\sigma }s\eta _{n}}{
	\sigma _{K_{l},l}^{2}-96C_{F,1}C_{\sigma }\eta _{n}}  \notag \\
&\leq &\frac{96\sqrt{2K_{l}}C_{F,1}C_{\sigma }s\eta _{n}}{c_{\sigma
	}^{2}-96C_{F,1}C_{\sigma }\eta _{n}}\leq \frac{136\sqrt{K_{l}}
	C_{F,1}C_{\sigma }\eta _{n}}{c_{\sigma }^{2}}  \notag \\
&\leq &136C_{F,2}\eta _{n},  \label{eq:CF2}
\end{eqnarray}%
where $C_{F,2}=\max_{l=0,1}\sqrt{K_{l}}C_{F,1}C_{\sigma }c_{\sigma }^{-2},$
and the third inequality holds due to Assumption \ref{ass:tune} and the
second last inequality is due to the fact that we can set $c_{F}$ to be
sufficiently small to ensure that $1-96\sqrt{2}C_{F,1}C_{\sigma
}(c_{F}+c_{F}^{2})c_{\sigma }^{-2}\geq \frac{96\sqrt{2}}{136}.$

Recall that $\widehat{V}_{l}^{(1)}=\sqrt{n}\widehat{\mathcal{V}}_{l}^{(1)}$
and $V_{l}=\sqrt{n}\mathcal{V}_{l}$, we have the desired result that $%
||V_{l}-\widehat{V}_{l}^{(1)}\widehat{O}_{l}^{(1)}||_{F}\leq 136C_{F,2}\sqrt{
	n}\eta _{n}.$ ${\tiny \blacksquare }$

\subsection{Proof of Theorem \protect\ref{thm:rowwisebound}}

\textbf{First, we prove the first result in the theorem.} Let $\Delta
_{i,l}=(\widehat{O}_{l}^{(1)})^{\top }\widehat{u}_{i,l}^{(1)}-u_{i,l}$ for $%
l=0,1$, and $\Delta _{iu}=(\Delta _{i,0}^{\top },\Delta _{i,1}^{\top
})^{\top }$. Denote
\begin{equation}
\widehat{\Lambda }_{n,ij}=\Lambda (\widehat{\tau }_{n}+%
\sum_{l=0}^{1}u_{i,l}^{\top }(\widehat{O}_{l}^{(1)})^{\top }\widehat{v}%
_{j,l}^{(1)}W_{l,ij}).  \label{eq:lambdahat}
\end{equation}%
Recall that $\Lambda _{n,ij}=\Lambda (\tau _{n}+\sum_{l=0}^{1}u_{i,l}^{\top
}v_{j,l}W_{l,ij})=\Lambda (\tau _{n}+\Theta _{0,ij}^{\ast }+\Theta
_{1,ij}^{\ast }W_{1,ij}).$ Let
\begin{equation}
\tilde{\Lambda}_{n,ij}=\Lambda (\dot{a}_{n,ij}),  \label{eq:lambdatilde}
\end{equation}%
where $\dot{a}_{n,ij}$ is an intermediate value that is between $\tau
_{n}+\Theta _{0,ij}^{\ast }+\Theta _{1,ij}^{\ast }W_{1,ij}$ and $\widehat{
	\tau }_{n}+\sum_{l=0}^{1}u_{i,l}^{\top }(\widehat{O}_{l}^{(1)})^{\top }%
\widehat{v}_{j,l}^{(1)}W_{l,ij}$. Define
\begin{equation*}
\widehat{\phi }_{ij}^{(1)}=%
\begin{bmatrix}
(\widehat{O}_{0}^{(1)})^{\top }\widehat{v}_{j,0}^{(1)} \\
(\widehat{O}_{1}^{(1)})^{\top }\widehat{v}_{j,1}^{(1)}W_{1,ij}%
\end{bmatrix}%
\ \text{and }\widehat{\Phi }_{i}^{(1)}=\frac{1}{n_{2}}\sum_{j\in I_{2},j\neq
	i}\widehat{\phi }_{ij}^{(1)}(\widehat{\phi }_{ij}^{(1)})^{\top }.
\end{equation*}%
Let $\widetilde{\Lambda }_{ij}^{(1)}(\mu )=\Lambda (\widehat{\tau }%
_{n}+\sum_{l=0}^{1}\mu _{l}^{\top }(\widehat{O}_{l}^{(1)})^{\top }\widehat{v}%
_{j,l}^{(1)}W_{l,ij})$ and $\ell _{ij}^{\left( 1\right) }\left( \mu \right)
=Y_{ij}\log (\widetilde{\Lambda }_{ij}^{(1)}(\mu ))$ $+(1-Y_{ij})\log (1-%
\widetilde{\Lambda }_{ij}^{(1)}(\mu )).$ Define $\widetilde{Q}%
_{in}^{(1)}(\mu )=\frac{-1}{n_{2}}\sum_{j\in I_{2},j\neq i}\ell
_{ij}^{\left( 1\right) }\left( \mu \right) .$ Then,
\begin{align}
0\geq & Q_{in,U}^{(0)}(\widehat{u}_{i,0}^{(1)},\widehat{u}%
_{i,1}^{(1)})-Q_{in,U}^{(0)}((\widehat{O}_{0}^{(1)})u_{i,0},(\widehat{O}%
_{1}^{(1)})u_{i,1})  \notag  \label{eq:Q} \\
=& \widetilde{Q}_{in}^{(1)}(u_{i,0}+\Delta _{i,0},u_{i,1}+\Delta _{i,1})-%
\widetilde{Q}_{in}^{(1)}(u_{i,0},u_{i,1})  \notag \\
\geq & \frac{-1}{n_{2}}\sum_{j\in I_{2},j\neq i}(Y_{ij}-\widehat{\Lambda }%
_{n,ij})(\widehat{\phi }_{ij}^{(1)})^{\top }\Delta _{iu}  \notag \\
& +\frac{1}{n_{2}}\sum_{j\in I_{2},j\neq i}\widehat{\Lambda }_{n,ij}(1-%
\widehat{\Lambda }_{n,ij})\left[ \exp (-|(\widehat{\phi }_{ij}^{(1)})^{\top
}\Delta _{iu}|)+|(\widehat{\phi }_{ij}^{(1)})^{\top }\Delta _{iu}|-1\right]
\notag \\
\geq & \frac{-1}{n_{2}}\sum_{j\in I_{2},j\neq i}(Y_{ij}-\widehat{\Lambda }%
_{n,ij})(\widehat{\phi }_{ij}^{(1)})^{\top }\Delta _{iu}+\frac{\underline{c}
	^{\prime }\zeta _{n}}{n_{2}}\sum_{j\in I_{2},j\neq i}\left[ \exp (-|(%
\widehat{\phi }_{ij}^{(1)})^{\top }\Delta _{iu}|)+|(\widehat{\phi }%
_{ij}^{(1)})^{\top }\Delta _{iu}|-1\right]  \notag \\
\geq & \frac{-1}{n_{2}}\sum_{j\in I_{2},j\neq i}(Y_{ij}-\widehat{\Lambda }%
_{n,ij})(\widehat{\phi }_{ij}^{(1)})^{\top }\Delta _{iu}+\frac{\underline{c}
	^{\prime }\zeta _{n}}{n_{2}}\sum_{j\in I_{2},j\neq i}\left[ \frac{((\widehat{
		\phi }_{ij}^{(1)})^{\top }\Delta _{iu})^{2}}{2}-\frac{|(\widehat{\phi }
	_{ij}^{(1)})^{\top }\Delta _{iu}|^{3}}{6}\right]
\end{align}%
where the second inequality is due to \citet[Lemma 1]{B10}, the third
inequality is due to the fact that $\exp (-t)+t-1\geq 0$ and Lemma \ref%
{lem:prelim}(2), the constant $\underline{c}^{\prime }$ is defined in Lemma %
\ref{lem:prelim}, and the last inequality is due to the fact that $\exp
(-t)+t-1\geq \frac{t^{2}}{2}-\frac{t^{3}}{6}$. The following argument
follows \cite{BCFH13}. Let
\begin{equation*}
F(\Delta _{iu})=\widetilde{Q}_{in}^{(1)}(u_{i,0}+\Delta
_{i,0},u_{i,1}+\Delta _{i,1})-\widetilde{Q}_{in}^{(1)}(u_{i,0},u_{i,1})+%
\frac{1}{n_{2}}\sum_{j\in I_{2},j\neq i}(Y_{ij}-\widehat{\Lambda }_{n,ij})(%
\widehat{\phi }_{ij}^{(1)})^{\top }\Delta _{iu},
\end{equation*}%
which is convex in $\Delta _{iu}$. Let
\begin{equation}
q_{in}=\inf_{\Delta }\frac{\left[ \frac{1}{n_{2}}\sum_{j\in I_{2},j\neq i}((
	\widehat{\phi }_{ij}^{(1)})^{\top }\Delta )^{2}\right] ^{3/2}}{\frac{1}{%
		n_{2} }\sum_{j\in I_{2},j\neq i}((\widehat{\phi }_{ij}^{(1)})^{\top }\Delta
	)^{3}}\quad \text{and}\quad \delta _{in}=\left[ \frac{1}{n_{2}}\sum_{j\in
	I_{2},j\neq i}((\widehat{\phi }_{ij}^{(1)})^{\top }\Delta _{iu})^{2}\right]
^{1/2}.  \label{eq:qin}
\end{equation}%
If $\delta _{in}\leq q_{in}$, then $\frac{1}{n_{2}}\sum_{j\in I_{2},j\neq
	i}((\widehat{\phi }_{ij}^{(1)})^{\top }\Delta _{iu})^{3}\leq \delta
_{in}^{2} $, and thus $F(\Delta _{iu})\geq \frac{\underline{c}^{\prime
	}\zeta _{n}}{3}\delta _{in}^{2}.$ On the other hand, if $\delta _{in}>q_{in}$%
, let $\tilde{\Delta}_{iu}=\frac{\Delta _{iu}q_{in}}{\delta _{in}}$, then $%
\left[\frac{1}{n_{2}}\sum_{j\in I_{2},j\neq i}((\widehat{\phi }%
_{ij}^{(1)})^{\top }\tilde{\Delta}_{iu})^{2}\right]^{1/2}\leq q_{in}.$ Then,
we have
\begin{equation*}
F(\Delta _{iu})=F(\frac{\delta _{in}\tilde{\Delta}_{iu}}{q_{in}})\geq \frac{
	\delta _{in}}{q_{in}}F(\tilde{\Delta}_{iu})\geq \frac{\underline{c}^{\prime
	}\zeta _{n}\delta _{in}}{3n_{2}q_{in}}\sum_{j\in \left[ n\right] ,j\neq i}((%
\widehat{\phi }_{ij}^{(1)})^{\top }\tilde{\Delta}_{iu})^{2}=\frac{\underline{
		c}^{\prime }\zeta _{n}q_{in}\delta _{in}}{3}.
\end{equation*}%
Therefore, by Lemma \ref{lem:phi},
\begin{equation}
F(\Delta _{iu})\geq \min \left( \frac{\underline{c}^{\prime }\zeta
	_{n}\delta _{in}^{2}}{3},\frac{\underline{c}^{\prime }\zeta _{n}q_{in}\delta
	_{in}}{3}\right) \geq \min \left( \frac{\underline{c}^{\prime }c_{\phi
	}\zeta _{n}\underline{c}||\Delta _{iu}||^{2}}{6},\frac{\underline{c}^{\prime
	}\zeta _{n}q_{in}\sqrt{c_{\phi }}||\Delta _{iu}||}{3\sqrt{2}}\right) .
\label{eq:Fdelta}
\end{equation}%
On the other hand, we have $|F(\Delta _{iu})|\leq \left\vert \frac{1}{n}%
\sum_{j\in I_{2},j\neq i}(Y_{ij}-\widehat{\Lambda }_{n,ij})(\widehat{\phi }%
_{ij}^{(1)})^{\top }\Delta _{iu}\right\vert \leq I_{i}+II_{i},$ where
\begin{equation*}
I_{i}=\left\vert \frac{1}{n}\sum_{j\in I_{2},j\neq i}\left( Y_{ij}-\Lambda
_{n,ij}\right) (\widehat{\phi }_{ij}^{(1)})^{\top }\Delta _{iu}\right\vert
\text{ and }II_{i}=\left\vert \frac{1}{n}\sum_{j\in I_{2},j\neq i}(\widehat{
	\Lambda }_{n,ij}-\Lambda _{n,ij})(\widehat{\phi }_{ij}^{(1)})^{\top }\Delta
_{iu}\right\vert .
\end{equation*}%
We aim to upper bound $I_{i}$ and $II_{i}$ uniformly in $i$ below.

We first bound $II_{i}$. Note that
\begin{align}
II_{i}\leq & \frac{1}{n_{2}}\sum_{j\in I_{2},j\neq i}\tilde{\Lambda}%
_{n,ij}(1-\tilde{\Lambda}_{n,ij})\biggl(|\widehat{\tau }_{n}-\tau
_{n}|+\sum_{l=0}^{1}\left\vert u_{i,l}^{\top }((\widehat{O}_{l}^{(1)})^{\top
}\widehat{v}_{j,l}^{(1)}-v_{j,l})W_{l,ij}\right\vert \biggr)|(\widehat{\phi }%
_{ij}^{(1)})^{\top }\Delta _{iu}|  \notag \\
\leq & \frac{2\overline{c}^{\prime }M(1+M_W)\zeta _{n}||\Delta _{iu}||}{
	n_{2}c_{\sigma }}\sum_{j\in I_{2},j\neq i}\left( |\widehat{\tau }_{n}-\tau
_{n}|+\sum_{l=0}^{1}\left\vert u_{i,l}^{\top }((\widehat{O}_{l}^{(1)})^{\top
}\widehat{v}_{j,l}^{(1)}-v_{j,l})W_{l,ij}\right\vert \right)  \notag \\
\leq & \frac{2\overline{c}^{\prime }M(1+M_W)\zeta _{n}||\Delta _{iu}||}{
	c_{\sigma }}\left[ 48C_{F,1}\eta _{n}+c_{II}\sum_{l=0}^{1}\frac{1}{n_{2}}%
\sum_{j\in I_{2},j\neq i}\left\Vert (\widehat{O}_{l}^{(1)})^{\top }\widehat{v%
}_{j,l}^{(1)}-v_{j,l}\right\Vert \right]  \notag \\
\leq & \frac{2\overline{c}^{\prime }M(1+M_W)\zeta _{n}||\Delta _{iu}||}{
	c_{\sigma }}\left[ 48C_{F,1}\eta _{n}+c_{II}\sum_{l=0}^{1}\frac{1}{\sqrt{
		n_{2}}}\left\Vert \widehat{V}_{l}^{(1)}\widehat{O}_{l}^{(1)}-V_{l}\right%
\Vert _{F}\right]  \notag \\
\leq & C_{II}||\Delta _{iu}||\zeta _{n}\eta _{n},  \label{eq:IIa}
\end{align}%
where $c_{II}=M(1+M_{W})$, $C_{II}=2\overline{c}^{\prime }M(1+M_W)\zeta
_{n}\left( 48C_{F,1}+136c_{II}C_{F,2}\right) c_{\sigma }^{-1}$, the first
inequality holds by the Taylor expansion, the second inequality holds by
Lemma \ref{lem:prelim}
\begin{align}
\max_{i,j\in I_{2},i\neq j}||\widehat{\phi }_{ij}^{(1)}||\leq & \max_{i,j\in
	I_{2},i\neq j}\left( ||\widehat{O}_{0}^{(1)}\widehat{v}_{j,0}^{(1)}||+M_{W}||%
\widehat{O}_{1}^{(1)}\widehat{v}_{j,1}^{(1)}||\right)  \notag
\label{eq:phi1} \\
\leq & 2M\sigma _{K_{0},0}^{-1}+2M_{W}M\sigma _{K_{1},1}^{-1}\leq
2M(1+M_{W})c_{\sigma }^{-1},
\end{align}%
the third inequality is due to Theorem \ref{thm:Frobeniusnorm} and the fact
that $||u_{i,l}^{\top }W_{l,ij}||\leq c_{II}$, the fourth inequality is due
to Cauchy's inequality, and the last inequality is due to Theorem \ref%
{thm:Frobeniusnorm}. Note that the constant $C_{II}$ does not depend on $i$,
the above upper bound for $II_{i}$ holds uniformly over $i$.

Next, we turn to the upper bound for $I_{i}$. Let $\mathcal{F}_{n}$ be the $%
\sigma $-field generated by $\{X_{i}\}_{i=1}^{n}\cup \{\varepsilon
_{ij}\}_{i\in I_{1},j\in \left[ n\right] ,j\neq i}\cup \{e_{ij}\}_{1\leq
	i,j\leq n}$ and $H_{ij}=(Y_{ij}-\Lambda _{n,ij})\widehat{\phi }_{ij}^{(1)}.$
Further note that, for $i \in I_2$, $\{\varepsilon _{ij}\}_{j\in I_{2},j\neq
	i}$ is independent of $\mathcal{F}_n$. Therefore, conditional on $\mathcal{F}%
_{n}$, $\left\{ H_{ij}\right\} _{j\in I_{2},j\neq i}$ only depends on $%
\{\varepsilon _{ij}\}_{j\in I_{2},j\neq i}$, and thus, is a sequence of
independent random vectors. Note that $I_{i}\leq ||\frac{1}{n_{2}}\sum_{j\in
	I_{2},j\neq i}H_{ij}||||\Delta _{iu}||. $ Let $H_{k,ij}$ be the $k$-th
coordinate of $H_{ij}$ where $k\in [ K_{0}+K_{1}]$ and
\begin{equation*}
\mathcal{A}_n=\{\max_{j\in I_{2}}||(\widehat{O}_{l}^{(1)})^{\top }\widehat{v}%
_{j,l}^{(1)}||\leq 2M\sigma _{K_{l},l}^{-1}\} \in \mathcal{F}_n.
\end{equation*}
By Lemma \ref{lem:prelim}, $\mathbb{P}(\mathcal{A}_n) \rightarrow 1$. Under $%
\mathcal{A}_n$ and Assumption \ref{ass:tune}, we have
\begin{equation}
\max_{1\leq i,j\leq n}|H_{k,ij}|\leq \left[ 2M(1+M_{W})c_{\sigma }^{-1}+1%
\right] ^{2}(1+\overline{c})\equiv C_{H}  \label{eq:CH}
\end{equation}%
and $\sum_{j\in I_{2},j\neq i}\mathbb{E}(H_{k,ij}^{2}|\mathcal{F}_{n})\leq
C_{H}\zeta _{n}n_{2}.$ Therefore, by the Bernstein inequality, for any $t>0$%
,
\begin{equation*}
\mathbb{P}\left( \max_{i\in I_{2}}\left\vert \sum_{j\in I_{2},j\neq
	i}H_{k,ij}\right\vert \geq n_{2}t\biggl|\mathcal{F}_{n}\right)1\{\mathcal{A}%
_n\} \leq \sum_{i\in I_{2}}2\exp \left( -\frac{\frac{n_{2}^{2}t^{2}}{2}}{%
	C_{H}\zeta _{n}n_{2}+\frac{C_{H}tn_{2}}{3}}\right) .
\end{equation*}%
Taking $t=4C_{H}\sqrt{\frac{\zeta _{n}\log n}{n}}$, we have
\begin{align*}
\mathbb{P}\left( \max_{i\in I_{2}}\frac{1}{n_{2}}\left\vert \sum_{j\in
	I_{2},j\neq i}H_{k,ij}\right\vert \geq t\biggl|\mathcal{F}_{n}\right)1\{%
\mathcal{A}_n\} \leq & 2n_{2}\exp \left( -\frac{\frac{16C_{H}^{2}\zeta
		_{n}\log nn_{2}^{2}}{2n}}{ C_{H}\zeta _{n}n_{2}+\frac{4C_{H}^{2}\sqrt{\frac{%
				\zeta _{n}\log n}{n}}n_{2}}{ 3}}\right) \\
& \leq 2n_{2}\exp \left( -\frac{8\log n}{7}\right) \leq n^{-1.1},
\end{align*}%
where the second inequality holds because $\log n/(n\zeta _{n})\leq c_{F}<1$
and $C_{H}>1$. Then, we have
\begin{align*}
\mathbb{P}\left( \max_{i\in I_{2}}\frac{1}{n_{2}}\left\vert \sum_{j\in
	I_{2},j\neq i}H_{k,ij}\right\vert \geq t\right) \leq & \mathbb{P}\left(
\max_{i\in I_{2}}\frac{1}{n_{2}}\left\vert \sum_{j\in I_{2},j\neq
	i}H_{k,ij}\right\vert \geq t,\mathcal{A}_n\right) + \mathbb{P}(\mathcal{A}%
_n^c) \\
\leq & \mathbb{E}\left[\mathbb{P}\left( \max_{i\in I_{2}}\frac{1}{n_{2}}%
\left\vert \sum_{j\in I_{2},j\neq i}H_{k,ij}\right\vert \geq t\biggl|%
\mathcal{F}_n\right)1\{\mathcal{A}_n\}\right] + \mathbb{P}(\mathcal{A}_n^c)
\\
\leq & n^{-1.1} + \mathbb{P}(\mathcal{A}_n^c) \rightarrow 0.
\end{align*}

This means
\begin{equation}
\max_{i\in I_{2}}I_{i}\leq \max_{i\in I_{2}}\frac{1}{n_{2}}\left\vert
\sum_{j\in I_{2},j\neq i}H_{k,ij}\right\vert \leq 4C_{H}\sqrt{\frac{\log
		n\zeta _{n}}{n}}~w.p.a.1.  \label{eq:Ia}
\end{equation}

Combining \eqref{eq:IIa} and \eqref{eq:Ia}, we have
\begin{equation}
|F(\Delta _{iu})|\leq (4C_{H}+C_{II})\zeta _{n}\eta _{n}||\Delta _{iu}||.
\label{eq:Fdelta2}
\end{equation}%
Then, \eqref{eq:Fdelta} and \eqref{eq:Fdelta2} imply
\begin{equation}
(4C_{H}+C_{II})\zeta _{n}\eta _{n}||\Delta _{iu}||\geq \min \left( \frac{
	\underline{c}c_{\phi }\zeta _{n}\Vert |\Delta _{iu}||^{2}}{6},\frac{
	\underline{c}\sqrt{c_{\phi }}\zeta _{n}q_{in}||\Delta _{iu}||}{3\sqrt{2}}%
\right) .  \label{eq:Fdelta3}
\end{equation}%
On the other hand, we have
\begin{equation*}
\liminf_{n}\min_{i\in I_{2}}\frac{\underline{c}^{\prime }\sqrt{c_{\phi }}
	\zeta _{n}q_{in}||\Delta _{iu}||}{3\sqrt{2}}\geq \frac{c_{\sigma }\underline{%
		c}^{\prime }c_{\phi }\zeta _{n}||\Delta _{iu}||}{24M(1+M_{W})}%
>(4C_{H}+C_{II})\zeta _{n}\eta _{n}||\Delta _{iu}||,
\end{equation*}%
where the first inequality holds by Lemma \ref{lem:q} and the second
inequality holds due to the fact that $c_F$ is sufficiently small so that
\begin{align*}
(4C_{H}+C_{II})(c_{F}+c_{F}^{2}) <\frac{\underline{c}^{\prime }c_{\phi
	}c_{\sigma }}{24M(1+M_{W})}.
\end{align*}

Therefore, \eqref{eq:Fdelta3} implies
\begin{equation}
||(\widehat{O}_{l}^{(1)})^{\top }\widehat{u}_{i,l}^{(1)}-u_{i,l}||\leq
||\Delta _{u}|| \leq \frac{6(4C_{H}+C_{II})}{\underline{c}c_{\phi }}\eta
_{n} \equiv C_{1}^{\ast }\eta _{n}~w.p.a.1.  \label{eq:C1star}
\end{equation}
Because the constant $C_{1}^{\ast }$ does not depend on index $i$, the above
inequality holds uniformly over $i\in I_{2}$.

\textbf{Now, we prove the second result in the theorem}. The proof follows
that of the first result with a notable difference: the regressors $\{%
\widehat{u}_{i,l}^{(1)}\}_{i\in I_{2},l=0,1}$ obtained from the previous
step are not independent of the observations $\left\{ Y_{ij}\right\} $ given
the covariates. Thus, the conditional Bernstein inequality argument above
cannot be used again. Recall that
\begin{equation*}
(\dot{v}_{j,0}^{(0,1)},\dot{v}_{j,1}^{(0,1)})=\argmin Q_{jn,V}^{(0)}(\nu
_{0},\nu _{1}),
\end{equation*}%
where $Q_{jn,V}^{(0)}(\nu )$ with $\nu =(\nu _{0}^{\top },\nu _{1}^{\top
})^{\top }$ is defined in Section \ref{sec:sslre}. Let
\begin{equation*}
\tilde{\Lambda}_{ij}^{\left( 0\right) }(\nu )=\Lambda (\widehat{\tau }%
_{n}+\sum_{l=0}^{1}\nu _{l}^{\top }(\widehat{O}_{l}^{(1)})^{\top }\widehat{u}%
_{i,l}^{(1)}W_{l,ij})
\end{equation*}
and
\begin{equation*}
\ell _{ij}^{\left( 0\right) }\left( \nu \right) =Y_{ij}\log (\tilde{\Lambda}%
_{ij}^{\left( 0\right) }(\nu ))+(1-Y_{ij})\log (1-\tilde{\Lambda}%
_{ij}^{\left( 0\right) }(\nu )).
\end{equation*}
Define $\widetilde{Q}_{jn,V}^{(0)}(\nu )=\frac{-1}{n_{2}}\sum_{j\in
	I_{2},j\neq i}\ell _{ij}^{\left( 1\right) }\left( \nu \right) .$ Then
\begin{equation*}
Q_{jn,V}^{(0)}(\nu _{0},\nu _{1})=\widetilde{Q}_{jn,V}^{(0)}((\widehat{O}%
_{0}^{(1)})^{\top }\nu _{0},(\widehat{O}_{1}^{(1)})^{\top }\nu _{1}).
\end{equation*}%
Recall that $\Lambda _{n,ij}=\Lambda (\tau _{n}+\sum_{l=0}^{1}u_{i,l}^{\top
}v_{j,l}W_{l,ij})=\Lambda (\tau _{n}+\Theta _{0,ij}^{\ast }+\Theta
_{1,ij}^{\ast }W_{1,ij}).$ Let $\dot{\Lambda}_{n,ij}=\Lambda (\widehat{\tau }%
_{n}+\sum_{l=0}^{1}v_{j,l}^{\top }(\widehat{O}_{l}^{(1)})^{\top }\widehat{u}%
_{i,l}^{(1)}W_{l,ij})$ and $\tilde{\Lambda}_{n,ij}=\Lambda (\dot{a}_{n,ij}),$
where $\dot{a}_{n,ij}$ is an intermediate value that is between $\tau
_{n}+\Theta _{0,ij}^{\ast }+\Theta _{1,ij}^{\ast }W_{1,ij}$ and $\widehat{
	\tau }_{n}+\sum_{l=0}^{1}v_{j,l}^{\top }(\widehat{O}_{l}^{(1)})^{\top }%
\widehat{u}_{i,l}^{(1)}W_{l,ij}$. Define%
\begin{equation*}
\dot{\psi}_{ij}=%
\begin{bmatrix}
(\widehat{O}_{0}^{(1)})^{\top }\widehat{u}_{i,0}^{(1)} \\
(\widehat{O}_{1}^{(1)})^{\top }\widehat{u}_{i,1}^{(1)}W_{1,ij}%
\end{bmatrix}%
\quad \text{and}\quad \dot{\Psi}_{j}=\frac{1}{n_{2}}\sum_{i\in I_{2},i\neq j}%
\dot{\psi}_{ij}(\dot{\psi}_{ij})^{\top }.
\end{equation*}%
Let $\Delta _{jv}\equiv (\Delta _{j,0}^{\top },\Delta _{j,1}^{\top })^{\top
},$ where $\Delta _{j,l}=(\widehat{O}_{l}^{(1)})^{\top }\dot{v}%
_{j,l}^{(0,1)}-v_{j,l}$ for $l=0,1.$ Then we have
\begin{align*}
0\geq & Q_{jn,V}^{(0)}(\dot{v}_{j,0}^{(0,1)},\dot{v}%
_{j,1}^{(0,1)})-Q_{jn,V}^{(0)}((\widehat{O}_{0}^{(1)})^{\top }v_{j,0},(%
\widehat{O}_{1}^{(1)})^{\top }v_{j,1}) \\
=& \widetilde{Q}_{jn,V}^{(0)}((\widehat{O}_{0}^{(1)})^{\top }\dot{v}%
_{j,0}^{(0,1)},(\widehat{O}_{1}^{(1)})^{\top }\dot{v}_{j,1}^{(0,1)})-%
\widetilde{Q}_{jn,V}^{(0)}(v_{j,0},v_{j,1}) \\
\geq & \frac{-1}{n}\sum_{i\in I_{2},i\neq j}(Y_{ij}-\dot{\Lambda}_{n,ij})(%
\dot{\psi}_{ij})^{\top }\Delta _{v}+\frac{\underline{c}^{\prime }\zeta _{n}}{
	n}\sum_{i\in I_{2},i\neq j}\left[ \frac{((\dot{\psi}_{ij})^{\top }\Delta
	_{v})^{2}}{2}-\frac{|(\dot{\psi}_{ij})^{\top }\Delta _{v}|^{3}}{6}\right] .
\end{align*}%
By the first result that $\max_{i\in I_{2}}||(\widehat{O}_{l}^{(1)})^{\top }%
\widehat{u}_{i,l}^{(1)}-u_{i,l}||\leq C_{1}^{\ast }\eta _{n}$, we have
\begin{equation*}
\max_{i\in I_{2}}||(\widehat{O}_{l}^{(1)})^{\top }\widehat{u}%
_{i,l}^{(1)}W_{l,ij}||\leq M_{W}\max_{i\in I_{2}}\left[ ||(\widehat{O}%
_{l}^{(1)})^{\top }\widehat{u}_{i,l}^{(1)}-u_{i,l}||+||u_{i,l}||\right] \leq
M_{W}(C_{1}^{\ast }\eta _{n}+M)<\infty .
\end{equation*}%
Therefore, similar to \eqref{eq:phi}, we have
\begin{align*}
||\dot{\Psi}_{j}-\Psi _{j}(I_{2})||\leq & \frac{2M_{W}(C_{1}^{\ast }\eta
	_{n}+M)}{n}\sum_{l=0}^{1}\sum_{i\in I_{2}}||(\widehat{O}_{l}^{(1)})^{\top }%
\widehat{u}_{i,l}^{(1)}-u_{i,l}|| \\
\leq & M_{W}(C_{1}^{\ast }\eta _{n}+M)C_{1}^{\ast }\eta _{n}~w.p.a.1.
\end{align*}%
As $c_{F}$ is sufficiently small so that $M_{W}(C_{1}^{\ast }\eta
_{n}+M)C_{1}^{\ast }(c_{F}+c_{F}^{2})\leq c_{\phi }/2$ can be ensured and
Assumption \ref{ass:phiprime} holds, we have $\min_{j\in [ n]}\lambda _{\min
}(\dot{\Psi}_{j})\geq c_{\phi }/2~w.p.a.1.$

Let
\begin{equation*}
F(\Delta _{jv})=\widetilde{Q}_{jn}^{(0)}(v_{j,0}+\Delta
_{j,0},v_{j,1}+\Delta _{j,1})-\widetilde{Q}_{jn}^{(0)}(v_{j,0},v_{j,1})+%
\frac{1}{n}\sum_{i\in I_{2},i\neq j}(Y_{ij}-\dot{\Lambda}_{n,ij})(\dot{\psi}%
_{ij})^{\top }\Delta _{jv}.
\end{equation*}%
Following the same argument in the proof of Theorem \ref{thm:rowwisebound},
we have
\begin{equation*}
F(\Delta _{jv})\geq \min \left( \frac{\underline{c}^{\prime }c_{\phi }\zeta
	_{n}\underline{c}||\Delta _{jv}||^{2}}{6},\frac{\underline{c}^{\prime }\zeta
	_{n}q_{jn}\sqrt{c_{\phi }}||\Delta _{jv}||}{3\sqrt{2}}\right) ,
\end{equation*}%
where $q_{jn}=\inf_{\Delta }\frac{\left[ \frac{1}{n_{2}}\sum_{i\in
		I_{2},i\neq j}((\dot{\psi}_{ij})^{\top }\Delta )^{2}\right] ^{3/2}}{\frac{1}{
		n_{2}}\sum_{i\in I_{2},i\neq j}((\dot{\psi}_{ij})^{\top }\Delta )^{3}}.$ For
the upper bound of $F(\Delta _{jv})$, we can show that
\begin{equation*}
F(\Delta _{jv})\leq \left\vert \frac{1}{n_{2}}\sum_{i\in I_{2},i\neq
	j}\left( Y_{ij}-\Lambda _{n,ij}\right) (\dot{\psi}_{ij})^{\top }\Delta
_{jv}\right\vert +\left\vert \frac{1}{n_{2}}\sum_{i\in I_{2},i\neq j}(\dot{
	\Lambda}_{n,ij}-\Lambda _{n,ij})(\dot{\psi}_{ij})^{\top }\Delta
_{jv}\right\vert \equiv \tilde{I}_{j}+\widetilde{II}_{j}.
\end{equation*}

We first bound $\widetilde{II}_{j}$. Following Lemma \ref{lem:prelim}(1), we
have
\begin{equation*}
||v_{j,l}^{\top }(\widehat{O}_{l}^{(1)})^{\top }\widehat{u}%
_{i,l}^{(1)}W_{l,ij}||\lesssim ||(\widehat{O}_{l}^{(1)})^{\top }\widehat{u}%
_{i,l}^{(1)}-u_{i,l}||+||u_{i,l}||\leq C<\infty .
\end{equation*}%
Then, by the same argument in the proof of Lemma \ref{lem:prelim}(2), we
have
\begin{equation*}
\overline{c}^{\prime }\zeta _{n}\geq \dot{\Lambda}_{n,ij}\geq \underline{c}%
^{\prime }\zeta _{n}\quad \text{and}\quad \overline{c}^{\prime }\zeta
_{n}\geq \tilde{\Lambda}_{n,ij}\geq \underline{c}^{\prime }\zeta _{n},
\end{equation*}%
for some constants $\infty >\overline{c}^{\prime }>\underline{c}^{\prime }>0$%
. Following \eqref{eq:IIa} and by noticing that $\frac{1}{n_{2}}\sum_{i\in
	I_{2},i\neq j}||(\widehat{O}_{l}^{(1)})^{\top }\widehat{u}%
_{i,l}^{(1)}-u_{i,l}||\leq C_{1}^{\ast }\eta _{n},$ we have
\begin{equation}
\widetilde{II}_{j}\leq C_{II}^{\prime }\zeta _{n}\eta _{n}||\Delta _{jv}||,
\label{eq:CIIprime}
\end{equation}%
for some constant $C_{II}^{\prime }>0$.

The analysis of $\widetilde{I}_{j}$ is different from that of $I_{i}$ as we
no longer have the independence between $\dot{\psi}_{ij}$ and $%
Y_{ij}-\Lambda _{n,ij}$ given $\{W_{1,ij}\}_{1\leq i<j\leq n}$. Instead, we
let $\psi _{ij}=%
\begin{bmatrix}
u_{i,0} \\
u_{i,1}W_{1,ij}%
\end{bmatrix}%
.$ Note that $\psi _{ij}$ is deterministic given $\{W_{1,ij}\}_{1\leq
	i<j\leq n}$. In addition, $\max_{i,j\in [ n],i\neq j}||\dot{\psi}_{ij}-\psi
_{ij}||\leq (1+M_{W})C_{1}^{\ast }\eta _{n}.$ Therefore,
\begin{equation*}
\tilde{I}_{j}\leq \left[ \left\Vert \frac{1}{n_{2}}\sum_{i\in I_{2},i\neq
	j}\left( Y_{ij}-\Lambda _{n,ij}\right) \psi _{ij}\right\Vert +\frac{1}{n_{2}}%
\sum_{i\in I_{2},i\neq j}\left\vert Y_{ij}-\Lambda _{n,ij}\right\vert ||\dot{
	\psi}_{ij}-\psi _{ij}||\right] ||\Delta _{jv}||.
\end{equation*}%
For the first term in the square brackets, by the conditional Bernstein
inequality given $\{W_{1,ij}\}_{1\leq i<j\leq n}$, we have
\begin{equation}
\max_{j\in [ n]}\left\Vert \frac{1}{n_{2}}\sum_{i\in I_{2},i\neq j}\left(
Y_{ij}-\Lambda _{n,ij}\right) \psi _{ij}\right\Vert \leq C_{H}^{\prime }%
\sqrt{\frac{\log n\zeta _{n}}{n}}~w.p.a.1,  \label{eq:CHprime}
\end{equation}%
where $C_{H}^{\prime }=4(1+\overline{c})^{2}\left[ C_{u}C_{\sigma
}(M_{W}+1)+1\right] ^{4}$. For the second term in the square brackets, we
have
\begin{align*}
\frac{1}{n_{2}}\sum_{i\in I_{2},i\neq j}\left\vert Y_{ij}-\Lambda
_{n,ij}\right\vert \cdot ||\dot{\psi}_{ij}-\psi _{ij}||\leq & \frac{
	(1+M_{W})C_{1}^{\ast }\eta _{n}}{n_{2}}\sum_{i\in I_{2},i\neq j}\left\vert
Y_{ij}-\Lambda _{n,ij}\right\vert \\
\leq & (1+M_{W})C_{1}^{\ast }\eta _{n}\left[ \frac{1}{n_{2}}\sum_{i\in
	I_{2},i\neq j}\left( Y_{ij}-\Lambda _{n,ij}\right) +\frac{2}{n_{2}}%
\sum_{i\in I_{2},i\neq j}\Lambda _{n,ij}\right] \\
\leq & (1+M_{W})C_{1}^{\ast }\eta _{n}\left( 4\overline{c}\sqrt{\frac{\zeta
		_{n}\log n}{n}}+2\overline{c}\zeta _{n}\right) \\
\leq & 3(1+M_{W})\overline{c}C_{1}^{\ast }\eta _{n}\zeta _{n},
\end{align*}%
where the second last inequality is due to the Bernstein inequality and
Assumption \ref{ass:tune}, and the last inequality holds because $4\sqrt{
	\frac{\log n}{n\zeta _{n}}}\leq 4c_{F}\leq 1.$

Combining the two estimates, we have uniformly in $j$ and
\begin{equation*}
\tilde{I}_{j}\leq \left( C_{H}^{\prime }\sqrt{\frac{\log n}{n\zeta _{n}}}%
+3(1+M_{W})\overline{c}C_{1}^{\ast }\right) \eta _{n}\zeta _{n}||\Delta
_{v}||\leq 4(1+M_{W})\overline{c}C_{1}^{\ast }\eta _{n}\zeta _{n}||\Delta
_{jv}||~w.p.a.1,
\end{equation*}%
where the last inequality holds because $c_F$ is sufficiently small so that $%
C_{H}^{\prime }(c_{F}+c_{F}^{2})\leq (1+M_{W})\overline{c}C_{1}^{\ast }$.

Combining the upper and lower bounds for $F(\Delta _{jv})$, we have,
w.p.a.1,
\begin{equation}
[ 4(1+M_{W})\overline{c}C_{1}^{\ast }+C_{II}^{\prime }]\eta _{n}\zeta
_{n}||\Delta _{jv}||\geq \min \left( \frac{\underline{c}^{\prime }c_{\phi
	}\zeta _{n}\underline{c}||\Delta _{jv}||^{2}}{6},\frac{\underline{c}^{\prime
	}\zeta _{n}q_{jn}\sqrt{c_{\phi }}||\Delta _{jv}||}{3\sqrt{2}}\right) .
\label{eq:Fdeltav}
\end{equation}%
By the same argument in Lemma \ref{lem:q}, we have
\begin{equation*}
q_{jn}\geq \inf_{\Delta }\sqrt{\frac{\frac{c_\sigma^2}{n}\sum_{i\in
			I_{2},i\neq j}(( \dot{\psi}_{ij})^{\top }\Delta )^{2}}{%
		16(1+M_{W})^{2}M^{2}||\Delta ||^{2}}}\geq \frac{c_{\sigma } \sqrt{c_{\phi}/2}%
}{4(1+M_{W})M}>0.
\end{equation*}%
In addition, because $c_{F}$ can be made sufficiently small to ensure $%
(4(1+M_{W})\overline{c}C_{1}^{\ast }+C_{II}^{\prime })(c_{F}+c_{F}^{2})<%
\frac{c_{\sigma }\underline{c}^{\prime }\sqrt{c_{\phi }}}{24 (1+M_{W})M},$
we have
\begin{align*}
(4(1+M_{W})\overline{c}C_{1}^{\ast }+C_{II}^{\prime })\eta _{n}\zeta
_{n}||\Delta _{jv}||& \leq (4(1+M_{W})\overline{c}C_{1}^{\ast
}+C_{II}^{\prime })(c_{F}+c_{F}^{2})\zeta _{n}||\Delta _{jv}|| \\
& <\frac{c_{\sigma }\underline{c}^{\prime }c_{\phi }\zeta _{n}||\Delta
	_{jv}||}{24(1+M_{W})M} \leq \frac{ \underline{c}^{\prime }\sqrt{c_{\phi }}%
	\zeta _{n}q_{jn}||\Delta _{jv}||}{3 \sqrt{2}}.
\end{align*}%
Then, \eqref{eq:Fdeltav} implies
\begin{equation}
||\Delta _{jv}||\leq \frac{6(4(1+M_{W})\overline{c}C_{1}^{\ast
	}+C_{II}^{\prime })}{\underline{c}^{\prime }c_{\phi }\underline{c}}\eta
_{n}\equiv C_{0,v}\eta _{n}~w.p.a.1.  \label{eq:C0v}
\end{equation}%
Note the constant $C_{0,v}$ on the right hand side does not depend on $j$ so
that the desired result holds uniformly over $j\in [ n]$. ${\tiny \
	\blacksquare }$

\subsection{Proof of Theorem \protect\ref{thm:iter}}

We can establish the desired results by induction. Given $\max_{j\in [ n]}||(%
\widehat{O}_{l}^{(1)})^{\top }\dot{v}_{j,l}^{(h-1,1)}-v_{j,l}||\leq
C_{h-1,v}\eta _{n}$ w.p.a.1, we can readily show that
\begin{equation*}
\max_{i\in [ n]}||(\widehat{O}_{l}^{(1)})^{\top }\dot{u}%
_{i,l}^{(h,1)}-u_{i,l}||\leq C_{h,u}\eta _{n}.
\end{equation*}%
Then, given $\max_{i\in [ n]}||(\widehat{O}_{l}^{(1)})^{\top }\dot{u}%
_{i,l}^{(h,1)}-u_{i,l}||\leq C_{h,u}\eta _{n}~w.p.a.1$, we can show that
\begin{equation*}
\max_{j\in [ n]}||(\widehat{O}_{l}^{(1)})^{\top }\dot{v}%
_{j,l}^{(h,1)}-v_{j,l}||\leq C_{h,v}\eta _{n}.
\end{equation*}%
As the regressors in both iteration steps have the uniform bound, the proof
of Theorem \ref{thm:iter} is similar to that of the second result in Theorem %
\ref{thm:rowwisebound}, and is thus omitted for brevity. ${\tiny \
	\blacksquare }$

\subsection{Proof of Theorem \protect\ref{thm:kmeans}}

\label{sec:thmkmeans} Let $v_{j}^{\ast }=\left( \frac{(\widehat{O}
	_{1}^{(1)}v_{j,1})^{\top }}{||\widehat{O}_{1}^{(1)}v_{j,1}||},\frac{(
	\widehat{O}_{1}^{(2)}v_{j,1})^{\top }}{||\widehat{O}_{1}^{(2)}v_{j,1}||}%
\right) ^{\top }$. Then we have
\begin{align}
||\overline{v}_{j}-v_{j}^{\ast }||& \leq \left\Vert \frac{\dot{v}
	_{j,1}^{(H,1)}}{||\dot{v}_{j,1}^{(H,1)}||}-\frac{\widehat{O}_{1}^{(1)}v_{j,1}%
}{||\widehat{O}_{1}^{(1)}v_{j,1}||}\right\Vert +\left\Vert \frac{\dot{v}
	_{j,1}^{(H,2)}}{||\dot{v}_{j,1}^{(H,2)}||}-\frac{\widehat{O}_{1}^{(2)}v_{j,1}%
}{||\widehat{O}_{1}^{(2)}v_{j,1}||}\right\Vert  \notag \\
& =\left\Vert \frac{(\widehat{O}_{1}^{(1)})^{\top }\dot{v}_{j,1}^{(H,1)}}{%
	||( \widehat{O}_{1}^{(1)})^{\top }\dot{v}_{j,1}^{(H,1)}||}-\frac{v_{j,1}}{
	||v_{j,1}||}\right\Vert +\left\Vert \frac{(\widehat{O}_{1}^{(2)})^{\top }
	\dot{v}_{j,1}^{(H,2)}}{||(\widehat{O}_{1}^{(2)})^{\top }\dot{v}
	_{j,1}^{(H,2)}||}-\frac{v_{j,1}}{||v_{j,1}||}\right\Vert  \notag \\
& \leq \frac{2\left\Vert (\widehat{O}_{1}^{(1)})^{\top }\dot{v}
	_{j,1}^{(H,1)}-v_{j,1}\right\Vert }{||(\widehat{O}_{1}^{(1)})^{\top }\dot{v}
	_{j,1}^{(H,1)}||}+\frac{2\left\Vert (\widehat{O}_{1}^{(2)})^{\top }\dot{v}
	_{j,1}^{(H,2)}-v_{j,1}\right\Vert }{||(\widehat{O}_{1}^{(2)})^{\top }\dot{v}
	_{j,1}^{(H,2)}||}  \notag \\
& \leq \frac{4C_{H,v}\eta _{n}}{||v_{j,1}||-C_{H,v}\eta _{n}}\leq
5C_{1}^{-1/2}C_{H,v}\eta _{n},  \label{eq:v}
\end{align}%
where the last inequality is due to the fact that $||v_{j,1}||\geq
C_{1}^{-1/2}$ and $C_{H,v}\eta _{n}\leq C_{H,v}(c_{F}+c_{F}^{2})\leq
C_{1}^{-1/2}/5$ as $c_{F}$ can be made sufficiently small. In addition, by
Lemma \ref{lem:theta}, for $z_{i}\neq z_{j}$,
\begin{align}
||v_{j}^{\ast }-v_{i}^{\ast }||=& \left[ \left\Vert \frac{\widehat{O}
	_{1}^{(1)}v_{i,1}}{||\widehat{O}_{1}^{(1)}v_{i,1}||}-\frac{\widehat{O}
	_{1}^{(1)}v_{j,1}}{||\widehat{O}_{1}^{(1)}v_{j,1}||}\right\Vert
^{2}+\left\Vert \frac{\widehat{O}_{1}^{(2)}v_{i,1}}{||\widehat{O}
	_{1}^{(2)}v_{i,1}||}-\frac{\widehat{O}_{1}^{(2)}v_{j,1}}{||\widehat{O}
	_{1}^{(2)}v_{j,1}||}\right\Vert ^{2}\right] ^{1/2}  \label{eq:vtrue} \\
=& \left[ \left\Vert \frac{v_{i,1}}{||v_{i,1}||}-\frac{v_{j,1}}{||v_{j,1}||}%
\right\Vert ^{2}+\left\Vert \frac{v_{i,1}}{||v_{i,1}||}-\frac{v_{j,1}}{
	||v_{j,1}||}\right\Vert ^{2}\right] ^{1/2}=2.
\end{align}

Given \eqref{eq:v} and \eqref{eq:vtrue}, the result of Theorem \ref%
{thm:kmeans} is a direct consequence of \citet[Theorem \rom{2}.3]{SWZ20}. In
particular, we only need to verify their Assumption 4 holds with $c_{1n}=2$,
$c_{2n}=5C_{1}^{-1/2}C_{H,v}\eta _{n}$, and $M=2$. Note when $c_{F}$ is
sufficiently small,
\begin{equation*}
2(5C_{1}^{-1/2}c_{1}^{1/2}C_{H,v}\eta _{n})^{1/2}\leq 2\left[
5C_{1}^{-1/2}c_{1}^{1/2}C_{H,v}(c_{F}+c_{F}^{2})\right] ^{1/2}\leq
K_{1}^{3/4}\sqrt{2}.
\end{equation*}%
Then their Assumption 4 holds as
\begin{eqnarray*}
	(2c_{2n}c_{1}^{1/2}+16K_{1}^{3/4}M^{1/2}c_{2n}^{1/2})^{2} &\leq
	&(17K_{1}^{3/4}M^{1/2}c_{2n}^{1/2})^{2}=1734K_{1}^{3/2}C_{1}^{-1/2}C_{H,v}
	\eta _{n} \\
	&\leq &1734K_{1}^{3/2}C_{1}^{-1/2}C_{H,v}(c_{F}+c_{F}^{2})\leq 2c_{1}
\end{eqnarray*}
when $c_{F}$ is sufficiently small. ${\tiny \blacksquare }$

\section{Some Technical Lemmas}

\label{sec:lems}

\begin{lem}
	\label{lem:op} Let $C_{\Upsilon }$ be an sufficiently large and fixed
	constant. Suppose that the assumptions in Theorem \ref{thm:Frobeniusnorm}
	hold. Then
	\begin{equation*}
	\max_{l=0,1}||\Upsilon _{l}||_{op}\leq C_{\Upsilon }M_{W}(\sqrt{\zeta _{n}n}+%
	\sqrt{\log n})~w.p.a.1.
	\end{equation*}
\end{lem}


\noindent \textbf{Proof. } Let $\mathcal{C}=\{X_{i}\}_{i=1}^{n}\cup
\{e_{ij}\}_{1\leq i<j\leq n}$ and $r_{n}=C_{\Upsilon }M_{W}\sqrt{\log
	(n)\zeta _{n}n}$ for some sufficiently large constant $C_{\Upsilon }$ whose
value will be determined later. In addition, we augment the $n_{1}\times n$
matrix $\Upsilon _{l}$ to a symmetric $n\times n$ matrix $\overline{\Upsilon
}_{l}$ with $(i,j)$-th entry
\begin{equation*}
\overline{\Upsilon }_{l,ij}=%
\begin{cases}
\Upsilon _{l,ij} & \text{if}\quad i\in I_{1},j=1,\cdots ,n \\
\Upsilon _{l,ji} & \text{if}\quad j\in I_{1},i\in [ n]/I_{1} \\
0 & \text{if}\quad i\notin I_{1},j\notin I_{1}.%
\end{cases}%
\end{equation*}%
Then, by construction, $||\Upsilon _{l}||_{op}\leq ||\overline{\Upsilon }%
_{l}||_{op}$. Therefore,
\begin{align*}
\mathbb{P}(\max_{l=0,1}||\Upsilon _{l}||_{op}\geq r_{n})\leq & 2\max_{l=0,1}%
\mathbb{P}(||\Upsilon _{l}||_{op}\geq r_{n})\leq 2\max_{l=0,1}\mathbb{E}%
\left[ \mathbb{P}(||\Upsilon _{l}||_{op}\geq r_{n}|\mathcal{C})\right] \\
\leq & 2\max_{l=0,1}\mathbb{E}\left[ \mathbb{P}(||\overline{\Upsilon }%
_{l}||_{op}\geq r_{n}|\mathcal{C})\right] .
\end{align*}%
Next, we bound $\mathbb{P}(||\overline{\Upsilon }_{l}||_{op}\geq r_{n}|%
\mathcal{C})$. Recall $\mathcal{I}_{1}=\{(i,j)\in I_{1}\times
I_{1},j>i\}\cup \{(i,j):i\in I_{1},j\notin I_{1}\}.$ Given $\mathcal{C}$,
the only randomness of $\overline{\Upsilon }_{l}$ comes from $\{\varepsilon
_{ij}\}_{(i,j)\in \mathcal{I}_{1} \times [n]}$, which is an i.i.d. sequence
of logistic random variables. In addition, $\{\varepsilon _{ij}\}_{(i,j)\in
	\mathcal{I}_{1} \times [n]}$ is independent of $\mathcal{C}$,
\begin{equation*}
\tilde{\sigma}^{2}\equiv \max_{i\in [ n]}\mathbb{E}\left( \sum_{l=1}^{n}%
\overline{\Upsilon }_{l,ij}^{2}|\mathcal{C}\right) \leq \max_{i\in [
	n]}\sum_{j=1}^{n}\Lambda _{n,ij}M_{W}^{2}\leq \overline{c}M_{W}^{2}n\zeta
_{n}
\end{equation*}%
and $|\overline{\Upsilon }_{l,ij}|\leq M_{W}.$ Then, by
\citet[Corollary
3.12 and Remark 3.13]{BH16}, there exists a universal constant $\tilde{c}$
such that
\begin{equation*}
\mathbb{P}\left( ||\overline{\Upsilon }_{l}||_{op}\geq 3\sqrt{2}\tilde{\sigma%
}+t|\mathcal{C}\right) \leq n\exp \left( -\frac{t^{2}}{\tilde{c}M_{W}^{2}}%
\right) .
\end{equation*}%
Choosing $t=3\sqrt{\tilde{c}}M_{W}$, we have
\begin{equation*}
2\mathbb{P}\left( ||\overline{\Upsilon }_{l}||_{op}\geq 3M_{W}\sqrt{2%
	\overline{c}n\zeta _{n}}+3\sqrt{\tilde{c}\log (n)}M_{W}|\mathcal{C}\right)
\leq n^{-1.1},
\end{equation*}%
and thus,
\begin{equation*}
||\overline{\Upsilon }_{l}||_{op}\leq 3M_{W}(\sqrt{2\overline{c}n\zeta _{n}}+%
\sqrt{\tilde{c}\log (n)})\leq C_{\Upsilon }M_{W}(\sqrt{n\zeta _{n}}+\sqrt{%
	\log (n)})~w.p.a.1.\text{ }{\tiny \blacksquare }
\end{equation*}%
%
%
%
%

\begin{lem}
	\label{lem:e} Suppose $M\geq t\geq 0$, then Then $\exp (-t)+t-1\geq \frac{%
		t^{2}}{4(M \vee \log(2))}.$
\end{lem}


\noindent \textbf{Proof. }First, suppose $M\geq \log (2)$. Let $f(t)=\exp
(-t)+t-1-\frac{t^{2}}{4M}.$ Then, $f^{\prime }(t)=1-\exp (-t)-\frac{t}{2M}$.
We want to show $f^{\prime }(t)\geq 0$ for $t\in [ 0,M]$. This implies that $%
\min_{t\in [ 0,M]}f(t)=f(0)=0$. Note that
\begin{equation*}
f^{\prime }(M)=0.5-\exp (-M)\geq 0.
\end{equation*}%
In addition, we note that $f^{\prime }(t)$ is concave so that for any $t\in
[ 0,M]$,
\begin{equation*}
f^{\prime }(t)\geq \frac{f^{\prime }(M)t}{M}\geq 0.
\end{equation*}%
This leads to the desired result.

Next, suppose $M<\log(2)$. Then, we have
\begin{align*}
\exp (-t)+t-1 \geq \frac{t^2}{2} - \frac{t^3}{6} \geq \frac{(3 - \log(2))t^2%
}{6} \geq \frac{t^2}{4 \log(2)}.
\end{align*}

This concludes the proof. ${\tiny \blacksquare }$


\begin{lem}
	\label{lem:prelim} Suppose that the Assumptions in Theorem \ref%
	{thm:Frobeniusnorm} hold. Then, w.p.a.1,
	
	\begin{enumerate}
		\item $\max_{j\in I_{2}}||(\widehat{O}_{l}^{(1)})^{\top }\widehat{v}%
		_{j,l}^{(1)}||\leq 2M\sigma _{K_{l},l}^{-1};$
		
		\item There exist some constants $\infty >\overline{c}^{\prime }>\underline{c%
		}^{\prime }>0$ such that
		\begin{equation*}
		\overline{c}^{\prime }\zeta _{n}\geq \widehat{\Lambda }_{n,ij}\geq
		\underline{c}^{\prime }\zeta _{n}\quad \text{and}\quad \overline{c}^{\prime
		}\zeta _{n}\geq \tilde{\Lambda}_{n,ij}\geq \underline{c}^{\prime }\zeta _{n},
		\end{equation*}%
		where $\widehat{\Lambda }_{n,ij}$ and $\tilde{\Lambda}_{n,ij}$ are defined
		in \eqref{eq:lambdahat} and \eqref{eq:lambdatilde}, respectively. 
	\end{enumerate}
\end{lem}


\noindent \textbf{Proof. 1.} Note that
\begin{align*}
||(\widehat{O}_{l}^{(1)})^{\top }\widehat{v}_{j,l}^{(1)}||=& ||\widehat{v}%
_{j,l}^{(1)}||\leq \hat{\sigma}_{K_{l},l}^{-1}||\widehat{\Sigma }_{l}^{(1)}%
\widehat{v}_{j,l}^{(1)}|| \\
=& n^{-1/2}\hat{\sigma}_{K_{l},l}^{-1}\left\Vert [(\widehat{\mathcal{U}}%
_{l}^{(1)})^{\top }\widehat{\Theta }_{l}^{(1)}]_{\cdot j}\right\Vert \leq
n^{-1/2}\hat{\sigma}_{K_{l},l}^{-1}\left\Vert [\widehat{\Theta }%
_{l}^{(1)}]_{\cdot j}\right\Vert \leq 2M\sigma _{K_{l},l}^{-1},
\end{align*}%
where the first equality holds because $\widehat{O}_{l}^{(1)}$ is unitary,
the second equality holds because
\begin{equation*}
n^{-1/2}(\widehat{\mathcal{U}}_{l}^{(1)})^{\top }\widehat{\Theta }_{l}^{(1)}=%
\widehat{\Sigma }_{l}^{(1)}\sqrt{n}(\widehat{\mathcal{V}}_{l}^{(1)})^{\top
}\equiv \widehat{\Sigma }_{l}^{(1)}(\widehat{V}_{l}^{(1)})^{\top },
\end{equation*}%
the second inequality holds because $||\widehat{\mathcal{U}}%
_{l}^{(1)}||_{op} \leq 1$, and the last inequality holds because $|\widehat{%
	\Theta }_{l,ij}|\leq M$ by construction and that by Theorem \ref%
{thm:Frobeniusnorm} and the fact that $c_F$ is sufficiently small so that $%
48C_{F,1}\eta_n \leq \sigma_{K_l,l}/2$, and thus,
\begin{align*}
|\hat{\sigma}_{K_{l},l}^{-1} - \sigma _{K_{l},l}^{-1}| \leq \frac{|\hat{%
		\sigma}_{K_{l},l} - \sigma _{K_{l},l}|}{\sigma _{K_{l},l}(\sigma _{K_{l},l}
	- |\hat{\sigma}_{K_{l},l} - \sigma _{K_{l},l}|)} \leq \sigma^{-1}
_{K_{l},l}~w.p.a.1.
\end{align*}

As the constant $M$ does not depend on $j$, the result holds uniformly over $%
j=1,\cdots ,n$.

\textbf{2.} By Theorem \ref{thm:Frobeniusnorm} and the previous result,
\begin{equation*}
\left\vert \widehat{\tau }_{n}+\sum_{l=0}^{1}u_{i,l}^{\top }(\widehat{O}%
_{l}^{(1)})^{\top }\widehat{v}_{j,l}^{(1)}W_{l,ij}-\tau _{n}\right\vert \leq
\left\vert \widehat{\tau }_{n}-\tau _{n}\right\vert +\left\vert
\sum_{l=0}^{1}u_{i,l}^{\top }(\widehat{O}_{l}^{(1)})^{\top }\widehat{v}%
_{j,l}^{(1)}W_{l,ij}\right\vert \leq 30C_{F,1}\eta _{n}+C,
\end{equation*}%
and thus, there exist some constants $\infty >\overline{c}^{\prime }>%
\underline{c}^{\prime }>0$ such that
\begin{equation*}
\overline{c}^{\prime }\zeta _{n}\geq \widehat{\Lambda }_{n,ij}\geq
\underline{c}^{\prime }\zeta _{n}.
\end{equation*}%
For the same reason, we have $\overline{c}^{\prime }\zeta _{n}\geq \tilde{%
	\Lambda}_{n,ij}\geq \underline{c}^{\prime }\zeta _{n}.$ ${\tiny \blacksquare
}$

\begin{lem}
	\label{lem:phi} Suppose Assumptions \ref{ass:dgp}--\ref{ass:phi} hold.
	Recall that
	\begin{equation*}
	\widehat{\Phi }_{i}^{(1)}=\frac{1}{n_{2}}\sum_{j\in I_{2},j\neq i}%
	\begin{bmatrix}
	(\widehat{O}_{0}^{(1)})^{\top }\widehat{v}_{j,0}^{(1)} \\
	(\widehat{O}_{1}^{(1)})^{\top }\widehat{v}_{j,1}^{(1)}W_{1,ij}%
	\end{bmatrix}%
	\begin{bmatrix}
	(\widehat{O}_{0}^{(1)})^{\top }\widehat{v}_{j,0}^{(1)} \\
	(\widehat{O}_{1}^{(1)})^{\top }\widehat{v}_{j,1}^{(1)}W_{1,ij}%
	\end{bmatrix}%
	^{\top }.
	\end{equation*}%
	Then, for the constant $c_{\phi }$ defined in Assumption \ref{ass:phi},
	\begin{equation*}
	\min_{i\in I_{2}}\lambda _{\min }(\widehat{\Phi }_{i}^{(1)})\geq c_{\phi
	}/2~w.p.a.1.
	\end{equation*}
\end{lem}


\noindent \textbf{Proof. }By Lemma \ref{lem:prelim}(1), $||(\widehat{O}%
_{l,U}^{(1)})^{\top }\widehat{v}_{j,l}^{(1)}||\leq 2M\sigma _{K_{l},l}^{-1}$
for $l=0,1$. Then, we have, w.p.a.1,
\begin{align}
||\widehat{\Phi }_{i}^{(1)}-\Phi _{i}(I_{2})||\leq & \frac{4M}{n_{2}}%
\sum_{l=0}^{1}\sum_{j\in I_{2}}\sigma _{K_{l},l}^{-1}||(\widehat{O}%
_{l}^{(1)})^{\top }\widehat{v}_{j,l}^{(1)}-v_{j,l}||  \notag  \label{eq:phi}
\\
\leq & 4M\sum_{l=0}^{1}\sigma _{K_{l},l}^{-1}n_{2}^{-1/2}||\widehat{V}_{l}%
\widehat{O}_{l}^{(1)}-V_{l}||_{F}  \notag \\
\leq & 1088\sqrt{2}MC_{F,2}c_{\sigma }^{-1} \eta _{n},
\end{align}%
where the second inequality holds due to Cauchy's inequality, and the last
inequality holds due to Theorem \ref{thm:Frobeniusnorm}. As $c_F$ is
sufficiently small so that $1088\sqrt{2}MC_{F,2}c_{\sigma }^{-1} (c_F +
c_F^2) \leq c_{\phi }/2$, we have, w.p.a.1,
\begin{equation*}
\min_{i\in I_{2}}\lambda _{\min }(\widehat{\Phi }_{i}^{(1)})\geq \min_{i\in
	I_{2}}\lambda _{\min }(\Phi _{i}(I_{2}))-\left( 544(\sqrt{K_{0}}+\sqrt{K_{1}}%
)C_{\sigma }MC_{F,1}c_{\sigma }^{-3}\right) \eta _{n}\geq c_{\phi }/2\text{ }%
{\tiny \blacksquare }
\end{equation*}%
%
%
%
%

\begin{lem}
	\label{lem:q} Let $q_{in}$ be defined in (\ref{eq:qin}). Suppose that
	Assumptions \ref{ass:dgp}--\ref{ass:phi} hold. Then,
	\begin{equation*}
	\liminf_{n}\min_{i\in I_{2}}q_{in}\geq \frac{\sqrt{c_{\phi }/2}c_{\sigma }}{%
		4M(1+M_{W})}>0~w.p.a.1,
	\end{equation*}%
	where $\underline{c}$ and $M$ are two constants in Assumption \ref{ass:phi}
	and Lemma \ref{lem:prelim}, respectively.
\end{lem}


\noindent \textbf{Proof. }Note
\begin{equation*}
q_{in}\geq \inf_{\Delta }\sqrt{\frac{c_{\sigma }^{2}\frac{1}{n_{2}}%
		\sum_{j\in I_{2},j\neq i}((\widehat{\phi }_{ij}^{(1)})^{\top }\Delta )^{2}}{%
		16M^{2}(1+M_{W})^{2}||\Delta ||^{2}}}\geq \frac{c_{\sigma
	}\liminf_{n}\min_{i\in I_{2}}\lambda _{\min }(\widehat{\Phi }_{i}^{(1)})}{%
	4M(1+M_{W})}\geq \frac{\sqrt{c_{\phi }/2}c_{\sigma }}{4M(1+M_{W})}>0,
\end{equation*}%
where the first inequality is due to Lemma \ref{lem:prelim}(1) and the
second inequality is due to Lemma \ref{lem:phi}. ${\tiny \blacksquare }$

%

\section{Proof of Theorem \protect\ref{thm:infer}}

\label{sec:infer_pf} Theorem \ref{thm:infer} is the direct consequence of %
\citet[Theorem 1]{G17}. Note that Assumptions 1--3 in \cite{G17} hold in our
setup. Although \cite{G17} requires that $W_{l,ij}=g_{l}(X_{i},X_{j})$, his
proof remains valid if we have $W_{l,ij}=g_{l}(X_{i},X_{j},e_{ij})$ for some
i.i.d. random variable $e_{ij}$ such that $e_{ij}=e_{ji}$ and $e_{ij}\perp
\!\!\!\perp (X_{i},X_{j},\varepsilon _{ij})$. In addition, Assumption
4(i)-(ii) in \cite{G17} hold as we have $n\zeta _{n}=\Omega (\log n)$. His
Assumption 4(iii) is the same as our Assumption \ref{ass:hessian}.\ ${\tiny %
	\blacksquare }$

\section{Proof of Theorem \protect\ref{thm:e2}}

\label{sec:e2_pf} Let $B=\text{vech}(B^{\ast })+u(n^{2}\zeta _{n})^{-1/2}$
for some $\mathcal{K}\times 1$ vector $u$. Then, by the change of variables,
we have $\hat{u}=\sqrt{n^{2}\zeta _{n}}(\widehat{B}-\text{vech}(B^{\ast }))$
and
\begin{equation*}
\hat{u}=\argmax_{u}\left[ Q_{n}\left( \text{vech}(B^{\ast })+u(n^{2}\zeta
_{n})^{-1/2}\right) -Q_{n}(\text{vech}(B^{\ast }))\right] .
\end{equation*}%
We divide the proof into two steps. In the first step, we show that for each
$u$,
\begin{equation}
Q_{n}\left( \text{vec}(B^{\ast })+u(n^{2}\zeta _{n})^{-1/2}\right) -Q_{n}(%
\text{vec}(B^{\ast }))+\upsilon _{n}^{\top }u-\frac{u^{\top }\mathcal{H}u}{2}%
=o_{p}(1),  \label{eq:step1}
\end{equation}%
where $\upsilon _{n}=O_{p}(1)$ and $\mathcal{H}$ is positive definite. Then,
by noticing that $Q_{n}\left( \text{vec}(B^{\ast })+u(n^{2}\zeta
_{n})^{-1/2}\right) $ is convex in $u$, we can apply the convexity lemma of
\cite{P91} and conclude that
\begin{equation}
\hat{u}-\mathcal{H}^{-1}\upsilon _{n}=o_{p}(1).  \label{eq:uhat}
\end{equation}%
In the step second, we derive the asymptotic distribution of $\mathcal{H}%
^{-1}\upsilon _{n}$.

\textbf{Step 1.} By Taylor expansion,
\begin{align*}
& Q_{n}\left( \text{vec}(B^{\ast })+u(n^{2}\zeta _{n})^{-1/2}\right) -Q_{n}(%
\text{vec}(B^{\ast })) \\
=& -\frac{1}{\sqrt{n^{2}\zeta _{n}}}\sum_{1\leq i<j\leq n}(Y_{ij}-\Lambda
_{n,ij})\omega _{ij}^{\top }u+\frac{1}{2}u^{\top }\frac{1}{n^{2}\zeta _{n}}%
\sum_{1\leq i<j\leq n}\Lambda _{n,ij}(\tilde{u})(1-\Lambda _{n,ij}(\tilde{u}%
))\omega _{ij}\omega _{ij}^{\top }u \\
\equiv & -\upsilon _{n}^{\top }u+\frac{1}{2}u^{\top }\mathcal{H}_{n}u,
\end{align*}%
where $\Lambda _{n,ij}=\Lambda _{n,ij}(0)$, $\tilde{u}$ is between $0$ and $%
u $, and the definitions of $\upsilon _{n}$ and $\mathcal{H}_{n}$ are
evident. By Assumption \ref{ass:H}, $\mathcal{H}_{n}\overset{p}{%
	\longrightarrow }\mathcal{H}$. In addition, $\mathbb{E}\upsilon _{n}=\mathbb{%
	E}(\mathbb{E}(\upsilon _{n}|\omega _{ij}))=0$ and $\text{Var}(\upsilon
_{n})<\infty $, implying that $\upsilon _{n}=O_{p}(1)$. Therefore, we have
established \eqref{eq:step1}, and thus \eqref{eq:uhat}.

\textbf{Step 2.} $\mathcal{H}$ is positive definite by Assumption \ref{ass:H}%
. Noting that, $\{\varepsilon _{ij}\}_{1\leq i<j\leq n}\perp \!\!\!\perp
\{W_{1,ij}\}_{1\leq i<j\leq n}$, and $\{\varepsilon _{ij}\}_{1\leq i<j\leq
	n} $ is independent across $(i,j)$, we have
\begin{equation*}
\frac{1}{n^{2}\zeta _{n}}\mathbb{E}\left[ (Y_{ij}-\Lambda _{n,ij})^{2}\omega
_{ij}\omega _{ij}^{\top }|\{W_{1,ij}\}_{1\leq i<j\leq n}\right] =\frac{1}{%
	n^{2}\zeta _{n}}\sum_{1\leq i<j\leq n}\Lambda _{n,ij}(1-\Lambda
_{n,ij})\omega _{ij}\omega _{ij}^{\top }\overset{p}{\longrightarrow }%
\mathcal{H},
\end{equation*}%
and for any $\epsilon >0$, there exists $n_{0}$ sufficiently large so that
for all $n\geq n_{0}$ and $k\in [ \mathcal{K}]$,
\begin{equation*}
\frac{1}{n^{2}\zeta _{n}}\sum_{1\leq i<j\leq n}\mathbb{E}\left[
(Y_{ij}-\Lambda _{n,ij})^{2}\omega _{k,ij}^{2}\mathbf{1}\{|(Y_{ij}-\Lambda
_{n,ij})^{2}\omega _{k,ij}^{2}|\geq \sqrt{n^{2}\zeta _{n}}\epsilon \}\right]
\leq M_{W}^{2}\mathbf{1}\{M_{W}^{2}\geq \sqrt{n^{2}\zeta _{n}}\epsilon \}=0,
\end{equation*}%
where $\omega _{k,ij}$ denotes the $k$-th element of $\omega _{ij}$.
Therefore, by the Lindeberg-Feller\ central limit theorem, $\upsilon
_{n}\rightsquigarrow \mathcal{N}(0,\mathcal{H})$ conditionally on $%
\{W_{1,ij}\}_{1\leq i<j\leq n}.$ As $\mathcal{H}$ is deterministic, the
above weak convergence holds unconditionally too. Therefore, $\hat{u}%
\rightsquigarrow \mathcal{N}(0,\mathcal{H}^{-1})=O_{p}(1).$ In addition, by
Assumption \ref{ass:H},
\begin{equation*}
\frac{1}{n^{2}\zeta _{n}}\widehat{\mathcal{H}}_{n}=\frac{1}{n^{2}\zeta _{n}}%
\sum_{1\leq i<j\leq n}\Lambda _{n,ij}(\hat{u})(1-\Lambda _{n,ij}(\hat{u}%
))\omega _{ij}\omega _{ij}^{\top }\overset{p}{\longrightarrow }\mathcal{H}.
\end{equation*}%
It follows that $\widehat{\mathcal{H}}_{n}^{-1/2}(\hat{B}-$vec$(B^{\ast
}))\rightsquigarrow \mathcal{N}(0,I_{\mathcal{K}}).$\ ${\tiny \blacksquare }$

\section{Algorithm for the Nuclear Norm Regularization \label{sec:algo}}

We apply the optimization algorithm proposed in \cite{CDCB13} to obtain the
nuclear norm penalized estimator given in (\ref{eq:optimization}). For any
given $r_{l}\geq K_{l}$ and $r_{l}\leq n$, $\Gamma _{l}$ can be written as $%
\Gamma _{l}=U_{l}V_{l}^{\top }$, where $U_{l}\in \mathbb{R}^{n\times r_{l}}$
and $V_{l}\in \mathbb{R}^{r_{l}\times n}$, for $l=0,...,p$. We consider the
optimization problem:%
\begin{equation}
Q_{n}^{(1)}(\Gamma )+\frac{\lambda _{n}^{(1)}}{2}\sum_{l=0}^{p}\gamma
_{l}(||U_{l}||_{F}^{2}+||V_{l}||_{F}^{2}),  \label{EQ:bilinear}
\end{equation}%
where $\Gamma =(\Gamma _{l},l=0,...,p$), and
\begin{equation*}
Q_{n}^{(1)}(\Gamma )=\sum_{i\in I_{1},j\in \left[ n\right] ,i\neq j}\left[
-Y_{ij}(W_{ij}^{\top }\Gamma _{ij})+\log \{1+\exp (W_{ij}^{\top }\Gamma
_{ij})\}\right] ,
\end{equation*}%
subject to $\Gamma _{l}=U_{l}V_{l}^{\top }$ for $l=0,...,p$. Let $\lambda
_{n}^{(1)}=C_{\lambda }(\sqrt{\zeta _{n}n}+\sqrt{\log n})$.

Let $\Gamma _{l}^{\ast }$ for $l=0,...,p$ be an optimal solution of (\ref%
{eq:optimization}) with rank$(\Gamma _{l}^{\ast })=K_{l}^{\ast }$. \cite%
{CDCB13} shows that any solution $\Gamma _{l}=U_{l}V_{l}^{\top }$ for $%
l=0,...,p$ of (\ref{EQ:bilinear}) with $r_{l}\geq K_{l}^{\ast }$ is a
solution of \eqref{eq:optimizationf}. Next we apply the Augmented Lagrange
Multiplier (ALM) method given in \cite{CDCB13} to solve (\ref{EQ:bilinear}).
The augmented Lagrangian function of (\ref{EQ:bilinear}) is
\begin{equation*}
Q_{n}^{(1)}(\Gamma )+\frac{\lambda _{n}^{(1)}}{2}\sum_{l=0}^{p}\gamma
_{l}(||U_{l}||_{F}^{2}+||V_{l}||_{F}^{2})+\sum_{l=0}^{p}\left\langle \Delta
_{l},\Gamma _{l}-U_{l}V_{l}^{\top }\right\rangle +\frac{\rho }{2}%
\sum_{l=0}^{p}||\Gamma _{l}-U_{l}V_{l}^{\top }||_{F}^{2},
\end{equation*}%
where $\Delta _{l}$ are Lagrange multipliers and $\rho $ is a penalty
parameter to improve convergence.

\begin{enumerate}
	\item At step $m+1$, for given $(U_{l}^{m},V_{l}^{m},\Delta _{l}^{m},\Theta
	^{m},l=0,...,p)$, $(\Gamma ^{m+1})$ minimizes%
	\begin{equation*}
	L_{n}(\Gamma )=Q_{n}^{(1)}(\Gamma )+\sum_{l=0}^{p}\left\langle \Delta
	_{l}^{m},\Gamma _{l}-U_{l}^{m}V_{l}^{m\top }\right\rangle +\frac{\rho }{2}%
	\sum_{l=0}^{p}||\Gamma _{l}-U_{l}^{m}V_{l}^{m\top }||_{F}^{2}+C.
	\end{equation*}%
	Moreover, for $i\in I_{1},j\in \left[ n\right] ,i\neq j$,
	\begin{equation*}
	\frac{\partial L_{n}(\Gamma )}{\partial \Gamma _{l,ij}}=(\mu
	_{ij}-Y_{ij})W_{l,ij}+\Delta _{l,ij}^{m}+\rho (\Theta
	_{l,ij}-V_{l,ij}^{m\top }U_{l,ij}^{m}),
	\end{equation*}%
	where $\mu _{ij}=\exp (\sum_{l=0}^{1}W_{l,ij}\Gamma _{l,ij})\{1+\exp
	(\sum_{l=0}^{1}W_{l,ij}\Gamma _{l,ij})\}^{-1}$, and
	\begin{equation*}
	\frac{\partial ^{2}L_{n}(\Gamma )}{\partial \Gamma _{l,ij}^{2}}=\mu
	_{ij}(1-\mu _{ij})W_{l,ij}^{2}+\rho ,
	\end{equation*}%
	\begin{equation*}
	\frac{\partial ^{2}L_{n}(\Gamma )}{\partial \Gamma _{l,ij}\Gamma _{l^{\prime
			},ij}}=\mu _{ij}(1-\mu _{ij})W_{l,ij}W_{l^{\prime },ij}\text{, for }l\neq
	l^{\prime }
	\end{equation*}%
	For $i=j\in I_{1}$,
	\begin{equation*}
	\frac{\partial L_{n}(\Gamma )}{\partial \Gamma _{l,ij}}=\Delta
	_{l,ij}^{m}+\rho (\Gamma _{l,ij}-V_{l,ij}^{m\top }U_{l,ij}^{m}),
	\end{equation*}%
	$\frac{\partial ^{2}L_{n}(\Gamma _{0},\Gamma _{1})}{\partial \Gamma
		_{l,ij}^{2}}=\rho $ and $\frac{\partial ^{2}L_{n}(\Gamma _{0},\Gamma _{1})}{%
		\partial \Gamma _{l,ij}\Gamma _{l^{\prime },ij}}=0$. Then,
	\begin{equation*}
	\Gamma ^{m+1}=-(\frac{\partial ^{2}L_{n}(\Gamma ^{m})}{\partial \Gamma
		_{ij}\Gamma _{ij}^{\top }})^{-1}(\frac{\partial L_{n}(\Gamma ^{m})}{\partial
		\Gamma _{ij}})+\Gamma ^{m},
	\end{equation*}%
	where $\Gamma _{ij}=(\Gamma _{0,ij},...,\Gamma _{p,ij})^{\top }$. Update $%
	\Gamma _{l,ij}^{^{m+1}}=\Gamma _{l,ij}^{^{m+1}}I\{|\Gamma
	_{l,ij}^{^{m+1}}|\leq \log n\}+\log nI\{|\Gamma _{l,ij}^{^{m+1}}|>\log n\}$.
	
	\item For given $(U_{l}^{m},V_{l}^{m},\Delta _{l}^{m},\Gamma ^{m+1},l=1,2)$,
	$U_{l}^{m+1}$ minimizes%
	\begin{equation*}
	\frac{\lambda _{n}^{(1)}}{2}\sum_{l=0}^{1}\gamma
	_{l}(||U_{l}||_{F}^{2}+||V_{l}^{m}||_{F}^{2})+\sum_{l=0}^{1}\left\langle
	\Delta _{l}^{m},\Gamma _{l}^{m+1}-U_{l}V_{l}^{m\top }\right\rangle +\frac{%
		\rho }{2}||\Gamma _{l}^{m+1}-U_{l}V_{l}^{m\top }||_{F}^{2}+C.
	\end{equation*}%
	Then
	\begin{equation*}
	U_{l}^{m+1}=(\Delta _{l}^{m}+\rho \Gamma _{l}^{m+1})V_{l}^{m}(\lambda
	_{n}^{(1)}\gamma _{l}I_{r_{l}}+\rho V_{l}^{m\top }V_{l}^{m})^{-1}.
	\end{equation*}%
	Similarly, $V_{l}^{m+1}=(\Delta _{l}^{m}+\rho \Gamma _{l}^{m+1})^{\top
	}U_{l}^{m+1}(\lambda _{n}^{(1)}\gamma _{l}I_{r_{l}}+\rho U_{l}^{m+1\top
	}U_{l}^{m+1})^{-1}.$
	
	\item Let $\Delta _{l}^{m+1}=$ $\Delta _{l}^{m}+\rho (\Theta
	_{l}^{m+1}-U_{l}^{m+1}V_{l}^{m+1\top }).$
	
	\item Let $\rho =\min (\rho \mu ,10^{20}).$
\end{enumerate}

\bibliographystyle{chicago}
\bibliography{CD-1}

\end{document}